%% file: INTENTIONAL_R02.tex
\documentclass{elsarticle}

\usepackage{makecell}
\usepackage{booktabs} 
\usepackage{color}
\usepackage{MnSymbol}
\usepackage{graphicx}
\usepackage{amsfonts}
\usepackage[linesnumbered,ruled,vlined,commentsnumbered]{algorithm2e}
\usepackage{pdflscape}
\usepackage{array}
\usepackage{multirow}
\usepackage{xcolor,colortbl}
\usepackage{enumitem}

\usepackage{url}

\usepackage{pifont}

\renewcommand{\sf}[1]{\textsf{\textup{#1}}}              

\newtheorem{example}{Example}
\newtheorem{definition}{Definition}
\newcommand{\mycomment}[1]{}            

\newcommand{\patrick}[1]{\textcolor{red}{#1}}
\newcommand{\panos}[1]{\textcolor{blue}{#1}}
\newcommand{\stefano}[1]{\textcolor{green}{#1}}

\newcommand{\silence}[1]{}

\def\result#1{\emph{\color{blue} #1}}

\newcommand{\inlineRem}[1]{}

\newcommand{\rem}[1]{}

\DeclareGraphicsExtensions{.pdf,.jpg,.png}

\newcommand{\tuple}[1]{\ensuremath{\langle #1 \rangle}}

\begin{document}

\begin{frontmatter}
\title{Beyond Roll-Up's and Drill-Down's: An Intentional Analytics Model to Reinvent OLAP (long-version)}


\author[uoi]{Panos Vassiliadis}
\ead{pvassil@cs.uoi.gr}
\address[uoi]{University of Ioannina, Ioannina, Hellas}

\author[ut]{Patrick Marcel}
\ead{patrick.marcel@univ-tours.fr}
\address[ut]{University of Tours, Tours, France}

\author[uob]{Stefano Rizzi}
\ead{stefano.rizzi@unibo.it}
\address[uob]{University of Bologna, Bologna, Italy}

\begin{abstract}
This paper structures a novel vision for OLAP by fundamentally redefining several of the pillars on which OLAP has been based for the last 20 years. We redefine OLAP queries, in order to move to higher degrees of abstraction from roll-up's and drill-down's, and we propose a set of novel \emph{intentional} OLAP operators, namely, \textsf{describe}, \textsf{assess}, \textsf{explain}, \textsf{predict}, and \textsf{suggest}, which express the user's need for results. We fundamentally redefine what a query answer is, and escape from the constraint that the answer is a set of tuples; on the contrary, we complement the set of tuples with \emph{models} (typically, but not exclusively, results of data mining algorithms over the involved data) that concisely represent the internal structure or correlations of the data. Due to the diverse nature of the involved models, we come up (for the first time ever, to the best of our knowledge) with a unifying framework for them, that places its pillars on the extension of each data cell of a cube with information about the models that pertain to it -- practically converting the small parts that build up the models to data that annotate each cell. We exploit this data-to-model mapping to provide \emph{highlights} of the data, by isolating data and models that maximize the delivery of new information to the user. We introduce a novel method for assessing the \emph{surprise} that a new query result brings to the user, with respect to the information contained in previous results the user has seen via a new interestingness measure. 
The individual parts of our proposal are integrated in a new data model for OLAP, which we call the \emph{Intentional Analytics Model}. We complement our contribution with a list of significant open problems for the community to address. 

\end{abstract}

\end{frontmatter}

\input{01_intro}

\input{10_overview}

\input{23_dashboards}
\input{30_interestingness}

\input{22_intentional}

\newpage
\input{50_experiments}

\newpage
\input{60_RW}
\input{70_conclusions}



\bibliographystyle{elsarticle-num}
\bibliography{biblio}

\newpage
\section*{APPENDIX}
\input{91_data-queries-model}

\input{92_algos-generatingData-model}

\input{93_models}

\input{94_highlights-dashboards}

\input{95_operators}


\end{document}

%% file: 01_intro.tex
\section{Introduction and overview}
How will business intelligence (BI) look like 10 years from now? What foundations should
academia build in order to \emph{rigorously} support the building
of tools, the optimization of OLAP sessions, and the training of
data scientists around a logical paradigm? In this paper, we revisit the foundations of OLAP in
an attempt to address the aforementioned questions.

To start with, it is worth to shortly revisit the evolution of analytical querying so far.
\begin{itemize}
  \item At the beginning of time, people would be working with relational queries and recordsets returned by these queries. This treatment of BI was very DBMS-oriented, as the focus of attention was on what the DBMS can do for the users \cite{DBLP:conf/icde/GrayBLP96}.  

  \item Then, both the scientific community and the industry understood that it is possible to 
  simplify the life of the business users, by providing a simpler view of the data to them, and hiding the complexities of the underlying database. So, users would deal (on-line) with cubes, rather than with traditional database data, which gives a very elegant simplification of the data to the user, as all the joins and aggregations are taken care by the system. The operators would be cube-oriented, one level of abstraction above the database operators and would involve so-called OLAP operators such as roll-ups, drill-downs, etc. (see for example  \cite{DBLP:conf/dawak/RomeroA07}).\footnote{The term \emph{abstraction} here does not imply that we move from the logical level of handling data to e.g., the conceptual or the goal level. The multidimensional modeling of data via cubes and lattice hierarchies is still a logical model of data. The term 'abstraction' is justified, though, as it practically provides a "neat", simplified representation of the data, independently of the complexities of their underlying structure and storage, which are hidden from the user.  }

  \item Rapidly, apart from simply querying data, efforts focused on facilitating easier ways to navigate the multidimensional data space. Research proposed advanced operators permitting \emph{discovery-driven analysis} via combinations of OLAP primitives  \cite{DBLP:conf/vldb/Sarawagi99, DBLP:conf/vldb/Sarawagi00,DBLP:conf/vldb/SatheS01}. More recently, different strategies in database exploration have been proposed as well, like keyword search over databases, presenting example tuples to infer the query, etc.
\end{itemize}
BI is now becoming more and more pervasive, which entails an increasing participation in the decision-making process of users with competence in the business domain but low ICT skills. This requires further investigation to provide users with even more effective and user-centered paradigms for analytical querying. As a further step in this direction, in this work we envision a new data model for OLAP, called \emph{Intentional Analytics Model}\footnote{To resolve any ambiguity with respect to the usage of the term 'model', we provide the following terminological clarifications. The Intentional Analytics Model is a \textbf{data model} in the sense of Tedd Codd's Turing Award speech \cite{DBLP:journals/cacm/Codd82}(a specification of structures, operations and constraints), and the long tradition of the database community, that produced the relational, object-relational, multidimensional, semi-structured data models. The Intentional Analytics Model is yet another data model in this series; thus, and we will employ the term 'data model' for it. Apart from the aforementioned terms, in all our deliberations, the term \textbf{model} when appearing without any other characterization, refers to concise representations of knowledge for the data, summarizing patterns and insights, that are typically results of KDD algorithms.}. Here, data are accompanied by knowledge insights and both of them are considered as first class citizens of the data model. Indeed, the user explores the information space by submitting \emph{intentions of information goals}, i.e., why she wants to discover relevant information rather than prescriptions of what data she needs, and receives both \emph{data and annotations of highly interesting subsets of them} as results. Under the hood, the intentions are mapped to traditional OLAP operations and knowledge discovery algorithms. In a sense, our data model can be seen as a particular case of database exploration that takes advantage of OLAP primitives and cubes to support higher-level data analysis.

\subsection{The revolution of Intentional Analytics}

In our Intentional Analytics Model we redefine what a query is, with respect to both what users ask the system, what the answer entails, and how this answer is computed:


\begin{itemize}
  \item \textbf{What a query is}. We start by redefining what a query is, by replacing the traditional query definition of \emph{which data are needed} with a specification of \emph{why} we need to explore the data space, so that query formulation is performed in a way that is closer to the users' analytical goals. Specifically, we propose to replace traditional OLAP operators with \emph{intentional operators}; this means that, instead of operating with cubes in terms of roll-ups and drill-downs, users will state their analytical goals over the cubes as intentions. For example, instead of saying ``drill down to store city'' or ``roll-up to product category'', the user might ask ``explain the drop in the sales of this product family'' or "assess whether the sales in a particular region are abnormal or not". 
  \item \textbf{What the answer to a query is}. We   argue we can no longer remain with plain cubes as the answers to queries. If we want to replace simple query answering with insight gaining, the answer to a query cannot be just data -- even if they are nicely packaged via fancy visualizations of textual storytelling. We believe that the answer to  an intentional query is a \emph{dashboard},  including (a) one or more cubes with the appropriate  visualization and data narrations,  (b) concise representations of knowledge hidden in the data, possibly obtained through automated mining of \emph{models} and patterns (e.g., via decision trees or regressions) and (c) \emph{highlights}, i.e., significant ``jewels'' hidden in the result that highlight parts of the data and model spaces that give significant insights to the user's intention.
  \item \textbf{Highlight mining}. Assuming that each intentional operation is accompanied by a set of knowledge extraction algorithms like outlier detection, regressions, correlations of measures and attributes, decision trees and other similar operations, one particular aspect of profound importance is how do we decide which of the models produced and which subset of the data and model space in particular is really standing out as a highlight. To assess the importance of findings, we assess them in terms of their \emph{significance}, using a \emph{subjective} interestingness measure which follows the framework of \cite{DBLP:conf/kdd/Bie11} for pattern exploration. 
  
\end{itemize}

%% file: 10_overview.tex
\subsection{The vision in a nutshell}\label{sec:overview}
We assume a typical OLAP setting \cite{DBLP:series/synthesis/2010Jensen} defined on a multidimensional space with \emph{cubes} holding the information for analysts and \emph{dimensions} providing a context for facts. This is especially
important if combined with the fact that dimension values come in
\emph{hierarchies} of \emph{levels}; therefore, every single fact can be
simultaneously placed in multiple hierarchically structured
contexts, providing thus the ability to analyze sets of facts from
multiple perspectives. The underlying \emph{data~sets} include \emph{measures} that are characterized with respect to these dimensions. \emph{Cube queries} involve measure aggregations at specific levels of granularity per dimension, along with filtering of data for specific values of interest. For a formal treatment of the data model of our approach, one can refer to Appendix of this paper (practically extending the data model of \cite{DBLP:journals/is/GkesoulisVM15}).

In our vision, an \emph{OLAP session} is a sequence of \emph{dashboards}  that the analyst sees, each with its own information, including data, charts and informative summaries of KPI performance. The sequence is produced by the actions of the analyst that changes the contents of the dashboard by requesting more information on the basis of a set of \emph{operations} made available to him by the tool. 

The main idea behind the transitions between the states of  session, which is obtained via the user operations, is that we move from a concrete data model of \emph{logical} operators like roll-up's and drill down's, to an \emph{intentional} data model where the user
expresses, in terms of operators, high-level requirements like
"explain a certain phenomenon", "predict the future values" and
these high-level requirements have to be \emph{automatically}
translated to specific OLAP and Data Mining algorithms
that will carry the answer. This can also facilitate greatly the
extraction of highlights, as the user's goal is explicitly stated
to the system.

\textbf{Intentional operators}. \result{In contrast to other data models where a
	user operation would practically be a query (relational or
	multidimensional), in our data model, a user operation characterizes the
	intention of the user with respect to her information need}. 

\begin{example}\label{ex:cubeAdult1}
Observe the cube depicted in Table~\ref{fig:cubeAdult}. This is a cube computed over a detailed data set on working hours and depicts the weekly working hours of people that (a) work with pay and (b) have completed a post-secondary education, grouped by their education level and work class. The columns of the result pertain to the values of the dimension education that demonstrated at the top row, and the rows pertain to the values of the dimension work class that are demonstrated at the leftmost columns. 

\begin{table}[ht]
\begin{center}
\begin{tabular}{|l||cccc|}
\hline 
Weekly Hrs       & Assoc & Post-grad & Some-college & University \\
\hline
\hline 
Gov      & 40.73 & 43.58     & 38.38        & 42.14      \\
Private  & 41.06 & 45.19     & 38.73        & 43.06      \\
Self-emp & 46.68 & 47.24     & 45.7         & 46.61   \\
\hline
\end{tabular}
\end{center}
\caption{Example of a cube $C^O$ that serves as the starting point of a user operation \label{fig:cubeAdult}}
\end{table}

The user studies this cube in a dashboard and has several opportunities to ask a subsequent query. We list the options that our data model equips the user with:

\begin{itemize}
\item A first remark the user makes can be that the observed table presents information in adequate detail with respect to the education categories but fails to \emph{\textsf{Describe}} in adequate detail the information with respect to the work class. Changing the level of detail or the focus of the presented information answers the question \emph{"Give me a different description of \textbf{what} the data tell us!"}.
\item A second possible exploration concerns the answer to the question posed to the system \emph{"Now that I know the situation, can you \emph{\textsf{Assess}} \textbf{how good the situation is} compared to a reference benchmark?"}. For example, the analyst might want to know how is the current status assessed when compared to the previous 10 years, or compared to "similar" countries, or equally interesting, how is the situation  assessed when compared to the goals that the state has put with respect to the working hours of people.
\item "\textbf{Why} is the situation as it is now?" Can you \emph{\textsf{Explain}} why things are in the current status? Is the number of working hours correlated to the educational level (observe the monotonicity with respect to the work class -- each row is increasing compared to its rows above it)? Or maybe it is correlated to a hidden variable? 
\item "\textbf{Henceforth} how will the situation be?" Can you \emph{\textsf{Predict}} how things will be in the near future? Are there regression or timeseries analysis models that can be employed to tell us what the future status will be, based on the current data?
\end{itemize}

\end{example}

\textbf{Dashboards as answers to queries}.  The states of a session are \emph{dashboards}. In the current state of practice, a dashboard is a pre-designed collection of charts and performance summaries, based on the results of several OLAP queries that are executed over the underlying data. \emph{The novelty of our proposal is founded on the idea that a dashboard, being the result of a user operation, includes (a) the data that answer the queries of a dashboard, (b) models that are concise representations of knowledge about these data, either extracted via machine learning algorithms or infused by the analyst in the form of KPI's, measure correlations or rules, and (c) highlights, which are important subsets of the knowledge and data artifacts that particularly address the user's intention}. The entire result is appropriately visualized and accompanied by textual descriptions (see \cite{DBLP:journals/is/GkesoulisVM15}  for a larger discussion on data narration). Therefore \textcolor{blue}{\emph{the dashboard, which is the ultimate answer to a user operation, replaces query answering by insight gaining, via the appropriate enrichment of query results with knowledge, annotations of importance and appropriate packaging}}.
In order to construct a dashboard, we envision several computations taking place. Here is the sequence of the performed actions:

\begin{enumerate}
	\item First, the queries of the state's dashboard are issued and their results, the \emph{generating data of the dashboard}, are computed. Any straightforward computations for extra, derived
	columns of the dashboard (e.g., $gain=price-cost$) are performed too.
	\item Then, the available data are fed to model-extraction algorithms for the
	computation of \emph{models} that abstract, summarize and provide patterns and insights for
	the data.  
	\item The potentially large amount of data and models computed has to be ranked and assessed on their interestingness for the analyst; the most important findings are classified as the dashboards \emph{highlights} to be used for providing the main insights and the main directions for future transitions by the analyst.
	\item The above are accompanied by visualization, text construction and reporting tasks that aim the process of understanding and communicating the main findings.
\end{enumerate}


\textbf{Models in a dashboard}. Whereas speedometers and charts are the current state of practice in the area of BI, our vision extends beyond that. The \emph{automatic} assessment and critical characterization of the presented data will be part of the {BI} of the near-future. See some simple cases based on the example of observing sales data of an international company:
\begin{itemize}
	\item Sales data will be \emph{automatically} characterized with respect to a decision tree that classifies them (e.g., as "successful", "risky", "potentially hazardous" etc).
	\item Sales per country will be \emph{automatically} clustered to reveal similarities and differences, as a first step towards understanding outliers and non-expected behavior.\rem{Rev.1.PDF: Like longview Brown and Ulrike Fischer. WTF?}
	\item Aggregate sales over significant periods will be fed into time series analysis and forecasting methods to \emph{automatically} detect trends, seasonalities and to deduce future values.
\end{itemize}
We consider the plugging of data analysis algorithms in the back-stage of a
dashboard as an indispensable part of {BI}.
These algorithms
can range from very simple ones (e.g., finding the top values of
a cuboid, or detecting whether a dimension value is systematically
related to top or bottom sales) to very complicated ones (like,
for example, outlier detection, dimensionality reduction, etc).
Most importantly, as the operation of the algorithms will likely
be as transparent as possible to the end user, their execution will
require an almost automatic tuning of their parameters. The
findings of these algorithms will be \emph{models} of the data
that are typically (not always) used to \emph{annotate the
	existing data} with characterizations and offer
\emph{focus~points} to the visualization of the dashboard
(forecasts, outliers, dimension values that dominate top or bottom
measures, $\ldots$). The models themselves give a multitude of results. However, some of these results indicate that a part of a dashboard's data are of important \emph{interestingness} value to the end user. Due to that, we  collectively refer to the
important results of the execution of these algorithms as \emph{highlights},
in an attempt to show that the aim is to enrich the current
data-intensive dashboards with knowledge that is worth exploring
or using for decision making.

\begin{example}\label{ex:overviewEx}
Assume now that the Ministry of Labor, based on the data of a previous census, has set-up some goals for the improvement of labor. Assume also that per combination of education type and workplace, a specific goal, say $Weekly~Working~Target$, has been assigned, saying that if the average number of weekly hours at work is in the area of [40-55] for any category, then this is $Expected$ behavior, whereas any other amount outside this domain is either $Low$, or $Excessive$. This is exactly what business analysts call a \emph{Key Performance Indicator (KPI)}

Then, assume that the analyst of the Ministry wishes to evaluate the situation based on these goals, and in fact, in more details than the aggregate summary of Example~\ref{ex:cubeAdult1}. The analyst issues the composition of two commands:\\
\newline
\noindent \textsf{Describe} the data of $C^O$ in more details by workplace;\\
\noindent \textsf{Assess} \emph{Hours~Per~Week} using \emph{Weekly~Working~Target}.\\

\noindent The results are then depicted in the cube $C^N$ of Table~\ref{tab:KPI}. The system has automatically performed the following actions for the analyst:
\begin{itemize}
	\item First, \textcolor{blue}{the necessary data are retrieved from the underlying database} and the new cube, say $C^N$ is computed. This is practically a drill-down, in traditional OLAP terminology. The data are depicted in the first three columns of Table~\ref{tab:KPI} (intentionally in non-pivot form for reasons to be made obvious right away).
	\item Second, \textcolor{blue}{the KPI, which is a very illustrative example of a model, assesses the data} by labeling them according to the measure values of the new cube $C^N$ (column "Assessment"). \emph{Observe that every cell of the cube is mapped to the respective value of the model!}
    \item Finally, in an effort to discover interesting part of the new cube, \textcolor{blue}{an interestingness assessment is performed}, in an attempt to answer the question: "what is really surprising for the user?" To this end, a simple discrepancy model is used to split data based on the label assigned, as illustrated by the two antagonistic components displayed in the right-most part of Table \ref{tab:KPI}. The highlight selection algorithm of the system selects the first of the two right-most columns, thus marking the cells with assessment $Low$ as the most interesting.
\end{itemize}



\begin{table}[tbh]
\centering
{\scriptsize 
\begin{tabular}{llll|c|l|c|l|c|}
\cline{5-5} \cline{7-7} \cline{9-9}
                                                    &                                       &                            &  & Assessment &  & Discrepancy &  & Discrepancy \\ \cline{1-3} \cline{5-5} \cline{7-7} \cline{9-9} 
\multicolumn{1}{|l|}{\multirow{6}{*}{Assoc}}        & \multicolumn{1}{l|}{Federal-gov}      & \multicolumn{1}{l|}{41.15} &  & Expected   &  & 0           &  & 1           \\ \cline{2-3} \cline{5-5} \cline{7-7} \cline{9-9} 
\multicolumn{1}{|l|}{}                              & \multicolumn{1}{l|}{Local-gov}        & \multicolumn{1}{l|}{41.33} &  & Expected   &  & 0           &  & 1           \\ \cline{2-3} \cline{5-5} \cline{7-7} \cline{9-9} 
\multicolumn{1}{|l|}{}                              & \multicolumn{1}{l|}{State-gov}        & \multicolumn{1}{l|}{39.09} &  & Low        &  & 1           &  & 0           \\ \cline{2-3} \cline{5-5} \cline{7-7} \cline{9-9} 
\multicolumn{1}{|l|}{}                              & \multicolumn{1}{l|}{Private}          & \multicolumn{1}{l|}{41.06} &  & Expected   &  & 0           &  & 1           \\ \cline{2-3} \cline{5-5} \cline{7-7} \cline{9-9} 
\multicolumn{1}{|l|}{}                              & \multicolumn{1}{l|}{Self-emp-inc}     & \multicolumn{1}{l|}{48.68} &  & Expected   &  & 0           &  & 1           \\ \cline{2-3} \cline{5-5} \cline{7-7} \cline{9-9} 
\multicolumn{1}{|l|}{}                              & \multicolumn{1}{l|}{Self-emp-not-inc} & \multicolumn{1}{l|}{45.88} &  & Expected   &  & 0           &  & 1           \\ \cline{1-3} \cline{5-5} \cline{7-7} \cline{9-9} 
\multicolumn{1}{|l|}{\multirow{6}{*}{Post-grad}}    & \multicolumn{1}{l|}{Federal-gov}      & \multicolumn{1}{l|}{43.86} &  & Expected   &  & 0           &  & 1           \\ \cline{2-3} \cline{5-5} \cline{7-7} \cline{9-9} 
\multicolumn{1}{|l|}{}                              & \multicolumn{1}{l|}{Local-gov}        & \multicolumn{1}{l|}{43.96} &  & Expected   &  & 0           &  & 1           \\ \cline{2-3} \cline{5-5} \cline{7-7} \cline{9-9} 
\multicolumn{1}{|l|}{}                              & \multicolumn{1}{l|}{State-gov}        & \multicolumn{1}{l|}{42.96} &  & Expected   &  & 0           &  & 1           \\ \cline{2-3} \cline{5-5} \cline{7-7} \cline{9-9} 
\multicolumn{1}{|l|}{}                              & \multicolumn{1}{l|}{Private}          & \multicolumn{1}{l|}{45.19} &  & Expected   &  & 0           &  & 1           \\ \cline{2-3} \cline{5-5} \cline{7-7} \cline{9-9} 
\multicolumn{1}{|l|}{}                              & \multicolumn{1}{l|}{Self-emp-inc}     & \multicolumn{1}{l|}{53.05} &  & Expected   &  & 0           &  & 1           \\ \cline{2-3} \cline{5-5} \cline{7-7} \cline{9-9} 
\multicolumn{1}{|l|}{}                              & \multicolumn{1}{l|}{Self-emp-not-inc} & \multicolumn{1}{l|}{43.39} &  & Expected   &  & 0           &  & 1           \\ \cline{1-3} \cline{5-5} \cline{7-7} \cline{9-9} 
\multicolumn{1}{|l|}{\multirow{6}{*}{Some-college}} & \multicolumn{1}{l|}{Federal-gov}      & \multicolumn{1}{l|}{40.31} &  & Expected   &  & 0           &  & 1           \\ \cline{2-3} \cline{5-5} \cline{7-7} \cline{9-9} 
\multicolumn{1}{|l|}{}                              & \multicolumn{1}{l|}{Local-gov}        & \multicolumn{1}{l|}{40.14} &  & Expected   &  & 0           &  & 1           \\ \cline{2-3} \cline{5-5} \cline{7-7} \cline{9-9} 
\multicolumn{1}{|l|}{}                              & \multicolumn{1}{l|}{State-gov}        & \multicolumn{1}{l|}{34.73} &  & Low        &  & 1           &  & 0           \\ \cline{2-3} \cline{5-5} \cline{7-7} \cline{9-9} 
\multicolumn{1}{|l|}{}                              & \multicolumn{1}{l|}{Private}          & \multicolumn{1}{l|}{38.73} &  & Low        &  & 1           &  & 0           \\ \cline{2-3} \cline{5-5} \cline{7-7} \cline{9-9} 
\multicolumn{1}{|l|}{}                              & \multicolumn{1}{l|}{Self-emp-inc}     & \multicolumn{1}{l|}{49.31} &  & Expected   &  & 0           &  & 1           \\ \cline{2-3} \cline{5-5} \cline{7-7} \cline{9-9} 
\multicolumn{1}{|l|}{}                              & \multicolumn{1}{l|}{Self-emp-not-inc} & \multicolumn{1}{l|}{44.03} &  & Expected   &  & 0           &  & 1           \\ \cline{1-3} \cline{5-5} \cline{7-7} \cline{9-9} 
\multicolumn{1}{|l|}{\multirow{6}{*}{University}}    & \multicolumn{1}{l|}{Federal-gov}      & \multicolumn{1}{l|}{43.38} &  & Expected   &  & 0           &  & 1           \\ \cline{2-3} \cline{5-5} \cline{7-7} \cline{9-9} 
\multicolumn{1}{|l|}{}                              & \multicolumn{1}{l|}{Local-gov}        & \multicolumn{1}{l|}{42.34} &  & Expected   &  & 0           &  & 1           \\ \cline{2-3} \cline{5-5} \cline{7-7} \cline{9-9} 
\multicolumn{1}{|l|}{}                              & \multicolumn{1}{l|}{State-gov}        & \multicolumn{1}{l|}{40.82} &  & Expected   &  & 0           &  & 1           \\ \cline{2-3} \cline{5-5} \cline{7-7} \cline{9-9} 
\multicolumn{1}{|l|}{}                              & \multicolumn{1}{l|}{Private}          & \multicolumn{1}{l|}{43.06} &  & Expected   &  & 0           &  & 1           \\ \cline{2-3} \cline{5-5} \cline{7-7} \cline{9-9} 
\multicolumn{1}{|l|}{}                              & \multicolumn{1}{l|}{Self-emp-inc}     & \multicolumn{1}{l|}{49.91} &  & Expected   &  & 0           &  & 1           \\ \cline{2-3} \cline{5-5} \cline{7-7} \cline{9-9} 
\multicolumn{1}{|l|}{}                              & \multicolumn{1}{l|}{Self-emp-not-inc} & \multicolumn{1}{l|}{44.44} &  & Expected   &  & 0           &  & 1           \\ \cline{1-3} \cline{5-5} \cline{7-7} \cline{9-9} 
\end{tabular}
}
\caption{Assessment with KPI \label{tab:KPI}}
\end{table}

\end{example}


\subsection{Contribution and outline}
The main contribution of this paper is that it structures a vision for the BI of the near future in terms of a data model, the Intentional Analytics Model, with novel concepts and operators. We aim our definitions to be broad enough, yet as precise as possible; at the same time, we want to link them as much as possible to the intentional nature of the next generation of BI tools, where the end-user requests information at a very high level and the system transforms these requests to concrete execution of algorithms in order to compute, visualize and comment data and important highlights among them as an answer to the information request made by the end-user.

\begin{enumerate}
\item \textcolor{blue}{We redefine what an OLAP query is} and place particular emphasis to the introduction of high-level intentions as the pillar of querying. We propose several \textcolor{blue}{intentional operators} addressing fundamental informational needs, like describe, assess, explain, predict and suggest to replace the existing data-centric state of the art operators like roll-up and drill down. 
\item \textcolor{blue}{We redefine what a query answer is} and we \textcolor{blue}{complement data with models} to produce, along with visualizations and textual commentaries (not covered in this paper), dashboards as answers to user queries.
\item As part of this fundamental change of what a query answer is, we address the problem of integrating an extensible, heterogeneous sets of information models (ranging from simple correlations, to clusters and decision trees) in a uniform framework. Similarly to prediction cubes \cite{DBLP:conf/vldb/ChenCLR05}, this is achieved \textcolor{blue}{by extending each cell of a cube with both data and model information that pertain to it} -- practically converting information on models (the members of each cluster, the paths of a decision tree, the expected values of a regression formula) to data that annotate each cell. This data-to-model mapping is proved very powerful in that it allows the information of models to be treated as part of each cell, independently of the model type that generated it.
\item \textcolor{blue}{We facilitate the comparison of alternative models in terms of their interestingness} via this integrated framework. We propose a simple method for assessing the significance of each model --practically, the surprise it brings to the user-- that is built upon the data-to-model mapping. Hence, we are able to compute highlights, independently of the model types used. 
\end{enumerate}
\silence{
Specifically:
\begin{enumerate}
\item We introduce an intentional analytics model for OLAP based on the concepts of dashboards and intentions, and propose 5 basic intention operators.
\item As a proof-of-concept for our model we provide a grammar for our intention operator, plus a list of logical operators and related highlights possibly corresponding to each intention operator.
\item We define a metric for assessing the significance of highlights, to be used for cost-based optimization of intention operators.
\end{enumerate}
Remarkably, our intentional language can be seen as a first step towards addressing some  of
the open challenges proposed by the research community like, e.g., in
\cite{DBLP:conf/sigmod/IdreosPC15}, namely {\em the lack of
declarative exploration languages to present and reason about
popular navigational idioms},
or the various challenges raised by  \cite{eichmann2016towards}
 around the benchmarking of interactive data exploration by measuring user's gain in terms of insights.
}

\noindent \textbf{Outline}. The paper is structured as follows. In Section~\ref{sec:state} we present background OLAP concepts and cube queries and 
we complement them with  fundamental concepts of our method, specifically, models and highlights. 
In Section~\ref{sec:interestingness}, we present our method for interestingness assessment and highlight selection. 
We present the intentional operators in Section~\ref{sec:transitions}. Related work is surveyed in Section~\ref{sec:RW}. We conclude with open roads for future work in Section~\ref{sec:conclusions}.

\mycomment{
\textbf{Why  bother?} }



%% file: 23_dashboards.tex
\section{Data, Models, Model Components, Highlights and Dashboards}\label{sec:state}
In this Section, we detail the fundamental concepts behind our proposal.  We believe that the traditional understanding of the multidimensional data model is not adequate any more, and the emphasis of this paper is on its extension with models, highlights and intentional operators; therefore, we confine ourselves to presenting a simplified version of the data model in this section, and refer the interested reader to the Appendix for its thorough formal definition. 

In our approach, each state of an OLAP session in the Intentional Analytics Model is a dashboard the user sees. A dashboard is ultimately based on the generating data provided by a finite collection of intentional queries, posed to the underlying database.  However, \textcolor{blue}{a sharp distinction from previous approaches is that we do not restrict ourselves to data but enrich them with a set of interesting findings, which come in terms of \emph{models}, i.e., results of data mining or machine learning algorithms applied over the data of a dashboard, and significant annotations of data with reference to the components of these models, to which we refer as \emph{highlights}}.

\subsection{Data, cubes and cube queries}
In this subsection we provide a concise formal background for modeling hierarchies, cubes, and queries. 

The following list provides a fundamental terminology for the subsequent discussions
\begin{itemize}
	\item We assume an environment structured as a multidimensional space. We assume that \emph{dimensions} provide a \emph{context} for facts \cite{DBLP:series/synthesis/2010Jensen}. A dimension is practically the active domain of attributes for facts, that is internally hierarchically structured. 
	\item Each dimension comes with a \emph{hierarchy of levels}. Each dimension (e.g., $StoreGeography$ or $SalesDate$)is a lattice of levels (e.g., $City$, $Prefecture$, $Country$, or $Day$, $Week$, $Month$, $Year$). Each level comes with an active domain of values and there are hierarchical mappings between values (e.g., the ancestor of city \textsf{Paris} at the country level is \textsf{France}). Domains have identifier attributes as well as other properties (e.g., a city can have population, surface, geolocation, etc). Being lattices, all hierarchies start with a common, lowest possible level of coarseness and, all its paths end up at a common highest level of coarseness (holding a single value, \textsf{all}).
	\item Facts are structured in \emph{cubes}. A cube is defined with respect to several dimensions, fixed at specific levels and also includes a number of \emph{measures} to hold the measurable aspects of its facts. Each record of a cube, also known as \emph{cell} is a point in the multidimensional space of the cube's dimensions hosting a set of measures. A \emph{detailed cube} is a cube having all its dimensions fixed at the lowest possible level. Cubes may also be enriched via derived measures, computed by applying functions (e.g., $profit$ is a derived measure computed as $price * qty - cost$).
	\item A \emph{subcube} is a subset of a cube derived by selecting a set of cells from a cube via a selection filter.
	\item A \emph{cube query} is a cube too, specified by (a) the detailed cube over which it is imposed, (b) a selection condition that isolates the
	facts that qualify for further processing, (c) the grouping levels, which determine the coarseness of the result, and
	(d) an aggregation over some or all measures of the cube that accompanies the grouping levels in the final result.
\end{itemize}

\begin{example}\label{ex:cubeAdultDS}
	Consider the detailed cube for the well known Adult (a.k.a census income) dataset referring to data from 1994 USA census. There are 8 dimensions (Age, Native Country, Education, Occupation, Marital status, Work class,  Race and Gender) in the  data set and a single measure, Hours per Week. Each dimension comes with a lowest possible level, which we denote as $L_0$. This detailed data set will be the basis  of our running example. Formally this detailed cube is a function $DS$: Dom($Age.L_0) \times \ldots \times Dom(Gender.L_0) \rightarrow Dom(Hours~per~Week)$, of schema $\tuple{\{$Age$, $\ldots$, $Gender$\},\{$Hours per Week$\}}$. 

\end{example}
\begin{example}\label{ex:cubeAdult2}
	The following cube query produces the cube of Table \ref{tab:cubeAdult2}: 
	

	$C^N=\tuple{DS,$
		
		Education.L3='Post-secondary' and Work\_class.L2='With-Pay',
		
		$\tuple{$ALL,ALL,L2,ALL,L0,ALL,ALL$},$
		
		$Avg(Hours~per~Week)}$
		
	
	\noindent where 
    the selection condition fixes Education to ’Post- Secondary’ (at level L3), and Work to ’With-Pay’ (at level L2), data is grouped by Education at level 2, and Work at level 0, and the Avg of Hours per Week is requested. 

	\begin{table}[ht]
		\begin{center}
        	\scriptsize{
			\begin{tabular}{|p{2cm}||cccc|}
				\hline
				Weekly Hrs&	Assoc&	Post-grad&	Some-college&	University\\
				\hline
				\hline
				Federal-gov&	41.15&	43.86	&40.31&	43.38\\
				Local-gov&	41.33&	43.96&	40.14&	42.34\\
				State-gov&	39.09& 42.96&	34.73&	40.82\\
				Private& 41.06&	45.19&	38.73&	43.06\\
				Self-emp-inc& 48.68&	53.05&	49.31&	49.91\\
				Self-emp-not-inc& 45.88&	43.39&	44.03&	44.44\\
				\hline
			\end{tabular}
            } 
		\end{center}
		\caption{A new cube $C^N$ as the output of the cube query of Example~\ref{ex:cubeAdult2} \label{tab:cubeAdult2}}
	\end{table}
	
	For the reader familiar with OLAP terminology, the new cube $C^N$ resulting from the query, is practically the result of a Drill-Down operation over the old cube $C^O$ of Example~\ref{ex:cubeAdult1}.
\end{example}

\subsection{Models}
Models are concise, information-rich knowledge artifacts \cite{DBLP:conf/ssdbm/TerrovitisVSBCM04} that allow users to
\begin{itemize}
\item
compute or predict values for measures that widen the users' view on the situation presented by the observed data;
\item 
document a-priori known, or discovered relationships hiding in the data;
\item
annotate data with respect to their status, based on a labeling scheme.
\end{itemize}

The space of possible models is vast as they range from simple functions (e.g., \sf{grossSales}  = \sf{qty} * \sf{price}) and measure correlations (e.g., the application of Kendal correlation to the pair [\sf{avgDailyTemperature}, \sf{amtIceCreamSold}]) to more elaborate schemes such as decision trees, clustering, etc. \rem{RV1.PDF: Should be formalized more. It is unclear what can be a model and what not}

\subsubsection{Model Types}
To create models we rely on an extensible palette of model types. Model types are molds for individual models. Essentially, model types are meta-concepts, used in the same fashion as data types are used for models of attributes in the relational data model, or complex types for object-valued attributes in the object-relational data model. Following the traditional terminology, the models that abide by the mold of a model type are called its 'instances'. \rem{RV1.PDF: At least give more examples} 

\begin{definition}[Model Type]
A \emph{model type} is defined by 
(i) a \textsf{name}, 
(ii) a signature for its input, including (ii') a complex-type attribute \textsf{model~parameters} with model-dependent parameters,
(iii) a signature for its output, as a list of model components (to be defined next) including 
(iii') a complex-type attribute \textsf{model~characterization} with statistical characteristics of the entire model. 

The semantics of a model type is not formally represented but rather intuitively implied by its name; this also implies the algorithm to be executed for the computation of models. The term ``signature'' implies a structuring in a list of named attributes (if needed, of complex type). 
\end{definition}


Observe that the definition, apart from  structuring the input and the output takes into consideration two important aspects. Concerning the input, we want the input signature to host (a) the attributes and parameters that participate in the feeding of the algorithm's execution
(e.g., the algorithm and the distance used for clustering) as well as (b) the binding choices that we make (e.g., how many values we want for a top-k selection). Concerning the output, apart from providing a structure (see model components in the sequel), we also want to measure characteristics of the output, as for example, objective quality measures like ARI for clustering, ROC for prediction, etc. 



\subsubsection{Models: roles, taxonomy, usage} 

\textbf{Models as instantiations of model types}. A model is determined by applying a model type to a cube. This requires binding the attributes in the input and output signatures to specific levels and/or measures in the cube schema. 
For instance, a decision tree that receives a generic set of attributes as input and relates them to a labeling attribute via an output tree structure is a model type; a decision tree on cube \sf{LastYearCustomerPurchases} over [\sf{location}, \sf{age}, \sf{income}] that characterizes each fact using a labeling attribute \sf{purchaseHeight} with values \{low, med, high\} is a model.

\begin{definition}[Model]
	A \emph{model} is an instance of a model type, with (i) a binding to a cube $C$ over which it is imposed, (ii) a binding of the input signature of the model type to levels/measures of $C$ and constants (including binding the \textsf{model~parameters}), (iii) the population of the output of the model type with model components, along with the computation of the model's \textsf{model~characterization} with statistical characteristics.
\end{definition}


\begin{example}\label{ex:model}
	Consider the cube $C^N$ of Example \ref{ex:cubeAdult2}. Table \ref{tab:models} shows two models of two different model types over this cube. 
	The first model is of type Rank  and the second one is of type KPI. The input binding of the first model is 
	$\tuple{$Hours per Week$}$. The input binding of the second model is 
	$\tuple{\{\tuple{[0,40),Low},\tuple{[40-55),Expected},\tuple{[55-],Excessive}\},$Hours per Week$}$. 
	The population of the outputs of the two models correspond respectively to columns Rank and Assessment of the table.  \\ 
	
	\begin{table}[htb]
		\centering
		{\scriptsize
			\begin{tabular}{ll|c|l|c|l|c|}
				\cline{3-3} \cline{5-5} \cline{7-7}
				&                  & Weekly Hrs&  & Rank &  & Assessm.      \\ \cline{1-3} \cline{5-5} \cline{7-7} 
				\multicolumn{1}{|p{1cm}|}{\multirow{6}{*}{Assoc}}        & Federal-gov      & 41.15        &  & 17    &  & Expected \\ \cline{2-3} \cline{5-5} \cline{7-7} 
				\multicolumn{1}{|c|}{}                              & Local-gov        & 41.33        &  & 16    &  & Expected \\ \cline{2-3} \cline{5-5} \cline{7-7} 
				\multicolumn{1}{|c|}{}                              & State-gov        & 39.09        &  & 22    &  & Low      \\ \cline{2-3} \cline{5-5} \cline{7-7} 
				\multicolumn{1}{|c|}{}                              & Private          & 41.06        &  & 18    &  & Expected \\ \cline{2-3} \cline{5-5} \cline{7-7} 
				\multicolumn{1}{|c|}{}                              & Self-emp-inc     & 48.68        &  & 4     &  & Expected \\ \cline{2-3} \cline{5-5} \cline{7-7} 
				\multicolumn{1}{|c|}{}                              & Self-emp-not-inc & 45.88        &  & 5     &  & Expected \\ \cline{1-3} \cline{5-5} \cline{7-7} 
				\multicolumn{1}{|p{1cm}|}{\multirow{6}{*}{Post-grad}}    & Federal-gov      & 43.86        &  & 10    &  & Expected \\ \cline{2-3} \cline{5-5} \cline{7-7} 
				\multicolumn{1}{|c|}{}                              & Local-gov        & 43.96        &  & 9     &  & Expected \\ \cline{2-3} \cline{5-5} \cline{7-7} 
				\multicolumn{1}{|c|}{}                              & State-gov        & 42.96        &  & 14    &  & Expected \\ \cline{2-3} \cline{5-5} \cline{7-7} 
				\multicolumn{1}{|c|}{}                              & Private          & 45.19        &  & 6     &  & Expected \\ \cline{2-3} \cline{5-5} \cline{7-7} 
				\multicolumn{1}{|c|}{}                              & Self-emp-inc     & 53.05        &  & 1     &  & Expected \\ \cline{2-3} \cline{5-5} \cline{7-7} 
				\multicolumn{1}{|c|}{}                              & Self-emp-not-inc & 43.39        &  & 11    &  & Expected \\ \cline{1-3} \cline{5-5} \cline{7-7} 
				\multicolumn{1}{|p{1cm}|}{\multirow{6}{*}{Some-coll.}} & Federal-gov      & 40.31        &  & 20    &  & Expected \\ \cline{2-3} \cline{5-5} \cline{7-7} 
				\multicolumn{1}{|l|}{}                              & Local-gov        & 40.14        &  & 21    &  & Expected \\ \cline{2-3} \cline{5-5} \cline{7-7} 
				\multicolumn{1}{|l|}{}                              & State-gov        & 34.73        &  & 24    &  & Low      \\ \cline{2-3} \cline{5-5} \cline{7-7} 
				\multicolumn{1}{|l|}{}                              & Private          & 38.73        &  & 23    &  & Low      \\ \cline{2-3} \cline{5-5} \cline{7-7} 
				\multicolumn{1}{|l|}{}                              & Self-emp-inc     & 49.31        &  & 3     &  & Expected \\ \cline{2-3} \cline{5-5} \cline{7-7} 
				\multicolumn{1}{|l|}{}                              & Self-emp-not-inc & 44.03        &  & 8     &  & Expected \\ \cline{1-3} \cline{5-5} \cline{7-7} 
				\multicolumn{1}{|p{1cm}|}{\multirow{6}{*}{Univ.}}    & Federal-gov      & 43.38        &  & 12    &  & Expected \\ \cline{2-3} \cline{5-5} \cline{7-7} 
				\multicolumn{1}{|l|}{}                              & Local-gov        & 42.34        &  & 15    &  & Expected \\ \cline{2-3} \cline{5-5} \cline{7-7} 
				\multicolumn{1}{|l|}{}                              & State-gov        & 40.82        &  & 19    &  & Expected \\ \cline{2-3} \cline{5-5} \cline{7-7} 
				\multicolumn{1}{|l|}{}                              & Private          & 43.06        &  & 13    &  & Expected \\ \cline{2-3} \cline{5-5} \cline{7-7} 
				\multicolumn{1}{|l|}{}                              & Self-emp-inc     & 49.91        &  & 2     &  & Expected \\ \cline{2-3} \cline{5-5} \cline{7-7} 
				\multicolumn{1}{|l|}{}                              & Self-emp-not-inc & 44.44        &  & 7     &  & Expected \\ \cline{1-3} \cline{5-5} \cline{7-7} 
			\end{tabular}
		}
		\caption{Two models over cube $C^N$ \label{tab:models} }
	\end{table}
	
\end{example}

\textbf{Role and purpose}.  A model is \emph{a concise representation of some knowledge about the data}. This knowledge can be some relationship between data attributes, some property or characterization of subsets of data,  or some computed value over the existing data. At the same time, despite its conciseness, typically a model also serves as \emph{an enrichment of the underlying data} -- in other words, each record of the data can be extended,  annotated, or, in any case, enriched with extra information by the model. 


\textbf{A Taxonomy of models}. In typical Machine Learning terminology, a model is a concise description of a data set that tries to ``fit'' the data in an accurate and semantically rich way; as such, it is driven by the data. Yet, some relevant and concise description of data (e.g., a formula on how measures interrelate, or some rule-based KPI) may as well be part of the domain knowledge the user has. Like in \cite{DBLP:journals/ai/LombardiMB17}, we use the term \emph{model} in both cases. So, in our approach models come in two flavors: (a) \emph{user-driven} models, where it is the analyst who defines a model for the relationship/labeling of a set of attributes, based on her a-priori domain knowledge, and (b) \emph{data-driven} models, where the analyst requests from the system to extract a model from a specified cube. 


Regardless of flavors, a model enriches a cube in one or more of the following ways, \panos{to which we refer as \emph{intentions} in this paper}: 
\begin{enumerate}
	\item
	\emph{Description}: the model describes the relationship between levels/measure of the cube(s), or between facts, or between existing and newly computed facts (e.g., customers are clustered according to their purchase frequency);
	\item
	\emph{Assessment}: the model characterizes each fact, or an entire cube, typically by comparing it to a baseline (e.g., the overall sales of TVs of this month are ``disappointing'' with reference to the average of last 3 years); 
	\item
	\emph{Explanation}: the model gives an explanation for some relevant observation by concisely representing hidden relationships among the levels/measures of the cube(s) (e.g., the purchase amount of a customer is mainly determined by her age and income);
	\item
	\emph{Prediction}: the model forecasts cube facts (e.g., sales during next Christmas period are expected to be 10\% higher than last year).
	\item
	\emph{Suggestion}: the model suggests the next query(s) in the analysis using a recommendation strategy (e.g., users who did a similar assessment of sales of TVs then saw sales of TVs in the neighboring countries).
\end{enumerate}

In Figure~\ref{fig:modelTax}, we concisely detail how alternative models (both user and data driven) are grouped for different intentions. For the sake of space, the figure does not include the two complex-type attributes that are present in all model types, 
namely: the complex-type  \textsf{model parameters} attribute
is omitted from the $Input$ signature column,
and the complex-type \textsf{model characterization} attribute is omitted from the $Output$ signature column. The term \emph{Name~Of~Measure} practically refers to the fact that models are applied over measures and thus, a parameter of which measure is going to be used for the application of the model is necessary.

\begin{figure*}[htb]
\begin{center}
{\scriptsize 
\begin{tabular}{p{2.2cm}ll}
Name & Input signature & Output signature \\
\hline
&\textbf{Model types for description}  &\\
Top-k & $\tuple{$Number Of Values$,$Name of Measure$}$ & $\tuple{$Rank$}$\\
Outlier & $\tuple{$Threshold$,$Name of Measure$}$ & $\tuple{$Outlierness$}$\\
Clustering & $\tuple{$Number Of Clusters$,$Name of Measure$}$ & 
$\tuple{$Cluster$_1, \ldots, $Cluster$_n,$Representative$}$
\\
Shrink & $\tuple{$Number Of Cells$,$Name of Measure$}$ & $\tuple{$Cell$_1, \ldots, $Cell$_n}$\\
Dominating Slice & $\tuple{$Name of Measure$}$& $\tuple{$DomSlice$_1, \ldots, $DomSlice$_n}$
%
\\
&&\\
&\textbf{Model types for assessment}  &\\
KPI & $\tuple{\{$Labeling~Rules$\},$Name of Measure$}$ & $\tuple{$Assessment$}$ \\ 
Function-based Benchmark & $\{function~parameters\}$ & $\tuple{$Discrepancy$}$\\
&&\\
&\textbf{Model types for explanation} &\\
Correlation & $\tuple{$Threshold$,$Name of Measure$}$ & $\tuple{$Participation$}$\\
Regression &  $\tuple{$Threshold$,$Name of Measure$}$ & $\tuple{$Discrepancy$}$\\
Decision tree & $\tuple{\{\tuple{$Range,Label$}\},\{\tuple{$Attributes$}\},$Name of Measure$}$ & 
$\tuple{$Label$}$
\\
Statistical test &  $\tuple{$Threshold$,$Name of Measure$}$& $\tuple{$Discrepancy$}$\\
&&\\
&\textbf{Model types for prediction} &\\
Auto-Regression &  $\tuple{$Threshold$,$Name of Measure$}$& $\tuple{$Discrepancy$}$\\
Time Series Decomposition & $\tuple{\{$Thresholds$\},$Name of Measure$}$&$\tuple{$Trend$,$Seasonality$, $Noise$}$\\
&&\\
&\textbf{Model types for suggestion} &\\
Content-based &  $\tuple{$Number Of Queries$}$& 
$\tuple{$QueryID$}$
\\
Collaborative & $\tuple{$Number Of Queries$}$& $\tuple{$QueryID$}$\\
Hybrid & $\tuple{$Number Of Queries$}$& $\tuple{$QueryID$}$\\
\end{tabular}
}
\end{center}
	\caption{A grouping of model types, organized per intention.}\label{fig:modelTax}


\end{figure*}

\silence{
\begin{table}[h!]
	\caption{A classification of common model types}
	\begin{center} \footnotesize
		\begin{tabular}{r|m{2cm}m{1.3cm}m{1.8cm}m{1.8cm}m{1.8cm}|}
			\multicolumn{1}{r}{} & \emph{Describe} & \emph{Assess} & \emph{Explain} & \emph{Predict} & \multicolumn{1}{c}{\emph{Suggest}} \\ \cline{2-6}
			User-driven & \makecell[l]{function;\\shrink \cite{DBLP:journals/jdwm/RizziGG15}} & function; KPI; correlation & measure correlation & & log-based recommendation \cite{DBLP:conf/eda/MarcelN11} \\\cline{2-6}
			Data-driven & top-k values; dominating row/col; clustering; time series decomposition; outliers &  & function; top-k values; clustering; decision tree; regression & regression; time series analysis & profile \& expectation-based recommendation \cite{DBLP:conf/eda/MarcelN11} \\ \cline{2-6}
		\end{tabular}
	\end{center}
	\label{tab:modeltypes}
\end{table}

} 

\textbf{How to work with models}. In terms of usage, the way of working with models is as follows: 

\begin{enumerate}
\item \emph{Model construction or retrieval}. Model construction is the step dedicated to taking the input data and extracting or assigning an abstraction of the 	relationships hidden in them. Be it the assignment of a function that computes a new measure, the choice of a time series analysis algorithm that splits a time series measure to 3 new measures (trend, periodicity, noise), or, the construction of a decision tree over the particular cube, the construction of a model is a representation of the relationships between the involved attributes. 
	
\item  	\emph{Application of the model to the data}. An extra step, which we introduce as a particularity of our method, is \emph{the linkage of the model to the data}. Model application, is the step that computes, for each tuple of the input data, the output measures that the model type carries. \textcolor{blue}{Each input tuple is then practically extended with a set of output attributes pertaining to the model}.
	
	\item \emph{Highlight extraction}. Highlight extraction is the step that focuses the interest  of the user to a subset of the annotated data. By reusing the output of the model, \textcolor{blue}{highlight extraction algorithms can pick potential "hidden jewels" or \emph{highlights}} out of the vastness of available data, and decide which of them are more significant for the user
\end{enumerate}


Of course, the steps can be blended for optimization purposes; here we separate them to illustrate their role. In the subsequent subsections, we will elaborate more on the structure and role of the output of a model and its linkage to the data as well as on the issue of highlights. Before that, however, we would like to address the following important issue.

\emph{Can we automate parameter tuning and model invocation}? Tuning the parameters for the invocation and application of a model type can range from one extreme, where everything is specified by the user, to the other extreme, where predefined values exist for every possible parameter. A middle-ground alternative is to consider a dynamic generation of models and tuning of parameters depending on the properties (size, content, etc.) of the cube the model is applied to. A statistical test can be used to decide whether a given model fits the data of the cube; for instance, the Hopkins statistics can be used to check for clustering tendency \cite{DBLP:conf/fuzzIEEE/BanerjeeD04} and decide whether clustering is worth testing on the cube. We note that an entire field of research, called \emph{meta-learning}, is devoted to answering problems like how to choose a learning algorithm based on data characteristics \cite{DBLP:journals/air/LemkeBG15}.  Practically, automated machine learning frameworks and tools (like e.g., auto-sklearn \cite{DBLP:conf/nips/FeurerKESBH15}) can be used to automatically select an algorithm and its parameters for a given dataset, given a computation cost.

\subsubsection{Model Components} 

\textbf{Model Components}. The output of a model is of particular importance. The elements of the output of each model are called \emph{output model components}. Specifically, we require that the output obeys a signature, meaning that the output is always structured as a list of attributes (it can be just one), each pertaining to a different component.
Examples of output model components include:
\begin{itemize}
	\item  A time series splits each of its points to 3 measurements, specifically error, trend and seasonality (practically creating 3 times series in the place of one, whose sum reconstructs the original one). 
	\item A clustering scheme includes a set of clusters, each coming with a centroid, as well as with an indication, for each tuple, of its participation to a specific cluster.
	\item A classification decision tree includes a tree structure, best expressed as the composition of a set of paths, leading to characterization classes; again, each class comprises a set of tuples in the underlying cube that pertain to it. 
    \item An outlier includes an outlier strength measurement per cell.
    \item A top-k or a ranking components includes selecting the uppermost $k$ values and annotating the rest as of no interest, or, respectively, the ranking of all cells in terms of their measure value. 
\end{itemize}



\textbf{Data-to-model mappings}. 
\emph{Is it possible to uniformly handle the heterogeneity of different model types?} Clearly, a cluster is inherently different from a decision tree or the formula for a trend. Is there a unifying common ground to cover them all?
\textcolor{blue}{\emph{The unifying essence of all the plethora of diverse model types is  that all of them are annotations of the original data}}. At the end of the day, every component of a complex model type (be it a cluster id, a path in a decision tree and a resulting class, a characterization of the top-k tuples, or a trend formula): (a) refers to a subset of the input data and vice-versa, and, (b) refers to the overall model via a part-of relationship. So, once a model of the underlying data is available, our solution to the problem is to provide a distinct identity to the components of a complex model type (here: as a distinct attribute) and \emph{annotate or characterize the data with respect to the model component that pertains to them}.
In fact, this step can be blended within the model extraction itself. Examples of such annotations follow:
\begin{itemize}
	\item Assuming a time series model that splits a time series to $trend$, $seasonality$ and $noise$, these attributes can be appended to the generating data 	set.
	\item Assuming a cluster model, the generating data can be annotated with the $id$ of the cluster to which they
	belong.
	\item Assuming a classification model, the input data can be labeled via an extra attribute with respect to the class(es) of the model to which they
	belong.
	\item Assuming a model of top-k values of a measure, the input data can be annotated with their rank, and whether they belong in the top-k set or not.
\end{itemize}

The above observations allow us to provide a \textbf{data-to-model mapping}. \emph{A notable property of our modeling is that we require  model components to be directly mapped and linked to their generating data in a bidirectional mapping, so that the end-user can navigate back and forth between cube cells and their models.} 



\textbf{Antagonism}. In addition to the components in its output, it is possible that the binding of a model induces \emph{antagonistic components} that provide an assignment of the cells of the cube to different components.
Examples of antagonistic model components include:
\begin{itemize}
	\item one component for each cluster of a clustering, to identify the participation of each cell to the cluster, and one component to identify the representative(s) of the cluster (e.g., medoid);
    \item one component for the outliers and one component for the non-outliers, based on a threshold on outlier strength measurement;
    \item one component for the top-k cells and one component for the non-top-k cells.
\end{itemize}

More frequently than not, this assignment to groups is a partition, i.e., each cell belongs to exactly one component. However, there are exceptions, like for example fuzzy clustering or fuzzy labeling. 
The ability to provide this assignment of cells to antagonistic components is fundamental to facilitate the highlight selection process. Take for example the case where we split the cells of the cube on the basis of a top-5 model (i.e., we set $Number~of~Values = 5$ at the binding of the top-k model type) and we have two  components, (a) one containing the top-5 cells, and, (b) another with the rest of the non-top-5 cells. The highlight selection process will then select which of these two components bears the greatest amount of new information to the user. Thus, the term antagonistic is justified, as the respective components antagonize to provide the maximum amount of surprise to the user. The antagonists can be either (a) components produced directly as the output of the model extraction algorithm, or (b) components derived from the regular output to serve the purpose of highlight extraction. As an example of the former case, consider any labeling algorithm (KPI, decision tree, or other), which by definition separates the cells into groups with the same label producing one component per label, so that these different components can antagonize with each other on which is the actual highlight. As an example of the latter case, consider the case of top-k cells, where the output component $Rank$ is used to derive two antagonists: Top-k and Non-Top-k.


%

Practically, we can define a model component as follows.
\begin{definition}[Model Component]
	A \emph{model component} is a named attribute containing either  the result of a model construction algorithm, or produced internally, as an induced attribute to be derived for highlight selection. The extent of a model component depends on the nature of the model. A model component can be annotated with its statistical characterizations via a \textsf{component~characterization} attribute.
    
\end{definition}	
The statistical strength of each component (the number of cells being outliers, or the cohesion of a cluster) is different than the one of the entire model. Here, each component carries it own statistical characteristics.


\begin{example}\label{ex:components}
While Table \ref{tab:models} of Example \ref{ex:cubeAdult2}
shows output components of two models over the cube $C^N$,
Table \ref{tab:components} shows the facts of cube $C^N$  together with two antagonistic components of model  Top-5, respectively attribute Top-5 and attribute Non-top-5, with their extents. 

\begin{table}[htb]
\centering
{\scriptsize
\begin{tabular}{llll|c|l|c|}
\cline{5-5} \cline{7-7}
                                                    &                                       &                            &  & Top-5 &  & Non-top-5 \\ \cline{1-3} \cline{5-5} \cline{7-7} 
\multicolumn{1}{|l|}{\multirow{6}{*}{Assoc}}        & \multicolumn{1}{l|}{Federal-gov}      & \multicolumn{1}{l|}{41.15} &  & 0     &  & 1         \\ \cline{2-3} \cline{5-5} \cline{7-7} 
\multicolumn{1}{|l|}{}                              & \multicolumn{1}{l|}{Local-gov}        & \multicolumn{1}{l|}{41.33} &  & 0     &  & 1         \\ \cline{2-3} \cline{5-5} \cline{7-7} 
\multicolumn{1}{|l|}{}                              & \multicolumn{1}{l|}{State-gov}        & \multicolumn{1}{l|}{39.09} &  & 0     &  & 1         \\ \cline{2-3} \cline{5-5} \cline{7-7} 
\multicolumn{1}{|l|}{}                              & \multicolumn{1}{l|}{Private}          & \multicolumn{1}{l|}{41.06} &  & 0     &  & 1         \\ \cline{2-3} \cline{5-5} \cline{7-7} 
\multicolumn{1}{|l|}{}                              & \multicolumn{1}{l|}{Self-emp-inc}     & \multicolumn{1}{l|}{48.68} &  & 1     &  & 0         \\ \cline{2-3} \cline{5-5} \cline{7-7} 
\multicolumn{1}{|l|}{}                              & \multicolumn{1}{l|}{Self-emp-not-inc} & \multicolumn{1}{l|}{45.88} &  & 1     &  & 0         \\ \cline{1-3} \cline{5-5} \cline{7-7} 
\multicolumn{1}{|l|}{\multirow{6}{*}{Post-grad}}    & \multicolumn{1}{l|}{Federal-gov}      & \multicolumn{1}{l|}{43.86} &  & 0     &  & 1         \\ \cline{2-3} \cline{5-5} \cline{7-7} 
\multicolumn{1}{|l|}{}                              & \multicolumn{1}{l|}{Local-gov}        & \multicolumn{1}{l|}{43.96} &  & 0     &  & 1         \\ \cline{2-3} \cline{5-5} \cline{7-7} 
\multicolumn{1}{|l|}{}                              & \multicolumn{1}{l|}{State-gov}        & \multicolumn{1}{l|}{42.96} &  & 0     &  & 1         \\ \cline{2-3} \cline{5-5} \cline{7-7} 
\multicolumn{1}{|l|}{}                              & \multicolumn{1}{l|}{Private}          & \multicolumn{1}{l|}{45.19} &  & 0     &  & 1         \\ \cline{2-3} \cline{5-5} \cline{7-7} 
\multicolumn{1}{|l|}{}                              & \multicolumn{1}{l|}{Self-emp-inc}     & \multicolumn{1}{l|}{53.05} &  & 1     &  & 0         \\ \cline{2-3} \cline{5-5} \cline{7-7} 
\multicolumn{1}{|l|}{}                              & \multicolumn{1}{l|}{Self-emp-not-inc} & \multicolumn{1}{l|}{43.39} &  & 0     &  & 1         \\ \cline{1-3} \cline{5-5} \cline{7-7} 
\multicolumn{1}{|l|}{\multirow{6}{*}{Some-college}} & \multicolumn{1}{l|}{Federal-gov}      & \multicolumn{1}{l|}{40.31} &  & 0     &  & 1         \\ \cline{2-3} \cline{5-5} \cline{7-7} 
\multicolumn{1}{|l|}{}                              & \multicolumn{1}{l|}{Local-gov}        & \multicolumn{1}{l|}{40.14} &  & 0     &  & 1         \\ \cline{2-3} \cline{5-5} \cline{7-7} 
\multicolumn{1}{|l|}{}                              & \multicolumn{1}{l|}{State-gov}        & \multicolumn{1}{l|}{34.73} &  & 0     &  & 1         \\ \cline{2-3} \cline{5-5} \cline{7-7} 
\multicolumn{1}{|l|}{}                              & \multicolumn{1}{l|}{Private}          & \multicolumn{1}{l|}{38.73} &  & 0     &  & 1         \\ \cline{2-3} \cline{5-5} \cline{7-7} 
\multicolumn{1}{|l|}{}                              & \multicolumn{1}{l|}{Self-emp-inc}     & \multicolumn{1}{l|}{49.31} &  & 1     &  & 0         \\ \cline{2-3} \cline{5-5} \cline{7-7} 
\multicolumn{1}{|l|}{}                              & \multicolumn{1}{l|}{Self-emp-not-inc} & \multicolumn{1}{l|}{44.03} &  & 0     &  & 1         \\ \cline{1-3} \cline{5-5} \cline{7-7} 
\multicolumn{1}{|l|}{\multirow{6}{*}{Univesity}}    & \multicolumn{1}{l|}{Federal-gov}      & \multicolumn{1}{l|}{43.38} &  & 0     &  & 1         \\ \cline{2-3} \cline{5-5} \cline{7-7} 
\multicolumn{1}{|l|}{}                              & \multicolumn{1}{l|}{Local-gov}        & \multicolumn{1}{l|}{42.34} &  & 0     &  & 1         \\ \cline{2-3} \cline{5-5} \cline{7-7} 
\multicolumn{1}{|l|}{}                              & \multicolumn{1}{l|}{State-gov}        & \multicolumn{1}{l|}{40.82} &  & 0     &  & 1         \\ \cline{2-3} \cline{5-5} \cline{7-7} 
\multicolumn{1}{|l|}{}                              & \multicolumn{1}{l|}{Private}          & \multicolumn{1}{l|}{43.06} &  & 0     &  & 1         \\ \cline{2-3} \cline{5-5} \cline{7-7} 
\multicolumn{1}{|l|}{}                              & \multicolumn{1}{l|}{Self-emp-inc}     & \multicolumn{1}{l|}{49.91} &  & 1     &  & 0         \\ \cline{2-3} \cline{5-5} \cline{7-7} 
\multicolumn{1}{|l|}{}                              & \multicolumn{1}{l|}{Self-emp-not-inc} & \multicolumn{1}{l|}{44.44} &  & 0     &  & 1         \\ \cline{1-3} \cline{5-5} \cline{7-7} 
\end{tabular}
}
\caption{Two antagonistic components of model type Top-k over cube $C^N$ \label{tab:components}}
\end{table}
\end{example}

\textcolor{blue}{\emph{The ability to annotate each cell of a cube with respect to a model component is of extreme importance and the driving force behind our definition, that practically models components as attributes of the relational data model. The possibility of integrating a vast space of heterogeneous models via a simple and uniform representation, which also facilitates a data-to-model mapping as suggested in \cite{DBLP:series/anis/Pedersen09}, allows us to practically treat models as data too and use them for addressing the user's information needs!}}


\subsection{Highlights}
As already mentioned, the set of
\emph{highlights} of the dashboard is a set of \emph{important	findings} that accompany the dashboard. These can be findings of
any nature, e.g., important outliers in the contents of the dashboard's data, all the tuples belonging to a certain class of a classification
scheme, the top or bottom values of a measure, etc. 

\emph{What is interesting for the user, however?} Are there
universal notions (esp., formulae) for interestingness \cite{DBLP:journals/csur/GengH06}?
Should we personalize interestingness for each
user on the grounds of a profile? Maybe interestingness is defined
by what everyone else found interesting? Or maybe interestingness
is fundamentally dependent upon the combination of  data and the
original intention the user had when he queried the data? We are
	mainly driven by the last option, without, of course,
disqualifying the others and in full comprehension that there is
quite some research effort before crystallizing to a specific
stance on the problem. Again, the holy grail here is to fully
automate the proactive highlighting of data of interest for the
user.

Whenever the result of a new user query is computed and new data, cuboids if you will,  are acquired to answer the query, one or more models are automatically computed. \result{The fundamental idea of our approach is that, ideally, one of the antagonistic components of one of these models is the most adequate to respond to the intention of the user. The determination of the quality that discriminates the most appropriate component, which we call interestingness,  depends on several factors, including its relevance to the original intention, its novelty (to which extent it reveals new information that was previously unknown to the user) or its surprise (to which extent it contradicts previous beliefs of the user)}. For example, assume the user is assessing a measure (e.g., \textsf{Sales}) with respect to a benchmark (e.g., \textsf{lastYearSales}) and the model annotates each cell by its difference to the respective cell of the previous year and measures the $z~score$ \emph{of the difference} over the population of differences. Then, the model can produce 3 components based on the ranges of $z~score$, e.g., (a) up to $\sigma$, (b) between $1 \cdot \sigma$ and  $2 \cdot \sigma$, and (c)  higher than  $2 \cdot \sigma$. The last component and the cells pertaining to it (those with $z~score$ higher than  $2 \cdot \sigma$) constitute 
the highlights of the operation.

\result{The essence of highlight selection is therefore the identification of a specific model component that maximizes the interestingness of the information delivered to the user by it with respect to her original intention. The highlight is, then, the combination of the component and the data that refer to it}.


The generalization of the above intuition to more sophisticated selection criteria is possible, of course. So, whereas here we select the top-1 component with respect to its interestingness, one can imagine schemes where the top-k are selected, or any component that surpasses an interestingness threshold. As already mentioned, the possibilities for defining highlight selection criteria are open and subject to lots of future work. In any case, we assume that we have (a) a scoring function $interestingness()$ that returns an interestingness score for each of the components of $M$, and, (b) a criterion $\phi_H$ to determine which component(s) qualify for highlights.

Another particular aspect that plugs into highlight selection is the idea of digging out the essence of a component. As every component annotates \emph{all} the data of the input cube, assume now that we have a criterion for selecting its "core data". Assume that each  model type has a selection criterion for the core data of its components, which we denote with $MC.coreElements$ that returns the set of elements that mostly pertain to the intuitive essence of a component, along with their respective cube cells, obtained via the 1:1 data-to-model mapping, which we denote $MC.coreCells$. 
To give a couple of concrete examples, here is a short list:
\begin{itemize}
    \item Assume a clustering model with clusters being modeled as bitmaps, having 1 for the data that pertain to it and 0 for the rest. Then, the cells annotated with 1 comprise the essence of the component.
    \item Assume a decision tree, with classification paths being its components. Each path is also a bitmap of 1 and 0 depending on whether each cell pertains or not this path. The cells annotated with 1 in a path component are its essential, core cells.
    \item Alternatively, assume a classification scheme where the cube cells are assigned a particular label within a single-component model. Then, the core cells of the component are the ones that abide by the criterion used ("expected" values if the criterion is non-outlierness, "low" or "high" values if the criterion is outlierness.
\end{itemize}

So overall, depending on the model type $T$ that induces a structure for its output components and the criterion $\phi_H$, for each component we can compute a subset of the core elements, as well as the respective core cells that define the essence of the component. Notably, in the case where all components are bitmaps (which is always the case for a component with a discrete domain of values), the computation of the core elements and cells is independent of the criterion $\phi_H$.



\begin{definition}[Highlight]
Given an intentional query $q$ issued over a cube $C^{O}$ of a dashboard, and resulting to a new cube $c$, a new model $M$ over cube $c$, with components $M.MC_1, \ldots, M.MC_k$, a criterion for highlight selection $\phi_H$ (on the basis of a component scoring function $interestingness$), then, a \emph{highlight} $h$ is a tuple including (a) $M.MC_I$, (b) the elements of $M.MC_I$ that are qualified as its core elements, and (c) the data of the new cube $C^{N}$ that pertain to it, via $MC_I.coreCells$, fulfilling the 1:1 mapping between (b) and (c) -- i.e., a triplet $h$ = $\{MC_I$, $MC_I.coreElements$, $MC_I.coreCells\}$ 
\end{definition}



In other words, there are many facets of a highlight: it is the most "interesting" component, and, at the same time, the data that pertain to it. 

\subsection{Packaging it all in a dashboard, or, what the answer to a query really is}

Having detailed models, model components and highlights, it is now time to integrate them into the big picture. We define dashboards as collections of cubes, coming along with their models and components. We refer to such a cube as an \emph{enhanced cube}. \rem{R1.PDF: 'Not in the def.' Unclear which of the two, possibly refers to enhanced cubes, never being mentioned before, or to the fact that dashboards have not been dealt with as sets of enhanced cubes. Maybe emph enhanced cubes inthe text before? But after we 're done with the formalities}
\begin{definition}[Enhanced Cube]
The triple of a cube $C$, its (set of) models $\mathbf{M}$, and its highlights is called an \emph{enhanced cube}. 
\end{definition}

\begin{definition}[Dashboard]
A \emph{dashboard} is a set of enhanced cubes.
\end{definition}

A dashboard comes with visualizations, generation of textual commentaries and possibly automatic compilation of reports. We consider the inherent integration of these aspects into the overall framework as part of the future work and refer the interested reader to \cite{DBLP:conf/dolap/VassiliadisM18} for a first discussion and \cite{DBLP:journals/is/GkesoulisVM15} for a review of related work.

%% file: 30_interestingness.tex
\section{Highlight Selection via a new Interestingness Measure}\label{sec:interestingness}



In this section, we present our approach for 
selecting significant highlights based on interestingness
assessment. We start by motivating the subjective interestingness measure we adopt for significance. We then detail the highlight selection algorithm. 
Finally, we reuse our running example to conclude the section with a concrete case of highlight selection.

\subsection{Interestingness measure}
Exploratory Data Analysis (EDA) \emph{aims to provide insights to users by presenting them with human-digestible pieces of "interesting" information about the data}~\cite{DBLP:conf/ida/Bie13}.
One particular EDA activity, Exploratory Data Mining (EDM), and mainly pattern-mining, has been particularly active in developing Interestingness Measures (IM) to filter the large set of artifacts resulting of the mining.
In this context, interestingness has been quantified mainly in an objective way, in the sense that measures are agnostic of variations among users.  
Only recently has the concept of subjective interestingness been formalized, first as a quantification of unexpectedness relying on the concept of a belief system. This belief system formalizes the beliefs of the data miner, to which mining artifacts are contrasted to determine their interestingness \cite{DBLP:conf/kdd/PadmanabhanT98}. 
In other words, the quantification of the subjective interestingness of an artifact is the result of its confrontation with a \emph{background model} related to the \emph{user's prior knowledge}.

We believe that, in our context of user-centered analytical querying, interestingness must be defined subjectively, in the same vein as it is for EDM. However, to the best of our knowledge, no such concept of subjective interestingness exists in the general landscape of Interactive Database Exploration \cite{DBLP:conf/sigmod/IdreosPC15}. In this paper we introduce a completely novel scheme for interestingness assessment, which (a) exploits the fact that the user is involved in sessions of OLAP intentional queries (thus, previously seen results constitute prior beliefs) to offer the desired subjectivity, (b) since exploration takes place within a strictly structured information space (i.e., the multidimensional space defined by hierarchies), let us relate these previously seen cells to the new ones, and (c) quite importantly, takes the models derived into consideration, while at the same time staying model-independent. The core of our framework is the assessment of \emph{surprise}, which is practically the difference of belief for the same subset of the multidimensional space, before and after an intentional query has been issued.

In our context, a user starts with a cube $C^O$, expresses an intention $q$ (e.g., Describe, Assess, etc.) that triggers the computation of new cube $C^N$, together with a set of $k$ models $M^1, \ldots, M^k$, each of them with their components,  $MC_1^1, \ldots, MC_m^k$. 
Instead of interactively presenting one artifact extracted from the data chosen after its confrontation to prior knowledge, we are interested in presenting one of the antagonistic components among $MC_1^1, \ldots, MC_m^k$, chosen as interesting after its confrontation with $C^O$. Therefore we measure surprise using an interestingness measure that quantifies for each antagonistic component the surprise brought by cube $C^N$ to cube $C^0$, as detailed next. The component with the highest surprise score is chosen to highlight the cells of $C^N$.



\subsection{Principle of highlight selection}
\begin{figure}[ht]
\includegraphics[width=\linewidth]{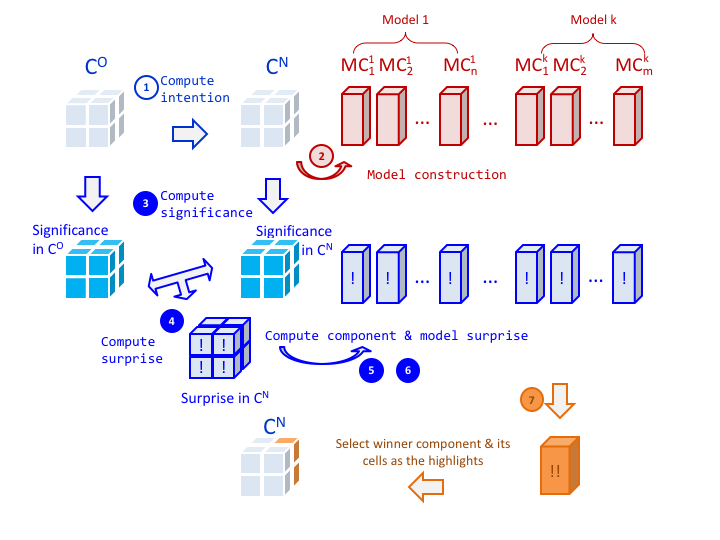} 
\caption{Principle of highlight selection \label{fig:highlightSection}}
\end{figure}

This generic principle of significance computation and highlight selection is depicted in Figure \ref{fig:highlightSection} and detailed in Algorithm \ref{algo:highlightSection}.
As displayed in Figure~\ref{fig:highlightSection}, this principle consists of the following steps. First, the intentional query is evaluated to get the new cube $C^N$ from $C^O$. Second, the models associated with the intention are instantiated and the components $MC_1^1, \ldots, MC_m^k$ are obtained. Subsequently, a significance score is computed locally for each of the two cubes (step 3), and these scores are contrasted, resulting in a surprise score for each cell of  $C^N$ (step 4). Then, each component among  $MC_1^1, \ldots, MC_m^k$ aggregates the surprise scores for the cells pertaining to it, resulting in a surprise score for this component (step 5). In a similar vein, we aggregate the component scores to compute surprise scores for entire models (Step 6). Finally, the component that maximizes the score is chosen for highlighting the cells of $C^N$ that pertain to it (step 7). 


\begin{algorithm}[ht]
 \KwData{\\
 cube $C^O$\\
 cube $C^N$\\
 set of models $\{M^1, \ldots, M^k\}$ with their components $MC_1^1, \ldots, MC_m^k$\\ 
 n:m  relationship $proxies$ between the cells of $C^O$ and $C^N$  \\
 function $significance$ for computing the significance of a cell\\
 function ${\cal D}$ for surprise characterization\\
 function ${\cal A}^{C}$ for computing component scores\\
 function ${\cal A}^{M}$ for computing model scores}
 \KwResult{\\
 a component among $MC_1^1, \ldots, MC_m^k$} 
 \For{all  cells $c^O$ in $C^O$}{
 compute $c^O.significance$
 }
 \For{all  cells $c^N$ in $C^N$}{
 compute $c^N.significance$
 }
 \For{all  cells $c^N$ in $C^N$}{
 compute $c^N.surprise={\cal D}(c^N.significance,c^N.proxies.significance)$
 }
 \For{all  components in $MC_1^1, \ldots, MC_m^k$}{
 compute $MC_j^i.surprise={\cal A}^{C}_{c^N\in MC_j^i}(c^N.surprise)$
 }
 \For{all  models in $M^1, \ldots, M^k$}{
 compute $M^i.surprise={\cal A}^{M}_{MC_j^i\in M^i}(MC_j^i.surprise)$ 
 }
 return the component maximizing $MC_j^i.surprise$\;
 \caption{Highlight selection \label{algo:highlightSection}}
\end{algorithm}

Algorithm \ref{algo:highlightSection} describes the highlight selection procedure. 
The algorithm  receives as input the two cubes $C^O$ and $C^N$, a set of model components $MC_1^1$, $\ldots$, $MC_m^k$ over $C^N$, and a set of functions depending on the intention. These functions include: (a) a relation $proxies$ that relate the cells of $C^N$ with the ones of $C^O$ (e.g., an ancestor relationship if the logical operation triggered by the intention is a drill-down), (b) a \emph{significance} function for assessing how significant a cell is (note that this function may use one measure value or more), (c) a function $\cal{D}$ for computing the surprise between two significance scores, and (d) two functions ${\cal A}^C$ and ${\cal A}^M$ for aggregating surprise scores by model components and models, respectively.
A few notes are due here:
\begin{itemize}
\item The relation $proxies$ is of $N$:$M$ nature to facilitate arbitrary relationships of old and new cells that can be produced via selections, roll-up's, drill-down's, or similar operations. In the typical scenario, the relationship will be annotated with a '1' in one of its two ends, facilitating easier computations of the subsequent steps. 
\item The $significance$ score is of great importance as it characterizes each cell with an objective importance (which is necessary in order to get the algorithm going). In our example that follows, significance is measured as outlierness, computed via a z-score; however, it is easy to conceive alternative objective significance scores, like, for example, the inverse ("typicality" if you will) to serve different intentions, the very same value of the measure, its rank, or others. Our framework is open-ended from this aspect for plugging more significance measures.   
\item The $surprise$ is the core of our method. Surprise exploits the fact that old and new cells are related via $proxies$, as they represent the same subspace of the entire multidimensional space. Thus, we can exploit this fact and contrast the prior and new belief (i.e., measure). For the rare case where the $proxies$ relationship is not straightforwardly defined for some cell, special care must be taken to contrast the cell's significance to some representative significance value  of the proxy cube (e.g., the mean or the average significance). Practically, this means that if there is no straightforward set of proxy cells (e.g., ancestor or descendants), we map the new cell to the entire previous cube. In the typical case, the ${\cal D}$ function can be a simple subtraction (as we envision it and use it here), or, again in an open-ended vein, another contrasting function (for example, one might consider the overall tendency of values in a subcube: if all values are increasing due to a trend, then the surprise s much less).
\item The aggregation functions ${\cal A}^{C}$ and ${\cal A}^{M}$ can be any desirable aggregate function. In our example that follows we use the average value, however, sum, max, min, or other can be used too.
\end{itemize}

The algorithm unveils as follows. In lines 1-2 (respectively 3-4), a score is computed for each cell of $C^O$ (respectively $C^N$) using the $significance$ function. In lines 5-6, a surprise score is computed for each cell of $C^N$, by contrasting their significance score with the score(s) in their relatives in $C^O$. In line 7-8, a surprise score is computed for each components, by aggregating the surprise scores of the cells participating to the component. In line 9-10, a surprise score is computed for each models, by aggregating the surprise scores of the components of this model. 



Before presenting a concrete example, we believe it is worth  presenting a summary of the novelty and merits of our framework:
\begin{itemize}
\item We offer \emph{an interestingness framework} that exploits the divine simplicity of the multidimensional data space of OLAP, as well as the existence of models along with the data -- and it is thus, \emph{appropriate for the new model of OLAP that we propose}.
\item We escape the trap of objective, non-contextualized interestingness measures and provide a \emph{subjective measure}, exploiting the transitions that OLAP queries via our operators offer, and based on the idea of prior belief (facts of the old cubes contrasted to the facts of new cubes). We base our approach to $surprise$ as the difference of belief for the same subset of the multidimensional space, \emph{obtained again by exploiting the relationship between old and new cells}.
\item We provide a method that exploits the data-to-model mappings of our model and is therefore \emph{independent of model types} (which we deem as a major feature of the framework)
\item We provide an \emph{open-ended framework} where new definitions for objective significance, subjective surprise, delta, and aggregate functions are always possible.
\end{itemize}
Note that we provide a \emph{bottom-up} method for computing highlights, starting from cells and ending up in models. The possibility of a \emph{top-down} method, is of course, an open issue, but falls outside the scope of this paper.


\begin{example}\label{ex:interestingness}
Recall cube $C^O$ of Example \ref{ex:cubeAdult1}. Assume the user issues the  intention  \textsf{with $C^O$ Describe Avg Working Hours by} giving more details for \textsf{workclass at the most detailed level}.
The first step in processing this intention results in evaluating the cube queries of Example \ref{ex:cubeAdult2} to drill-down to cube $C^N$, recalled below in Table \ref{tab:cn}.

The highlight selection algorithm is called with cubes $C^O$, $C^N$, a set of models and their model components and a set of functions. Regarding the models and model components, in this example, for the sake of brevity, we consider only two model components of two different models. They are displayed mapped on the cube $C^N$ in Table \ref{tab:cn}: (i) the top-5 cells (the 5 cells in yellow, note that this component is also presented in Example \ref{ex:components} in binary form) and (ii) the outliers greater than two standard deviation (the 2 cells in blue). The cell in green participates to both components. This example shows how the algorithm chooses among these two components.

The functions are as follows. The $proxies$ relationship allows to find in cube $C^O$ the ancestor of a cell of $C^N$.
The $significance$ function used to compute the significance of each cell is the z-score, i.e., the number of standard deviations the value of this cell is from the mean of all the values of the aggregate cube.
Function ${\cal D}$ is the difference and function ${\cal A}^C$ (respectively function ${\cal A}^M$) is the average.


\begin{table}[ht]
\begin{center}
\begin{tabular}{|l||cccc|}
\hline
$C^N$&	Assoc&	Post-grad&	Some-coll.&	Univ.\\
\hline
\hline
Federal-gov&	41.15&	43.86	&40.31&	43.38\\
Local-gov&	41.33&	43.96&	40.14&	42.34\\
State-gov&	39.09&	42.96&	\cellcolor{cyan}34.73&	40.82\\
Private&	41.06&	45.19&	38.73&	43.06\\
Self-emp-inc&	\cellcolor{yellow}48.68&	\cellcolor{green}53.05&	\cellcolor{yellow}49.31&	\cellcolor{yellow}49.91\\
Self-emp-not-inc&	\cellcolor{yellow}45.88&	43.39&	44.03&	44.44\\
\hline
\end{tabular}
\end{center}
\caption{Cube $C^N$ \label{tab:cn}}
\end{table}

The algorithm starts by computing the z-score for the cells of $C^O$, which results in the scores displayed in Table \ref{tab:zscoresCO}. 
Then the same is done for $C^N$, resulting in Table \ref{tab:zscoresCN}.

\begin{table}[ht]
\begin{center}
\begin{tabular}{lllll}
z-cores of $C^O$ & Assoc       & Post-grad   & Some-coll. & Univ.  \\
Gov       & 0.8167 & 0.1039    & 1.5759       & 0.3613     \\
Private   & 0.7101 & 0.6240    & 1.4628       & 0.0641     \\
Self-emp  & 1.1053 & 1.2862    & 0.7888       & 1.0827   
\end{tabular}
\end{center}
\caption{Significance scores of the cells in Cube  $C^O$ \label{tab:zscoresCO}}
\end{table}


\begin{table}[ht]
\begin{center}
\begin{tabular}{lllll}
z-scores of $C^N$       & Assoc       & Post-grad   & Some-coll. & Univ.  \\
Federal-gov      & 0.554 & 0.123     & 0.764        & 0.003      \\
Local-gov        & 0.509 & 0.148     & 0.806        & 0.257      \\
State-gov        & 1.069 & 0.102     & 2.158        & 0.636      \\
Private          & 0.576 & 0.456     & 1.159        & 0.077      \\
Self-emp-inc     & 1.328 & 2.420     & 1.485        & 1.635      \\
Self-emp-not-inc & 0.628 & 0.006     & 0.166        & 0.268     
\end{tabular}
\end{center}
\caption{Significance scores of the cells in Cube $C^N$ \label{tab:zscoresCN}}
\end{table}

The surprise score of each cell $c^N$ of $C^N$ is computed as the difference of its z-score with the one of its ancestor in $C^O$, resulting in Table \ref{tab:surpriseCN}. The score of each component is computed by averaging the surprise scores of the cells participating to this component. In our example, this score is (0.222 + 1.134 + 0.697 + 0.552 + 0.447)/5 = 0.61 for the top-5 cells and (1.134 + 0.582)/2 = 0.85 for outliers, meaning that in this example, the two cells of $C^N$ with to extreme values will be highlighted.

\begin{table}[ht]
\begin{center}
\begin{tabular}{lllll}
surprise in $C^N$ & Assoc   & Post-grad   & Some-coll. & Univ.  \\
Federal-gov       & 0.263 & 0,019     & 0.812        & 0.358      \\
Local-gov         & 0.308 & 0,044     & 0.770        & 0.105      \\
State-gov         & 0.252 & 0,002     & \cellcolor{cyan}0.582        & 0.275      \\
Private           & 0.134 & 0,168     & 0.304        & 0.013      \\
Self-emp-inc      & \cellcolor{yellow}0.222 & \cellcolor{green}1.134     & \cellcolor{yellow}0.697        & \cellcolor{yellow}0.552      \\
Self-emp-not-inc  & \cellcolor{yellow}0.477 & 1.280     & 0.623        & 0.814     
\end{tabular}
\end{center}
\caption{Surprise scores of the cells in Cube $C^N$ \label{tab:surpriseCN}}
\end{table}

\end{example}

%% file: 22_intentional.tex
\section{Intentions}\label{sec:transitions}

In this section we discuss the deeper essence of the \emph{intentional} nature of our proposal: user operations or, equivalently, transitions between the states of an OLAP session, i.e., dashboards. The main idea is that we move from a declarative model of \emph{logical} operators, like roll-up and drill down, to an \emph{intentional analytics} model where the user expresses high-level requirements like ``explain a certain phenomenon'' or ``predict the future values'', and these high-level requirements are \emph{automatically} translated into specific logical operators, models, and highlights that will carry the answer.
To this end we provide a set of \emph{intentional operators}; the term operator refers to an algebraic, template representation of an operation that can be applied over any cube, whereas the term \emph{intentional query} refers to a concrete instantiation of the operator, over a specific cube $C$ of a dashboard.  

Before detailing the operators, we need to detail the process that takes place once a user submits an intentional query to the system. The process is generic and the semantics of the process of query execution are identical for any operator (although, naturally, an optimizer can be constructed in order to mix or prune the steps of the process to achieve a faster execution).
The process of query execution includes the following steps:
\begin{enumerate}
 \item\emph{Data acquisition}. During this step, the system translates the intentional operator used in the query to a logical one, which is executed on $C^{old}$ to retrieve the necessary data for subsequent tasks in the form of a new cube $C^{new}$. Note that, depending on the expression of the user's intention, the same intention operator can be translated into different logical operators.

 \item \emph{Model construction}. A set of model types (in the trivial case, just one) are applied to $C^{new}$; in the case of models that are mined from the underlying data, the corresponding extraction algorithm is fired so as to obtain the model.  The cube $C^{new}$ is practically extended with the model components of the resulting models -- remember that each component comes with one or more attributes as its output, with one value per cell of the cube for each attribute (thus model components are linked to the cube's data as new measures).

\item \emph{Highlight selection}. The following step involves computing the significance of cells, model components and components. The reason is that we have fired several models to annotate the new data and we treat this as an antagonistic race between them, to decide which one is the most informative for the user.  The algorithms on the selection of the best model for the intentional query are detailed in Section~\ref{sec:interestingness} and here we give a very short overview, in order to facilitate the discussion of the intentional operators in this section. For each cell of cube $C^{new}$ a \emph{significance} score is computed by applying a significance evaluation algorithm to a specific subset of its measures, depending on the intentional operator that is applied. We aggregate significance scores for model components and models, based on the participation of cube cells to them. Based on these scores, we can pick (a) the model component with the maximum significance score, (b) the model that contains it and (c) its corresponding cells as the highlights of the new cube.




\item \emph{Packaging}. Once all data, models and highlights have been computed, and the system has picked the most significant configuration of them to add to the dashboard, the appropriate visual and textual packaging takes place. Despite its importance, the details of this task are outside the scope of this paper, and we do not elaborate further. We refer the reader to \cite{DBLP:journals/is/GkesoulisVM15}  for more information on related work and simple techniques for this task.

\end{enumerate}


The language we propose includes five intention operators:
\begin{itemize}
\item \sf{Describe}, which provides an answer to the user asking ``show me my business''. This is done by describing one or more cube measures (e.g., \sf{revenue} in a sales cube), possibly focused on one or more dimension members (e.g., food product category and March 2018), either at some given granularity (e.g., \sf{storeNation}) or using a given number of clusters or producing a result with a given maximum size.
\item \sf{Assess}, which provides an answer to the user asking ``is my business good?''. The goal is to judge one or more cube measures, possibly focused on one or more dimension members, with reference to some baseline (e.g., with reference to past values of the same measure, or to its values for other members, or to some benchmark) and using some KPI for comparison.
\item \sf{Explain}, which provides an answer to the user asking ``why is this happening?''. This is done by revealing some hidden information that is not part of the dashboard the user is observing, for instance in the form of a significant correlation between two cube measures or using a decision tree that classifies facts based on level members.
\item \sf{Predict}, which provides an answer to the user asking ``what will my business be like in the future?''. This is done by showing data not in the original cubes, but derived from them for instance with time-series analysis or regression.
\item \sf{Suggest}, which provides an answer to the user asking ``where should I look next?''. The goal here is to show data similar to those the current user, or similar users, have been interested in, for instance using collaborative query recommendation approaches. 
\end{itemize}

In the sequel, we introduce these operators in more detail.



\subsection{Describe}

The \sf{describe} operator is invoked to enrich the user's dashboard with more data that are currently missing; the user's intention is to know something more about a set of facts. The general syntax for invoking this operators is shown below (in extended Backus-Naur form):

\sf{with} \emph{cube} \sf{describe} \emph{measure} \{, \emph{measure}\} [\sf{for} \emph{subcube}] [\sf{by} (\{\emph{level}\}  $|$ \sf{size} \emph{integer})] 

In practice \sf{describe} can be invoked using either a generic signature or a specific one; in all cases, it specifies the cube $c$ on which the operator should be applied, the measures of $c$ which have to be described, and optionally a subcube of $c$ on which to focus. In the following we assume for simplicity that a single measure $m$ and a subcube consisting of a single slice on dimension member $v$ is specified. 

The first signature of \sf{Describe} refers to a specific measure, $m$ and, \emph{optionally}, to a dimension member, say $v$
\[\sf{with } c \sf{ describe } m \sf{ for }  v\]
The goal of this invocation is to facilitate focusing on a specific subset of the data space without changing the level of abstraction. In both this and the subsequent variants, the \sf{for} clause can optionally be added to focus the execution on a subset of the cells of $c$ that pertain to the specific member (practically applying a filter that retains only the cells with this value).

A second signature of \sf{Describe} includes a \sf{by} clause that comes in more than one variants. The \sf{by} clause results in a change of granularity which can come by drilling down to more detailed data or by abstracting to coarser descriptions of the cube. These coarser descriptions, in turn, can be computed either by rolling up, or by reducing the cube to a specified size that includes only its most characteristic cells (which, in turn, can be done either by clustering or applying the shrink operator \cite{DBLP:journals/dke/GolfarelliGR14}).

The first of the abstraction-altering variants is
\[\sf{with } c \sf{ describe } m \sf{ for } v \sf{ by } l\]
where $l$ is a level. Again, the \sf{ for } clause is optional. Note also that the previous variant of \sf{ Describe }  is a special case of this one.

This invocation practically instructs the system to execute a cube query.

\begin{enumerate}
	\item \emph{Data acquisition}. Data are obtained by the specification of a cube query, including a filter on $v$ (selection in relational algebra, or slice-n-dice in OLAP terminology), a projection of measure $m$ (projection in relational algebra) and a change of abstraction dictated by $l$ (roll-up or drill down, depending on where the current cube is located) 

	\item \emph{Model construction}. Several models are applicable to support the invocation of the operator, specifically: (a) find the top-k values of $m$ and highlight the facts yielding the top value, or (b) find the dominating row/column of $c$ for the values of $m$ and highlight its facts, or (c) detect the outliers for $m$ and highlight the outlier facts with the highest score.

	\item \emph{Highlight selection}: the generic highlight selection algorithm mentioned in the beginning of this section is applied over the cells of the cube, using the measure $m$ for significance assessment. For the case of \sf{describe}, the cells are divided in antagonizing components like topk vs non-topk, dominant vs non-dominant etc, based on their value of $m$. Then, the component that is scored by the algorithm as the most interesting, along with its respective cells, are selected  as highlights.
\end{enumerate}

The second variant is
\[\sf{with } c \sf{ describe } m \sf{ by size } k\]
where $k$ is an integer. 
This variant, after data acquisition, requires to either apply a clustering algorithm to detect $k$ clusters and highlight the medoids, or to apply the shrink operator \cite{DBLP:journals/jdwm/RizziGG15} to reduce the result to $k$ cells. In the first case, the output of the model is a set of attributes, including (a) one attribute per cluster, where each cell marks its participation to its respective cluster, and (b) an attribute to track the cluster medoids. Remember that for each cell of the cube we have a mapping to the respective attributes of the model; so, in the case of clusters each cell is "annotated" with a bit vector that tells us to which cluster the cell participates and whether it is also its medoid or not. In the second case the output of the model is a set of $k$ cells, each summarizing a set of cube cells yielding similar values; each cell is annotated with the approximation introduced by the shrinking.





\begin{example}
Consider the cube $C^O$ of Example \ref{ex:cubeAdult1}, and the intention: 

\[\textsf{with $C^O$ Describe Hours per Week by WorkClass.L0} \]

Processing this intention results in the cube $C^N$ in Table \ref{tab:cubeWithHL} where two cells have been automatically highlighted to display the two most significant outliers. 
In the data acquisition step, the cube query of Example \ref{tab:cubeAdult2} is derived from the expression of the intention and evaluated to produce $C^N$. Specifically, the \textsf{by Work} clause of the expression indicates that the logical Drill-down operation is to performed over cube $C^O$ to obtain $C^N$. 
In the model construction  step, the model types associated with the \textsf{Describe} intention (see Table \ref{fig:modelTax}) are instantiated to produce models computed over $C^N$, and their respective components are  mapped to the cube. Table \ref{tab:outliers} illustrates this for outliers detection, where outliers are detected using the Grubbs test, i.e., by computing for each measure value the number of standard deviations they are from the mean of all values. The two components are produced with the binding $\tuple{2,$Hours per Week$}$ where 2 indicates that outliers measure values whose score is above two standard deviations from the mean.
In the last step, the highlight extraction algorithm is called with z-score for significance computation (function significance) and difference for surprise computation (function ${\cal D}$). 
Details of the computation can be found in Example \ref{ex:interestingness} of Section \ref{sec:interestingness}.
The algorithm outputs component Outliers, that achieves the best surprise score, resulting in highlighting the two outliers shown in Table \ref{tab:cubeWithHL}.\rem{Table \ref{tab:outliers} is out of bounds at double column!}

\begin{table}[ht]
\begin{center}
\begin{tabular}{|l||cccc|}
\hline
Weekly Hrs &	Assoc&	Post-grad&	Some-coll.&	Univ.\\
\hline
\hline
Federal-gov&	41.15&	43.86	&40.31&	43.38\\
Local-gov&	41.33&	43.96&	40.14&	42.34\\
State-gov&	39.09&	42.96&	\cellcolor{yellow}34.73&	40.82\\
Private&	41.06&	45.19&	38.73&	43.06\\
Self-emp-inc&	48.68&	\cellcolor{yellow}53.05&	49.31&	49.91\\
Self-emp-not-inc&	45.88&	43.39&	44.03&	44.44\\
\hline
\end{tabular}
\end{center}
\caption{Output of  intention: \textsf{with $C^O$ Describe Hours per Week by Work}\label{tab:cubeWithHL}}
\end{table}

\begin{table}[htb]
\centering
{\scriptsize 
\begin{tabular}{llll|c|l|c|l|c|}
\cline{5-5} \cline{7-7} \cline{9-9}
                                                    &                                       &                            &  & Outlierness &  & Outliers &  & Non-outliers \\ \cline{1-3} \cline{5-5} \cline{7-7} \cline{9-9} 
\multicolumn{1}{|l|}{\multirow{6}{*}{Assoc}}        & \multicolumn{1}{l|}{Federal-gov}      & \multicolumn{1}{l|}{41.15} &  & -0.55       &  & 0        &  & 1            \\ \cline{2-3} \cline{5-5} \cline{7-7} \cline{9-9} 
\multicolumn{1}{|l|}{}                              & \multicolumn{1}{l|}{Local-gov}        & \multicolumn{1}{l|}{41.33} &  & -0.50       &  & 0        &  & 1            \\ \cline{2-3} \cline{5-5} \cline{7-7} \cline{9-9} 
\multicolumn{1}{|l|}{}                              & \multicolumn{1}{l|}{State-gov}        & \multicolumn{1}{l|}{39.09} &  & -1.06       &  & 0        &  & 1            \\ \cline{2-3} \cline{5-5} \cline{7-7} \cline{9-9} 
\multicolumn{1}{|l|}{}                              & \multicolumn{1}{l|}{Private}          & \multicolumn{1}{l|}{41.06} &  & -0.57       &  & 0        &  & 1            \\ \cline{2-3} \cline{5-5} \cline{7-7} \cline{9-9} 
\multicolumn{1}{|l|}{}                              & \multicolumn{1}{l|}{Self-emp-inc}     & \multicolumn{1}{l|}{48.68} &  & 1.327       &  & 0        &  & 1            \\ \cline{2-3} \cline{5-5} \cline{7-7} \cline{9-9} 
\multicolumn{1}{|l|}{}                              & \multicolumn{1}{l|}{Self-emp-not-inc} & \multicolumn{1}{l|}{45.88} &  & 0.628       &  & 0        &  & 1            \\ \cline{1-3} \cline{5-5} \cline{7-7} \cline{9-9} 
\multicolumn{1}{|l|}{\multirow{6}{*}{Post-grad}}    & \multicolumn{1}{l|}{Federal-gov}      & \multicolumn{1}{l|}{43.86} &  & 0.123       &  & 0        &  & 1            \\ \cline{2-3} \cline{5-5} \cline{7-7} \cline{9-9} 
\multicolumn{1}{|l|}{}                              & \multicolumn{1}{l|}{Local-gov}        & \multicolumn{1}{l|}{43.96} &  & 0.148       &  & 0        &  & 1            \\ \cline{2-3} \cline{5-5} \cline{7-7} \cline{9-9} 
\multicolumn{1}{|l|}{}                              & \multicolumn{1}{l|}{State-gov}        & \multicolumn{1}{l|}{42.96} &  & -0.10       &  & 0        &  & 1            \\ \cline{2-3} \cline{5-5} \cline{7-7} \cline{9-9} 
\multicolumn{1}{|l|}{}                              & \multicolumn{1}{l|}{Private}          & \multicolumn{1}{l|}{45.19} &  & 0.455       &  & 0        &  & 1            \\ \cline{2-3} \cline{5-5} \cline{7-7} \cline{9-9} 
\multicolumn{1}{|l|}{}                              & \multicolumn{1}{l|}{Self-emp-inc}     & \multicolumn{1}{l|}{\textcolor{red}{53.05}} &  & 2.419       &  & \textcolor{red}{1}        &  & 0            \\ \cline{2-3} \cline{5-5} \cline{7-7} \cline{9-9} 
\multicolumn{1}{|l|}{}                              & \multicolumn{1}{l|}{Self-emp-not-inc} & \multicolumn{1}{l|}{43.39} &  & 0.005       &  & 0        &  & 1            \\ \cline{1-3} \cline{5-5} \cline{7-7} \cline{9-9} 
\multicolumn{1}{|l|}{\multirow{6}{*}{Some-college}} & \multicolumn{1}{l|}{Federal-gov}      & \multicolumn{1}{l|}{40.31} &  & -0.76       &  & 0        &  & 1            \\ \cline{2-3} \cline{5-5} \cline{7-7} \cline{9-9} 
\multicolumn{1}{|l|}{}                              & \multicolumn{1}{l|}{Local-gov}        & \multicolumn{1}{l|}{40.14} &  & -0.80       &  & 0        &  & 1            \\ \cline{2-3} \cline{5-5} \cline{7-7} \cline{9-9} 
\multicolumn{1}{|l|}{}                              & \multicolumn{1}{l|}{State-gov}        & \multicolumn{1}{l|}{\textcolor{red}{34.73}} &  & -2.15       &  & \textcolor{red}{1}        &  & 0            \\ \cline{2-3} \cline{5-5} \cline{7-7} \cline{9-9} 
\multicolumn{1}{|l|}{}                              & \multicolumn{1}{l|}{Private}          & \multicolumn{1}{l|}{38.73} &  & -1.15       &  & 0        &  & 1            \\ \cline{2-3} \cline{5-5} \cline{7-7} \cline{9-9} 
\multicolumn{1}{|l|}{}                              & \multicolumn{1}{l|}{Self-emp-inc}     & \multicolumn{1}{l|}{49.31} &  & 1.485       &  & 0        &  & 1            \\ \cline{2-3} \cline{5-5} \cline{7-7} \cline{9-9} 
\multicolumn{1}{|l|}{}                              & \multicolumn{1}{l|}{Self-emp-not-inc} & \multicolumn{1}{l|}{44.03} &  & 0.165       &  & 0        &  & 1            \\ \cline{1-3} \cline{5-5} \cline{7-7} \cline{9-9} 
\multicolumn{1}{|l|}{\multirow{6}{*}{Univesity}}    & \multicolumn{1}{l|}{Federal-gov}      & \multicolumn{1}{l|}{43.38} &  & 0.003       &  & 0        &  & 1            \\ \cline{2-3} \cline{5-5} \cline{7-7} \cline{9-9} 
\multicolumn{1}{|l|}{}                              & \multicolumn{1}{l|}{Local-gov}        & \multicolumn{1}{l|}{42.34} &  & -0.25       &  & 0        &  & 1            \\ \cline{2-3} \cline{5-5} \cline{7-7} \cline{9-9} 
\multicolumn{1}{|l|}{}                              & \multicolumn{1}{l|}{State-gov}        & \multicolumn{1}{l|}{40.82} &  & -0.63       &  & 0        &  & 1            \\ \cline{2-3} \cline{5-5} \cline{7-7} \cline{9-9} 
\multicolumn{1}{|l|}{}                              & \multicolumn{1}{l|}{Private}          & \multicolumn{1}{l|}{43.06} &  & -0.07       &  & 0        &  & 1            \\ \cline{2-3} \cline{5-5} \cline{7-7} \cline{9-9} 
\multicolumn{1}{|l|}{}                              & \multicolumn{1}{l|}{Self-emp-inc}     & \multicolumn{1}{l|}{49.91} &  & 1.635       &  & 0        &  & 1            \\ \cline{2-3} \cline{5-5} \cline{7-7} \cline{9-9} 
\multicolumn{1}{|l|}{}                              & \multicolumn{1}{l|}{Self-emp-not-inc} & \multicolumn{1}{l|}{44.44} &  & 0.268       &  & 0        &  & 1            \\ \cline{1-3} \cline{5-5} \cline{7-7} \cline{9-9} 
\end{tabular}
}
\caption{Outlier model and components for cube $C^N$ \label{tab:outliers}}
\end{table}
\end{example}

The \sf{ Describe } operator practically covers the operators \sf{ FocusOn } and \sf{ Abstract } of the short version of this paper~\cite{DBLP:conf/dolap/VassiliadisM18}, both in terms of data acquired and models fired (top-k values, clustering). 


\subsection{Assess}

How do we handle the case when the user wants to tell the system: \emph{"please tell me how good, bad, normal, unexpected, \ldots is the situation I observe for this particular (sub)cube or cell?"}. From the philosophical point of view, among many definitions, the closest to our fully automated mentality suggests that assessment boils down to the case where \emph{"results are assessed in relation to some predetermined goal"} \cite{Hansen2005}.  In our case, this practically means that we compare the observed status (for us: observed cube) to a possible benchmark that defines an expected value, and automatically label the divergence of the attained to the desired performance (for us: measure). 
In any case, the overall idea of assessment requires (a) a \emph{benchmark} 
against which the current performance is going to be compared, (b) the 
actual execution of the \emph{comparison} (e.g., a simple 
difference, or the difference of the z-scores, in the case of simple measures), and (c) a characterization 
of the result of the comparison either via explicitly specified rules 
(as in a KPI) or via automatically computed "outlierness" measures 
(e.g., a z-score).

We resort to the invocation of \emph{benchmark models} for describing measures. Benchmark models are models that can be linked to the observed cube, by relating each cell of the observed cube with an "expected" value. Remember that models come \emph{always} with output measures, therefore we can place the role of a benchmark that tells us what the expected performance should be as the output of a model.
\begin{definition}[Benchmark Model]
	Given a cube $c$ with a measure $m$, a benchmark model $b$ for $c.m$ is any model having a \emph{benchmark} measure $b.m^{comp}$ in its output that extends each cell of the cube with a new  value that is to be contrasted to the respective value of $m$. An extra \emph{discrepancy} measure of the benchmark model, $b.d$, is reserved to store the result of this comparison.
\end{definition}

The invocation of the \sf{ Assess } operator follows the syntax 

\sf{ with } \emph{cube} \sf{ assess } \emph{ measure }  \{,~\emph{measure}\}  [\sf{ for } \emph{subcube}]  \sf{ using } \emph{benchmark~model}  \{,~\emph{benchmark~model}\}

Again, the syntax of the operation can include a selection on specific slices of the cube, via the \sf{ for } clause, and specific measures. The interesting part involves the specification of benchmark models. We envision an open,  extensible list of benchmark models for a cube:

\begin{itemize}
	\item A predefined goal for each cell (i.e., via the retrieval of the respective KPI)
	\item Any query that returns a cube with the same coordinates with the observed cube, and any measure (data- or function-based) that can be contrasted to the observed measure of the cube, via a 1:1 mapping of cells
	\item A benchmark model for the past performance of a cube via the invocation of a $lastKValues(cube.measure)$ operator that computes an aggregate statistic over the last $k$ values for each cell of the cube
	\item A benchmark model that combines the performance of all peers of the observed cube (i.e., find siblings a-la \emph{ Cinecubes' put-in-context} operator)
	\item A benchmark model that translates the general context of the observed cube (via a roll-up action) to its expected value
	\item Any predefined golden standard peer, like, e.g., comparing a stock value to the S\&P 500 index, or the performance of a specific EU country over a 
	certain measure against the European average
	\item A benchmark model that involves computing a forecasted/expected value via a forecasting function that involves other/past measure values e.g.,. $expectedValue(m_{1}$) = $f(m_{2}, \ldots , m_{k})$ )
\end{itemize}
The list is open to additions of course, but the main message is that, for each cell of a (sub)cube, we provide an expected value via a benchmark model.

The semantics of the invocation of  \sf{Assess} is as follows:

\begin{enumerate}
	\item \emph{Data acquisition}: the system obtains the necessary data via (a) the appropriate cube query that prescribed (sub)cube, and (b) the retrieval or computation of the prescribed benchmark model's output, along with a 1:1 mapping of cells to it.
	\item \emph{Model construction}: the models used per se, and the values that they generate for each of the cells being assessed, along with a \emph{discrepancy} model component tracing the difference of a model's output from the measure value, per cell.
	\item \emph{Highlight selection}: we apply the generic highlight selection procedure to the \emph{discrepancy} measure and obtain the model component with the maximum aggregate discrepancy from the measure of the cube, along with its cells, as the highlights of the new cube.
	

\end{enumerate}

\begin{example}\label{ex:assess}



Consider the cube $C^N$ = $\tuple{DS$, $education.L3='Post-secondary'$ $and~work\_class.L2='With-Pay',$ $\tuple{$ALL,ALL,L2,ALL,L0,ALL,ALL$},$~
$Avg(Hours~per~Week)}$
\noindent fixing Education to ’Post-Secondary’ (at level L3), and Work to ’With-Pay’ (at level L2), and grouping by Education at level 2, and Work at level 1. Assume the user wishes to check to what extent working hours per week  for Female deviate from  the data of $C^N$. Doing so can be done with the following intention:

\[ \textsf{with $C^N$ assess Hours per Week using $q_{Female}$} \]

where $q_{Female}$ is a benchmark cube query drilling down from $C^N$ to the L0 level of dimension Gender and selecting Female, i.e.,:

$\tuple{DS,$

Gender.L0 = 'Female' and education.L3 = 'Post-secondary' and work\_class.L2 = 'With-Pay',

$\tuple{$ALL,ALL,L2,ALL,L0,ALL,L0$},$

$Avg(Hours~per~Week)}$

The data acquisition step consists of executing $q_{Female}$. In the model construction step, a discrepancy model is used to compute the difference between average hours per week for Females and the overall average hours per week of $C^N$, splitting the results in two components $MC^-$ and $MC^+$ according to the sign of the difference. 
In the highlight selection step, the highlight selection algorithm is called with the following parameters:
\begin{itemize}
\item the cube $C^N$
\item the cube $C^F$ retrieved by $q_{Female}$ 
\item components $MC^-$ and $MC^+$ 
\item function $proxies(x)$ maps a fact $C^F$ of $C^F$ to the fact $C^N$ of $C^N$ having the same coordinates
\item function $significance(x)$ returns the value of  measure Hours per Week of fact $x$ 
\item function $surprise(x,y)=y-x$
\item function ${\cal A}^C(x)=|C^N|-sum(x)$ where sum is the traditional sum aggregation function.
\end{itemize}


With these parameters, the algorithm picks  component $MC^+$ i.e., the component having the highest count of deviations, according to the aforementioned highlight selection rule.\rem{Used to be MC- with the least deviation. RV1.PDF said "why the least?" and he is right. Changed it to MC+}  \rem{Table~\ref{tab:female} is too large for 2 columns. SR: We can take care of that if and when we are asked for a camera ready}



\begin{table}[htb]
\centering
{\scriptsize 
\begin{tabular}{llll|c|l|c|l|c|l|c|}
\cline{5-5} \cline{7-7} \cline{9-9} \cline{11-11}
                                                    &                                       &                            &  & Gender= &  & Discre- &  &  &  &  \\                                                  &                                       &                            &  & Female &  & pancy &  & $MC^-$ &  & $MC^+$ \\ \cline{1-3} \cline{5-5} \cline{7-7} \cline{9-9} \cline{11-11} 
\multicolumn{1}{|l|}{\multirow{6}{*}{Assoc}}        & \multicolumn{1}{l|}{Federal-gov}      & \multicolumn{1}{l|}{41.15} &  & 40.66         &  & 0.49        &  & 0  &  & 1  \\ \cline{2-3} \cline{5-5} \cline{7-7} \cline{9-9} \cline{11-11} 
\multicolumn{1}{|l|}{}                              & \multicolumn{1}{l|}{Local-gov}        & \multicolumn{1}{l|}{41.33} &  & 37.61         &  & 3.72        &  & 0  &  & 1  \\ \cline{2-3} \cline{5-5} \cline{7-7} \cline{9-9} \cline{11-11} 
\multicolumn{1}{|l|}{}                              & \multicolumn{1}{l|}{State-gov}        & \multicolumn{1}{l|}{\textcolor{red}{39.09}} &  & 39.36         &  & -0.27       &  & \textcolor{red}{1}  &  & 0  \\ \cline{2-3} \cline{5-5} \cline{7-7} \cline{9-9} \cline{11-11} 
\multicolumn{1}{|l|}{}                              & \multicolumn{1}{l|}{Private}          & \multicolumn{1}{l|}{41.06} &  & 38.05         &  & 3.1         &  & 0  &  & 1  \\ \cline{2-3} \cline{5-5} \cline{7-7} \cline{9-9} \cline{11-11} 
\multicolumn{1}{|l|}{}                              & \multicolumn{1}{l|}{Self-emp-inc}     & \multicolumn{1}{l|}{48.68} &  & 42.07         &  & 6.61        &  & 0  &  & 1  \\ \cline{2-3} \cline{5-5} \cline{7-7} \cline{9-9} \cline{11-11} 
\multicolumn{1}{|l|}{}                              & \multicolumn{1}{l|}{Self-emp-not-inc} & \multicolumn{1}{l|}{45.88} &  & 38.47         &  & 7.41        &  & 0  &  & 1  \\ \cline{1-3} \cline{5-5} \cline{7-7} \cline{9-9} \cline{11-11} 
\multicolumn{1}{|l|}{\multirow{6}{*}{Post-grad}}    & \multicolumn{1}{l|}{Federal-gov}      & \multicolumn{1}{l|}{\textcolor{red}{43.86}} &  & 47.76         &  & -3.9        &  & \textcolor{red}{1}  &  & 0  \\ \cline{2-3} \cline{5-5} \cline{7-7} \cline{9-9} \cline{11-11} 
\multicolumn{1}{|l|}{}                              & \multicolumn{1}{l|}{Local-gov}        & \multicolumn{1}{l|}{43.96} &  & 43.83         &  & 0.13        &  & 0  &  & 1  \\ \cline{2-3} \cline{5-5} \cline{7-7} \cline{9-9} \cline{11-11} 
\multicolumn{1}{|l|}{}                              & \multicolumn{1}{l|}{State-gov}        & \multicolumn{1}{l|}{42.96} &  & 40.14         &  & 2.82        &  & 0  &  & 1  \\ \cline{2-3} \cline{5-5} \cline{7-7} \cline{9-9} \cline{11-11} 
\multicolumn{1}{|l|}{}                              & \multicolumn{1}{l|}{Private}          & \multicolumn{1}{l|}{45.19} &  & 41.55         &  & 3.64        &  & 0  &  & 1  \\ \cline{2-3} \cline{5-5} \cline{7-7} \cline{9-9} \cline{11-11} 
\multicolumn{1}{|l|}{}                              & \multicolumn{1}{l|}{Self-emp-inc}     & \multicolumn{1}{l|}{53.05} &  & 48.73         &  & 4.32        &  & 0  &  & 1  \\ \cline{2-3} \cline{5-5} \cline{7-7} \cline{9-9} \cline{11-11} 
\multicolumn{1}{|l|}{}                              & \multicolumn{1}{l|}{Self-emp-not-inc} & \multicolumn{1}{l|}{43.39} &  & 38.28         &  & 5.11        &  & 0  &  & 1  \\ \cline{1-3} \cline{5-5} \cline{7-7} \cline{9-9} \cline{11-11} 
\multicolumn{1}{|l|}{\multirow{6}{*}{Some-college}} & \multicolumn{1}{l|}{Federal-gov}      & \multicolumn{1}{l|}{40.31} &  & 38.25         &  & 2.06        &  & 0  &  & 1  \\ \cline{2-3} \cline{5-5} \cline{7-7} \cline{9-9} \cline{11-11} 
\multicolumn{1}{|l|}{}                              & \multicolumn{1}{l|}{Local-gov}        & \multicolumn{1}{l|}{40.14} &  & 35.45         &  & 4.69        &  & 0  &  & 1  \\ \cline{2-3} \cline{5-5} \cline{7-7} \cline{9-9} \cline{11-11} 
\multicolumn{1}{|l|}{}                              & \multicolumn{1}{l|}{State-gov}        & \multicolumn{1}{l|}{34.73} &  & 34.01         &  & 0.72        &  & 0  &  & 1  \\ \cline{2-3} \cline{5-5} \cline{7-7} \cline{9-9} \cline{11-11} 
\multicolumn{1}{|l|}{}                              & \multicolumn{1}{l|}{Private}          & \multicolumn{1}{l|}{38.73} &  & 34.86         &  & 3.87        &  & 0  &  & 1  \\ \cline{2-3} \cline{5-5} \cline{7-7} \cline{9-9} \cline{11-11} 
\multicolumn{1}{|l|}{}                              & \multicolumn{1}{l|}{Self-emp-inc}     & \multicolumn{1}{l|}{49.31} &  & 43.96         &  & 5.35        &  & 0  &  & 1  \\ \cline{2-3} \cline{5-5} \cline{7-7} \cline{9-9} \cline{11-11} 
\multicolumn{1}{|l|}{}                              & \multicolumn{1}{l|}{Self-emp-not-inc} & \multicolumn{1}{l|}{44.03} &  & 36.57         &  & 7.46        &  & 0  &  & 1  \\ \cline{1-3} \cline{5-5} \cline{7-7} \cline{9-9} \cline{11-11} 
\multicolumn{1}{|l|}{\multirow{6}{*}{University}}    & \multicolumn{1}{l|}{Federal-gov}      & \multicolumn{1}{l|}{43.38} &  & 42.41         &  & 0.97        &  & 0  &  & 1  \\ \cline{2-3} \cline{5-5} \cline{7-7} \cline{9-9} \cline{11-11} 
\multicolumn{1}{|l|}{}                              & \multicolumn{1}{l|}{Local-gov}        & \multicolumn{1}{l|}{42.34} &  & 41.66         &  & 0.68        &  & 0  &  & 1  \\ \cline{2-3} \cline{5-5} \cline{7-7} \cline{9-9} \cline{11-11} 
\multicolumn{1}{|l|}{}                              & \multicolumn{1}{l|}{State-gov}        & \multicolumn{1}{l|}{40.82} &  & 38.95         &  & 1.87        &  & 0  &  & 1  \\ \cline{2-3} \cline{5-5} \cline{7-7} \cline{9-9} \cline{11-11} 
\multicolumn{1}{|l|}{}                              & \multicolumn{1}{l|}{Private}          & \multicolumn{1}{l|}{43.06} &  & 39.45         &  & 3.61        &  & 0  &  & 1  \\ \cline{2-3} \cline{5-5} \cline{7-7} \cline{9-9} \cline{11-11} 
\multicolumn{1}{|l|}{}                              & \multicolumn{1}{l|}{Self-emp-inc}     & \multicolumn{1}{l|}{49.91} &  & 44.83         &  & 5.08        &  & 0  &  & 1  \\ \cline{2-3} \cline{5-5} \cline{7-7} \cline{9-9} \cline{11-11} 
\multicolumn{1}{|l|}{}                              & \multicolumn{1}{l|}{Self-emp-not-inc} & \multicolumn{1}{l|}{44.44} &  & 39.04         &  & 5.4         &  & 0  &  & 1  \\ \cline{1-3} \cline{5-5} \cline{7-7} \cline{9-9} \cline{11-11} 
\end{tabular}
}
\caption{Assessment wrt Gender='Female'} \label{tab:female}
\end{table}

\end{example}

One can also envision an invocation of the operator without a prescribed benchmark and the assessment with various alternative models, appropriately selected -- however, for the moment we stick to a well specified benchmark. The operator \sf{ Assess } extends and covers the operator \sf{ Compare }  of the short version of this paper~\cite{DBLP:conf/dolap/VassiliadisM18}, which practically carries the same intention. The operator \sf{ Verify } of the short version of this paper~\cite{DBLP:conf/dolap/VassiliadisM18} that involves comparing the behavior of the observed cube with its broader context (e.g., a rolled-up cube) can similarly be facilitated by the current operator \sf{ Assess } using the broader context as a benchmark model.



\subsection{Explain}
The hardest possible task that the user can ask the system to do is to instruct it \emph{"please explain to me why I see what I am seeing!"}.  Explanation is fundamentally the delineation of causation for an observed phenomenon. Practically, explanation answers the question \emph{Why?}  for an observed phenomenon, by providing a causal model for it. In the case where a person is performing the explanation, and, even more typically, in the case of scientific explanations, several alternative causal models can be constructed and, out of them the person picks the one that most satisfactorily aligns with the observed data \cite {Mayes2018}. To avoid the well known confusion between causation and correlation, and since determining causation requires a too sophisticated process that cannot be automated, the term 'explanation' in this paper refers at the automated revelation of \emph{hidden~correlations} and information that are not directly observable as parts of the dashboard.

In the preliminary version of this paper~\cite{DBLP:conf/dolap/VassiliadisM18}, we introduced a pair of operators, \sf{ Analyze } which was intended to provide a collection of data for a cube at a more detailed level and \sf{ Explain } for more elaborate (but again data-oriented) actions that try to highlight interesting subsets or aggregations of data. Here, we collectively group these alternatives under the new operator \sf{ Explain } that provides an emphasis to models, rather than data for the explanation of phenomena.

We discriminate two kinds of phenomena that need explanation (thus producing two variants of the operator's invocation). In the first  case, "explain" means: give me an explanation model for the measure I am observing (e.g., "why do the \emph{Sales} for \emph{Rhodes} have a value of 2500?"). In the second case, "explain" means: give me an explanation model for the discrepancy from my benchmark model ("why are the \emph{Sales} for \emph{Rhodes} down by 1000 units compared to last year?"), which practically refers to comparing the models for two different cubes.

In accordance with the aforementioned theory on explanation, our operator requires (a) a phenomenon (in the form either of a simple measure, or of two comparable measures) for which we ask, (b) a question "why?", as well as (c) the provision of "the most fit" model as an answer to the "why?" question. Again, the process for explaining practically requires the acquisition of the necessary data, the computation of one or more explanation models, and the assessment of its results by evaluating the important model components.\\


The simplest invocation of the \sf{Explain} operator follows the syntax 

\sf{ with } \emph{cube} \sf{ explain } \emph{measure}   [\sf{ for } \emph{subcube} ]  \sf{ using } \emph{explanation~model (attribute~list)} \{, \emph{explanation~model (attribute~list)}\}  \\

\noindent where: we start with a $cube$ (possibly extended with derived measures); we restrict our focus to a specific $measure$; define a $subcube$ via the appropriate selections (if the \sf{ for } part is absent, we refer to the entire cube); and use one or more explanation models over a set of attributes (for example, the attribute to test correlation with $measure$, or the attribute list over which a linear regression will be tested, or the attribute needed to be used as a factor in a hypothesis testing, or the attributes needed to perform decision tree analysis, etc.).


There are several possibilities for models that can be used as explanatory means:
\begin{itemize}
	\item Performing statistical tests between different descriptions at different levels of granularity of the same multidimensional space (i.e., via roll-up's or drill down's) to establish an analogy between their statistical measures via, e.g., t-tests or F-tests 
    
	\item Correlating the measure with one (or a vector of )  correlation measure(s) listed in an attribute-list of other measures and (attributes of) dimension levels
	
	\item The extraction of a regression formula that relates it with an attribute-list of other measures and (attributes of) dimension levels and 	its comparison to the result of this formula
	
	\item A decision tree that classifies the measure with respect to an attribute-list of other measures and (attributes of) dimension levels
	
\end{itemize}

Deciding highlights, in an automated way is not straightforward and this is due to the fact that the split of cells in different components is not always straightforward. For example, although any labeling scheme has a natural way to split cells by target labels, correlation / regression schemes are not directly separable. However, as we can still annotate individual cells with their contribution to the overall correlation or with the discrepancy of the actual to the expected value, we still have (a) model component measures per cell and (b) the possibility of separating them in different classes via thresholds (e.g., cells that deviate too much from the expected value, or cells whose concordant tuples are too many in a Kendall correlation).

The semantics of the execution of the invocation of the \sf{ Explain } operator are as follows:
\begin{enumerate}
	\item \emph{Data Acquisition}: we obtain the data for the 	cell/(sub)cube
	\item \emph{Model construction}: we compute the specified \emph{explanation models} over the specified attribute list. Depending on the model, the output comprises several computed measures (model components) which annotate the cells of the cube.
	\item \emph{Highlight selection}: we apply the generic highlight selection procedure to the measures produced by the binding of parameters to the model type and obtain the model component with the maximum surprise, along with its cells, as the highlights of the new cube.

\end{enumerate}

The second variant of the operator utilizes a \emph{Comparison Cube}. A comparison cube is just another,  previously specified, cube with the same schema, against which the current (sub)cube is to be contrasted. The execution differs from the above, in the sense that  the explanation model is going to be computed for both cubes, and it is their differences that have to be presented. 

\sf{ with } \emph{cube} \sf{ explain } \emph{measure}   [\sf{ for } \emph{subcube} ]  \sf{ using } \emph{explanation~model (attribute~list)} \{, \emph{explanation~model (attribute~list)}\} \sf{ against }  \emph{comparison~cube}\\


The essence of the operator is the demonstration to the user of the differences in the models of the antagonizing cubes. This is of course specific to the model type. For example, the difference in correlation is just a numerical value, whereas the difference in a decision tree is a set of paths, along with the change in the strength measures per path. 

\begin{example}
Assume the user used to be in cube $C^O$ and nagivates to cube $C^N$ via a describe operation that drills down from $Work.L_1$ to $Work.L_2$. Then, suppose the user, after observing the new data, wonders "why is my data in gov like this?" with the following hypothesis in mind: does drilling down to $C^N$ drastically change the variance of data? One alternative is that variance remains approximately the same  (null hypothesis), and the other is that it does not (alternative hypothesis).
Practically, the explanation here is "if my data are like this, this is because variance at a finest granularity is somehow propagated to the level I am seeing now". 
Then, we have the intention:
\[ \textsf{with $C^N$ explain Hours per Week for Work.L1 = gov using F-test (work\_class.L0)}\] 
\[ \textsf{against describe $C^0$ for Work.L1 = gov}\]\\

This means that the answer to the user's intention is the subset of $C^N$ corresponding to $gov$ that highlights how cells behave with respect to the mean of $C^O$ for $gov$. Data acquisition is the drill down to $C^N$ for $gov$. Model construction is the application of the F-test between the two cubes (for gov), and computing a discrepancy for each cell value to the mean of $C^O$ (for gov), resulting in two components (Table \ref{tab:ftest}): (a) one for cells deviating by more than one standard deviation, and, (b) its complement. Note that each component can be interpreted as corresponding to one of the hypotheses (e.g., component "$>stdev$" corresponds to the alternative hypothesis).
Highlight selection is done by calling the highlight selection algorithm with the following parameters:\rem{Table is too large for double column}
\begin{itemize}
\item the cube obtained by applying the selection L1.work\_class=gov over $C^O$, denoted $C^O_{gov}$
\item the cube obtained by applying the selection L1.work\_class=gov over $C^N$, denoted $C^N_{gov}$
\item components "$<stdev$" and "$>stdev$"
\item function $proxies(x)$ that maps each fact $C^N$ of $C^N_{gov}$ to its ancestor in $C^O_{gov}$
\item in this case, we use two different significance functions:  $significance_{C^N_{gov}}(x)$ returns the value of $x$ for the cells of $C^N_{gov}$, and $significance_{C^O_{gov}}(x)$  returns the mean of the cells of $C^O_{gov}$. 
\item function $surprise(x,y)=y-x$
\item function ${\cal A}^C(x)=|C^N|-sum(x)$ where sum is the traditional sum aggregation function.
\end{itemize}
One could have easily used a single significance function, e.g., a z-score. We opted for this setup in order to illustrate that the specific settings for the involved functions can vary. Determining whether there exist  significance / surprise / aggregation functions outperforming the others with "universal" applicability is of course a matter of future research.
\begin{table}[htb]
\centering
{\scriptsize 
\begin{tabular}{llll|c|l|c|l|c|l|c|}
\cline{5-5} \cline{7-7} \cline{9-9} \cline{11-11}
                                                    &                                  &                            &  & F-test &  & Discrepancy &  & \textgreater stdev &  & \textless stdev \\ \cline{1-3} \cline{5-5} \cline{7-7} \cline{9-9} \cline{11-11} 
\multicolumn{1}{|l|}{\multirow{3}{*}{Assoc}}        & \multicolumn{1}{l|}{Federal-gov} & \multicolumn{1}{l|}{41.15} &  & 0.907  &  & -0.02       &  & 0                  &  & 1               \\ \cline{2-3} \cline{5-5} \cline{7-7} \cline{9-9} \cline{11-11} 
\multicolumn{1}{|l|}{}                              & \multicolumn{1}{l|}{Local-gov}   & \multicolumn{1}{l|}{41.33} &  & 0.907  &  & 0.15        &  & 0                  &  & 1               \\ \cline{2-3} \cline{5-5} \cline{7-7} \cline{9-9} \cline{11-11} 
\multicolumn{1}{|l|}{}                              & \multicolumn{1}{l|}{State-gov}   & \multicolumn{1}{l|}{39.09} &  & 0.907  &  & -2.08       &  & 0                  &  & 1               \\ \cline{1-3} \cline{5-5} \cline{7-7} \cline{9-9} \cline{11-11} 
\multicolumn{1}{|l|}{\multirow{3}{*}{Post-grad}}    & \multicolumn{1}{l|}{Federal-gov} & \multicolumn{1}{l|}{\textcolor{red}{43.86}} &  & 0.907  &  & 2.68        &  & \textcolor{red}{1}                  &  & 0               \\ \cline{2-3} \cline{5-5} \cline{7-7} \cline{9-9} \cline{11-11} 
\multicolumn{1}{|l|}{}                              & \multicolumn{1}{l|}{Local-gov}   & \multicolumn{1}{l|}{\textcolor{red}{43.96}} &  & 0.907  &  & 2.78        &  & \textcolor{red}{1}                  &  & 0               \\ \cline{2-3} \cline{5-5} \cline{7-7} \cline{9-9} \cline{11-11} 
\multicolumn{1}{|l|}{}                              & \multicolumn{1}{l|}{State-gov}   & \multicolumn{1}{l|}{42.96} &  & 0.907  &  & 1.78        &  & 0                  &  & 1               \\ \cline{1-3} \cline{5-5} \cline{7-7} \cline{9-9} \cline{11-11} 
\multicolumn{1}{|l|}{\multirow{3}{*}{Some-college}} & \multicolumn{1}{l|}{Federal-gov} & \multicolumn{1}{l|}{40.31} &  & 0.907  &  & -0.86       &  & 0                  &  & 1               \\ \cline{2-3} \cline{5-5} \cline{7-7} \cline{9-9} \cline{11-11} 
\multicolumn{1}{|l|}{}                              & \multicolumn{1}{l|}{Local-gov}   & \multicolumn{1}{l|}{40.14} &  & 0.907  &  & -1.03       &  & 0                  &  & 1               \\ \cline{2-3} \cline{5-5} \cline{7-7} \cline{9-9} \cline{11-11} 
\multicolumn{1}{|l|}{}                              & \multicolumn{1}{l|}{State-gov}   & \multicolumn{1}{l|}{\textcolor{red}{34.73}} &  & 0.907  &  & -6.44       &  & \textcolor{red}{1}                  &  & 0               \\ \cline{1-3} \cline{5-5} \cline{7-7} \cline{9-9} \cline{11-11} 
\multicolumn{1}{|l|}{\multirow{3}{*}{University}}   & \multicolumn{1}{l|}{Federal-gov} & \multicolumn{1}{l|}{43.38} &  & 0.907  &  & 2.20        &  & 0                  &  & 1               \\ \cline{2-3} \cline{5-5} \cline{7-7} \cline{9-9} \cline{11-11} 
\multicolumn{1}{|l|}{}                              & \multicolumn{1}{l|}{Local-gov}   & \multicolumn{1}{l|}{42.34} &  & 0.907  &  & 1.16        &  & 0                  &  & 1               \\ \cline{2-3} \cline{5-5} \cline{7-7} \cline{9-9} \cline{11-11} 
\multicolumn{1}{|l|}{}                              & \multicolumn{1}{l|}{State-gov}   & \multicolumn{1}{l|}{40.82} &  & 0.907  &  & -0.35       &  & 0                  &  & 1               \\ \cline{1-3} \cline{5-5} \cline{7-7} \cline{9-9} \cline{11-11} 
\end{tabular}
}
\caption{Explanation for \textsf{with $C^O$ explain Hours per Week for gov using F-test (work\_class.L0)}\label{tab:ftest}}
\end{table}


\end{example}

As already mentioned, the current operator covers both the \sf{ Analyze }  and the \sf{ Explain } operators of \cite{DBLP:conf/dolap/VassiliadisM18}. As open problems, one can envision the usage of all the attributes of the data set, in a setup of the operator where the attribute-list is missing for the query specification. One can also envision a specification of the operator without any explicit assignment of explanatory model (in which case all possible models are computed and subsequently assessed).  Yet, this leaves open the specification of attributes to be considered for each individual case -- to avoid the complexity, we have simply focused on the case where a specific explanation in terms of both model and attributes is requested.

\subsection{Predict}
After having observed what the current situation of the state of affairs is, and after an analyst has assessed it by comparing it to relevant benchmarks and tried to explain it by understanding the hidden correlations behind the participating variables, the next possible task is to pose the question \emph{“can you please tell me how the measurement will evolve in the next period?”}

Prediction of a future variable is typically based on the idea of using previous measurements of a time series to forecast the following measurements. The main observable of a time series is a measure, recorded at different time points as it evolves over time~\cite{forecasting14}. There is, of course, the case where a regression model is used, to relate the observed measure to the values of other variables, and, as already mentioned, this is a very powerful explanatory tool. But typically, forecasting for time series is grouped in two different types of tasks. The first type of task has to do with relating the prediction to the past via a plethora of alternative values, ranging from very naive ones (like averaging, or using drift) to elaborate ones, like exponential smoothing and even more sophisticated ones, like ARIMA models that combine the autoregression of relating the future to past values with the moving average models that predict the future based on the past prediction errors. The second type of task requires splitting the time series to three components, specifically (a) trend (for the long term behavior of the series), (b) seasonality (for repeating behaviors) and (c) error –-or noise- for the very local discrepancies of each time point from the combination of the other two factors.

The operator Predict requires (a) a measure evolving with respect to (b) a time dimension and (c) a predictive model that computes the predicted value. The simplest syntax of the operator is:\\

\sf{ with } \emph{cube} \sf{ predict next } \emph{k} \sf{ points of } \emph{measure}  [\sf{ for } \emph{subcube}] \sf{ over } \emph{time~dimension} \sf{ using } \emph{predictive~model}\\

\noindent where the predictive model can be any method of time series implemented by the system (STL, exponential smoothing, ARIMA, etc)~\cite{forecasting14} to compute the next k values of the measure.

The semantics of the execution is as follows:
\begin{itemize}
\item \emph{Data Acquisition}: we obtain the data for the (sub)cube and sort them by the values of the time dimension.

\item \emph{Model construction}: we compute the predictive model; depending on the model the output can be (a) the expected values (one per input cell) along with a vector of the forecasted $k$ points, as well as the error of the prediction, or (b) a set of measures (trend, season, noise), and of course, the k forecasted points.

\item \emph{Highlight selection}: by definition, the $k$ forecasted values are the highlights of the operation.
\end{itemize}
One can also envision the execution of multiple predictive models along with a highlight selection with respect to their error levels.

\begin{example}
Assume that we have available the time evolution of a certain measure. Although the Adult data set does not have such a measure, assume that we work with the data depicted in the  first two columns in the left-hand side of Table~\ref{tab:predict}. The data are actually the OECD countries average weekly work hours, for all declared employment. This is the old cube $C^O$. The data do not demonstrate any seasonality, as all tests confirm, so the resulting time series is practically the sum of trend and noise. We have performed a Loess trend extraction to the time series, that gave us trend and noise, and subsequently, we used autoregression over the trend to predict the next 5 years. The two antagonizing components are (a) the old values and (b) the new ones, and since, we are actually predicting values, by definition the highlight is the component with the new values.  


\begin{table}[h]
\centering
{\scriptsize 
\begin{tabular}{|l|c|l|c|l|c|}
\cline{1-2} \cline{4-4} \cline{6-6}
Year & Weekly hrs &  & Known &  & Predicted \\ \cline{1-2} \cline{4-4} \cline{6-6} 
2000 & 40.52      &  & 1     &  & 0         \\ \cline{1-2} \cline{4-4} \cline{6-6} 
2001 & 38.80      &  & 1     &  & 0         \\ \cline{1-2} \cline{4-4} \cline{6-6} 
2002 & 38.58      &  & 1     &  & 0         \\ \cline{1-2} \cline{4-4} \cline{6-6} 
2003 & 38.48      &  & 1     &  & 0         \\ \cline{1-2} \cline{4-4} \cline{6-6} 
2004 & 38.40      &  & 1     &  & 0         \\ \cline{1-2} \cline{4-4} \cline{6-6} 
2005 & 38.43      &  & 1     &  & 0         \\ \cline{1-2} \cline{4-4} \cline{6-6} 
2006 & 38.39      &  & 1     &  & 0         \\ \cline{1-2} \cline{4-4} \cline{6-6} 
2007 & 38.20      &  & 1     &  & 0         \\ \cline{1-2} \cline{4-4} \cline{6-6} 
2008 & 38.07      &  & 1     &  & 0         \\ \cline{1-2} \cline{4-4} \cline{6-6} 
2009 & 37.78      &  & 1     &  & 0         \\ \cline{1-2} \cline{4-4} \cline{6-6} 
2010 & 37.77      &  & 1     &  & 0         \\ \cline{1-2} \cline{4-4} \cline{6-6} 
2011 & 37.67      &  & 1     &  & 0         \\ \cline{1-2} \cline{4-4} \cline{6-6} 
2012 & 37.66      &  & 1     &  & 0         \\ \cline{1-2} \cline{4-4} \cline{6-6} 
2013 & 37.53      &  & 1     &  & 0         \\ \cline{1-2} \cline{4-4} \cline{6-6} 
2014 & 37.59      &  & 1     &  & 0         \\ \cline{1-2} \cline{4-4} \cline{6-6} 
2015 & 37.60      &  & 1     &  & 0         \\ \cline{1-2} \cline{4-4} \cline{6-6} 
2016 & 37.53      &  & 1     &  & 0         \\ \cline{1-2} \cline{4-4} \cline{6-6} 
\textcolor{red}{2017} & \textcolor{red}{37.73}      &  & 0     &  & \textcolor{red}{1}         \\ \cline{1-2} \cline{4-4} \cline{6-6} 
\textcolor{red}{2018} & \textcolor{red}{37.86}      &  & 0     &  & \textcolor{red}{1}         \\ \cline{1-2} \cline{4-4} \cline{6-6} 
\textcolor{red}{2019} & \textcolor{red}{37.95}      &  & 0     &  & \textcolor{red}{1}         \\ \cline{1-2} \cline{4-4} \cline{6-6} 
\textcolor{red}{2020} & \textcolor{red}{38.02}      &  & 0     &  & \textcolor{red}{1}         \\ \cline{1-2} \cline{4-4} \cline{6-6} 
\textcolor{red}{2021} & \textcolor{red}{38.07}      &  & 0     &  & \textcolor{red}{1}         \\ \cline{1-2} \cline{4-4} \cline{6-6} 
\end{tabular}
}
\caption{Predicting Weekly Hours \label{tab:predict}}
\end{table}
\end{example}

\subsection{Suggest}

The \textsf{Suggest} intention allows to answer to questions like: "where else should I be looking now?", i.e., questions asked when the phenomenon to be analyzed is not clear in the user's mind,  the overall analysis has not yet focused on some particular restricted zone of the dataset, or the user simply thinks there is more to do to investigate the phenomenon. The user expects the system to answer her intention by recommending one or more queries, by using a given recommendation strategy.

The intention here is essentially to benefit from the expertise of other users, or to let the system automatically steer the current user towards zones in the dataset that are relevant, based on the data. 
Therefore invoking the \textsf{Suggest} intention requires a recommendation model that indicates what type of suggestions is sought and what recommendation strategy will be used to compute it. 

The invocation of the \textsf{Suggest} operator follows the syntax: 

\[\sf{ with } \emph{cube} \sf{ suggest }  \sf{ using } \emph{ recommendation~model } \]

The \textsf{using} clause is optional, and if the recommendation model is omitted, different alternative recommendation strategies are tried.

We  distinguish between the following traditional recommendation model types:
 \begin{itemize}
 \item content-based: suggested queries are computed based on the data of the analyzed dataset and the data viewed by the user in the  current exploration. Models of this type include 
  YMAL  \cite{DBLP:journals/vldb/DrosouP13}, or the discovery driven operators proposed by  Cariou et al. \cite{DBLP:conf/dawak/CariouCDGGK08} and 
 Sarawagi  \cite{DBLP:conf/vldb/Sarawagi00}. The latter operator, called  INFORM,
 applies entropy maximization to lead the user to surprising parts of the cube given the user's current exploration.
 
 \item collaborative: suggested queries are computed based  on a query log and the beginning of the current exploration. Models of this type include collaborative query recommender systems that essentially differ  in the way they compute similarity between the current exploration and past explorations, like those described  in \cite{DBLP:conf/dawak/Sapia00,DBLP:conf/dawak/AufaureKMRV13,DBLP:journals/dss/AligonGGMR15}, to list a few.

\item hybrid: combination of the two above types. QueRIE is an example of a hybrid query recommender system \cite{DBLP:journals/tkde/EirinakiAPS14}, where a "mixing factor" is used to determine the importance of the content-based strategy with respect to the collaborative one. 
 \end{itemize}
 

Conceptually, and independently of its type, a recommender system can be seen as a prediction system that computes a score of interest for some  queries (the candidate recommendations), rank them, and suggest the query(ies) having the highest score. This score, together with classical quality measures associated with recommendations (diversity, serendipity, etc.), participates in the characterization of the recommendation.

The semantics of the invocation of \textsf{Suggest} is as follows:
\begin{itemize}
\item Data acquisition: the system obtains the necessary data via (a) accessing outside data needed by the recommendation model (query logs, user profiles, etc.),
(b) the appropriate cube queries to retrieve the cubes corresponding to queries that are the candidate recommendations. \rem{PV: Not for now, but we might try to think of sth that avoids actually computing the queries. In the HL selection.(i) we say the intention of the queries. But here, it is clear that we compute them. But, that's an issue not for now. }

\item Model construction: in this step, all cube queries executed during the first step receive a score corresponding to the application of the recommendation model(s) used. 
Each of them is turned into a model component.

\item Highlight selection: the goal of this step is to select the final recommendation, i.e., among all the components computed in the previous step (each representing a candidate recommendation), the one achieving the best prediction score. Displaying this recommendation under the form of a highlight is made by i) providing as output of the intention the union of all cubes corresponding to candidate recommendations, and ii) highlighting in this cube the cells of the component corresponding to the recommendation.
\end{itemize}

\begin{example}
Consider our running example with the Adult dataset. Assume a user starts analyzing the cube and only knows  the global average of this dataset, 40.93, shown in a cube named $C$. The user invokes the following Suggest intention:

\[\sf{ with } \emph{C} \sf{ suggest }  \sf{ using } \emph{ INFORM } \]

where INFORM is a version of the INFORM operator 
\cite{DBLP:conf/vldb/Sarawagi00}\footnote{In this version of the INFORM operator, Avg is used as the aggregation function.}.

Assume that the INFORM model finds in the dataset that the two cubes of Example \ref{ex:cubeAdult1} and \ref{ex:cubeAdult2}  are the most informative in the sense of its internal scoring mechanism (precisely, the Kullback-Leibler divergence between actual data and cubes where the global average is uniformly distributed), scoring respectively 0.0034 and 0.0058.

The highlight selection algorithm is then called with:
\begin{itemize}
\item the initial cube  $C$,
\item the output cube is the union of cubes $C^O$ and $C^N$ of Examples \ref{ex:cubeAdult1} and  \ref{ex:cubeAdult2},
\item each of these cubes correspond to one of the components, 
\item the proxies relationship maps each cell of
$C^O$ and $C^N$ to the corresponding cell of $C$,
\item significance(x)  returns the score of the recommendation for  $C^O$ and $C^N$ and 0 for the cells of $C$,
\item $surprise(x,y)=x-y$,
\item ${\cal A}^C(x)$ outputs e.g., $max(x)$ to get the component's score.
\end{itemize}

The algorithm outputs the component corresponding to the query producing the cube $C^N$.
\begin{table}[]
\centering
{\scriptsize 
\begin{tabular}{llll|c|l|c|l|c|}
\cline{5-5} \cline{7-7} \cline{9-9}
     &          &        &  & Score &  & Candidate 1 &  & Candidate 2 \\      \cline{1-3} \cline{5-5} \cline{7-7} \cline{9-9} 
\multicolumn{1}{|l|}{\multirow{9}{*}{Assoc}}        & \multicolumn{1}{l|}{Federal-gov}      & \multicolumn{1}{l|}{\textcolor{red}{41.15}} &  & 0.0058 &  & 0           &  & \textcolor{red}{1}           \\ \cline{2-3} \cline{5-5} \cline{7-7} \cline{9-9} 
\multicolumn{1}{|l|}{}                              & \multicolumn{1}{l|}{Local-gov}        & \multicolumn{1}{l|}{\textcolor{red}{41.33}} &  & 0.0058 &  & 0           &  & \textcolor{red}{1}           \\ \cline{2-3} \cline{5-5} \cline{7-7} \cline{9-9} 
\multicolumn{1}{|l|}{}                              & \multicolumn{1}{l|}{State-gov}        & \multicolumn{1}{l|}{\textcolor{red}{39.09}} &  & 0.0058 &  & 0           &  & \textcolor{red}{1}           \\ \cline{2-3} \cline{5-5} \cline{7-7} \cline{9-9} 
\multicolumn{1}{|l|}{}                              & \multicolumn{1}{l|}{Gov}              & \multicolumn{1}{l|}{40.73} &  & 0.0034 &  & 1           &  & 0           \\ \cline{2-3} \cline{5-5} \cline{7-7} \cline{9-9} 
\multicolumn{1}{|l|}{}                              & \multicolumn{1}{l|}{Private}          & \multicolumn{1}{l|}{\textcolor{red}{41.06}} &  & 0.0058 &  & 0           &  & \textcolor{red}{1}           \\ \cline{2-3} \cline{5-5} \cline{7-7} \cline{9-9} 
\multicolumn{1}{|l|}{}                              & \multicolumn{1}{l|}{Private}          & \multicolumn{1}{l|}{41.06} &  & 0.0034 &  & 1           &  & 0           \\ \cline{2-3} \cline{5-5} \cline{7-7} \cline{9-9} 
\multicolumn{1}{|l|}{}                              & \multicolumn{1}{l|}{Self-emp-inc}     & \multicolumn{1}{l|}{\textcolor{red}{48.68}} &  & 0.0058 &  & 0           &  & \textcolor{red}{1}           \\ \cline{2-3} \cline{5-5} \cline{7-7} \cline{9-9} 
\multicolumn{1}{|l|}{}                              & \multicolumn{1}{l|}{Self-emp-not-inc} & \multicolumn{1}{l|}{\textcolor{red}{45.88}} &  & 0.0058 &  & 0           &  & \textcolor{red}{1}           \\ \cline{2-3} \cline{5-5} \cline{7-7} \cline{9-9} 
\multicolumn{1}{|l|}{}                              & \multicolumn{1}{l|}{Self-emp}         & \multicolumn{1}{l|}{46.68} &  & 0.0034 &  & 1           &  & 0           \\ \cline{1-3} \cline{5-5} \cline{7-7} \cline{9-9} 
\multicolumn{1}{|l|}{\multirow{9}{*}{Post-grad}}    & \multicolumn{1}{l|}{Federal-gov}      & \multicolumn{1}{l|}{\textcolor{red}{43.86}} &  & 0.0058 &  & 0           &  & \textcolor{red}{1}           \\ \cline{2-3} \cline{5-5} \cline{7-7} \cline{9-9} 
\multicolumn{1}{|l|}{}                              & \multicolumn{1}{l|}{Local-gov}        & \multicolumn{1}{l|}{\textcolor{red}{43.96}} &  & 0.0058 &  & 0           &  & \textcolor{red}{1}          \\ \cline{2-3} \cline{5-5} \cline{7-7} \cline{9-9} 
\multicolumn{1}{|l|}{}                              & \multicolumn{1}{l|}{State-gov}        & \multicolumn{1}{l|}{\textcolor{red}{42.96}} &  & 0.0058 &  & 0           &  & \textcolor{red}{1}           \\ \cline{2-3} \cline{5-5} \cline{7-7} \cline{9-9} 
\multicolumn{1}{|l|}{}                              & \multicolumn{1}{l|}{Gov}              & \multicolumn{1}{l|}{43.58} &  & 0.0034 &  & 1           &  & 0           \\ \cline{2-3} \cline{5-5} \cline{7-7} \cline{9-9} 
\multicolumn{1}{|l|}{}                              & \multicolumn{1}{l|}{Private}          & \multicolumn{1}{l|}{\textcolor{red}{45.19}} &  & 0.0058 &  & 0           &  & \textcolor{red}{1}           \\ \cline{2-3} \cline{5-5} \cline{7-7} \cline{9-9} 
\multicolumn{1}{|l|}{}                              & \multicolumn{1}{l|}{Private}          & \multicolumn{1}{l|}{45.19} &  & 0.0034 &  & 1           &  & 0           \\ \cline{2-3} \cline{5-5} \cline{7-7} \cline{9-9} 
\multicolumn{1}{|l|}{}                              & \multicolumn{1}{l|}{Self-emp-inc}     & \multicolumn{1}{l|}{\textcolor{red}{53.05}} &  & 0.0058 &  & 0           &  & \textcolor{red}{1}           \\ \cline{2-3} \cline{5-5} \cline{7-7} \cline{9-9} 
\multicolumn{1}{|l|}{}                              & \multicolumn{1}{l|}{Self-emp-not-inc} & \multicolumn{1}{l|}{\textcolor{red}{43.39}} &  & 0.0058 &  & 0           &  & \textcolor{red}{1}           \\ \cline{2-3} \cline{5-5} \cline{7-7} \cline{9-9} 
\multicolumn{1}{|l|}{}                              & \multicolumn{1}{l|}{Self-emp}         & \multicolumn{1}{l|}{47.24} &  & 0.0034 &  & 1           &  & 0           \\ \cline{1-3} \cline{5-5} \cline{7-7} \cline{9-9} 
\multicolumn{1}{|l|}{\multirow{9}{*}{Some-college}} & \multicolumn{1}{l|}{Federal-gov}      & \multicolumn{1}{l|}{\textcolor{red}{40.31}} &  & 0.0058 &  & 0           &  & \textcolor{red}{1}           \\ \cline{2-3} \cline{5-5} \cline{7-7} \cline{9-9} 
\multicolumn{1}{|l|}{}                              & \multicolumn{1}{l|}{Local-gov}        & \multicolumn{1}{l|}{\textcolor{red}{40.14}} &  & 0.0058 &  & 0           &  & \textcolor{red}{1}           \\ \cline{2-3} \cline{5-5} \cline{7-7} \cline{9-9} 
\multicolumn{1}{|l|}{}                              & \multicolumn{1}{l|}{State-gov}        & \multicolumn{1}{l|}{\textcolor{red}{34.73}} &  & 0.0058 &  & 0           &  & \textcolor{red}{1}           \\ \cline{2-3} \cline{5-5} \cline{7-7} \cline{9-9} 
\multicolumn{1}{|l|}{}                              & \multicolumn{1}{l|}{Gov}              & \multicolumn{1}{l|}{38.38} &  & 0.0034 &  & 1           &  & 0           \\ \cline{2-3} \cline{5-5} \cline{7-7} \cline{9-9} 
\multicolumn{1}{|l|}{}                              & \multicolumn{1}{l|}{Private}          & \multicolumn{1}{l|}{\textcolor{red}{38.73}} &  & 0.0058 &  & 0           &  & \textcolor{red}{1}           \\ \cline{2-3} \cline{5-5} \cline{7-7} \cline{9-9} 
\multicolumn{1}{|l|}{}                              & \multicolumn{1}{l|}{Private}          & \multicolumn{1}{l|}{38.73} &  & 0.0034 &  & 1           &  & 0           \\ \cline{2-3} \cline{5-5} \cline{7-7} \cline{9-9} 
\multicolumn{1}{|l|}{}                              & \multicolumn{1}{l|}{Self-emp-inc}     & \multicolumn{1}{l|}{\textcolor{red}{49.31}} &  & 0.0058 &  & 0           &  & \textcolor{red}{1}           \\ \cline{2-3} \cline{5-5} \cline{7-7} \cline{9-9} 
\multicolumn{1}{|l|}{}                              & \multicolumn{1}{l|}{Self-emp-not-inc} & \multicolumn{1}{l|}{\textcolor{red}{44.03}} &  & 0.0058 &  & 0           &  & \textcolor{red}{1}           \\ \cline{2-3} \cline{5-5} \cline{7-7} \cline{9-9} 
\multicolumn{1}{|l|}{}                              & \multicolumn{1}{l|}{Self-emp}         & \multicolumn{1}{l|}{45.7}  &  & 0.0034 &  & 1           &  & 0           \\ \cline{1-3} \cline{5-5} \cline{7-7} \cline{9-9} 
\multicolumn{1}{|l|}{\multirow{9}{*}{Univesity}}    & \multicolumn{1}{l|}{Federal-gov}      & \multicolumn{1}{l|}{\textcolor{red}{43.38}} &  & 0.0058 &  & 0           &  & \textcolor{red}{1}           \\ \cline{2-3} \cline{5-5} \cline{7-7} \cline{9-9} 
\multicolumn{1}{|l|}{}                              & \multicolumn{1}{l|}{Local-gov}        & \multicolumn{1}{l|}{\textcolor{red}{42.34}}&  & 0.0058 &  & 0           &  & \textcolor{red}{1}           \\ \cline{2-3} \cline{5-5} \cline{7-7} \cline{9-9} 
\multicolumn{1}{|l|}{}                              & \multicolumn{1}{l|}{State-gov}        & \multicolumn{1}{l|}{\textcolor{red}{40.82}} &  & 0.0058 &  & 0           &  & \textcolor{red}{1}           \\ \cline{2-3} \cline{5-5} \cline{7-7} \cline{9-9} 
\multicolumn{1}{|l|}{}                              & \multicolumn{1}{l|}{Gov}              & \multicolumn{1}{l|}{42.14} &  & 0.0034 &  & 1           &  & 0           \\ \cline{2-3} \cline{5-5} \cline{7-7} \cline{9-9} 
\multicolumn{1}{|l|}{}                              & \multicolumn{1}{l|}{Private}          & \multicolumn{1}{l|}{\textcolor{red}{43.06}} &  & 0.0058 &  & 0           &  & \textcolor{red}{1}           \\ \cline{2-3} \cline{5-5} \cline{7-7} \cline{9-9} 
\multicolumn{1}{|l|}{}                              & \multicolumn{1}{l|}{Private}          & \multicolumn{1}{l|}{43.06} &  & 0.0034 &  & 1           &  & 0           \\ \cline{2-3} \cline{5-5} \cline{7-7} \cline{9-9} 
\multicolumn{1}{|l|}{}                              & \multicolumn{1}{l|}{Self-emp-inc}     & \multicolumn{1}{l|}{\textcolor{red}{49.91}} &  & 0.0058 &  & 0           &  & \textcolor{red}{1}           \\ \cline{2-3} \cline{5-5} \cline{7-7} \cline{9-9} 
\multicolumn{1}{|l|}{}                              & \multicolumn{1}{l|}{Self-emp-not-inc} & \multicolumn{1}{l|}{\textcolor{red}{44.44}} &  & 0.0058 &  & 0           &  & \textcolor{red}{1}           \\ \cline{2-3} \cline{5-5} \cline{7-7} \cline{9-9} 
\multicolumn{1}{|l|}{}                              & \multicolumn{1}{l|}{Self-emp}         & \multicolumn{1}{l|}{46.61} &  & 0.0034 &  & 1           &  & 0           \\ \cline{1-3} \cline{5-5} \cline{7-7} \cline{9-9} 
\end{tabular}
}
\caption{Suggesting 
\label{my-label}}
\end{table}
\end{example}

%% file: 50_experiments.tex
\section{Experiments}\label{sec:experiments}

We have implemented a Cube Query Engine as a research prototype to accommodate our proposal. We call our system Delian Cube Engine (to honor the mathematical Delian Problem); the code is publicly available as Free Open Source Software at \url{https://github.com/pvassil/DelianCubeEngine}.

\begin{figure*}[h]
	\centering
	\includegraphics[width=0.99\linewidth]{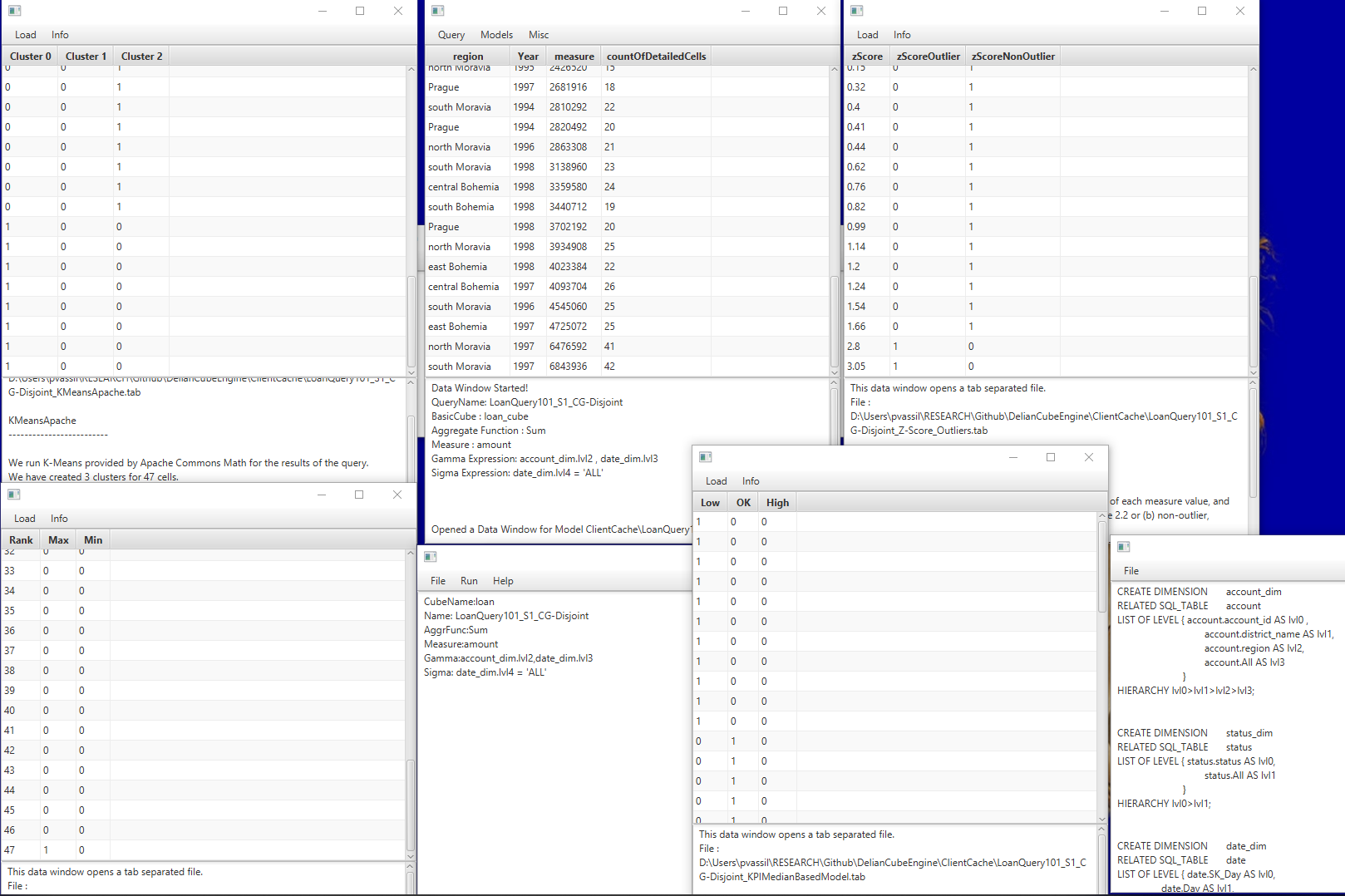}
	\caption{Delian Cube Engine: A dimension and cube description (bottom right), a cube query and its result (bottom and top middle, respectively) and a set of models and their model components as bitmaps.}\label{fig:queryNmodels}
\end{figure*}

Our Delian Cube Engine operates on top of a relational DBMS to support query answering and allows the registration of dimension hierarchies and cubes, in order to facilitate the answering of aggregate queries along the lines prescribed in the respective definition, involving selections on the levels, aggregations (and, internally, implicit joins of the fact to the dimension tables, in order to correctly construct the respective relational query). Practically, the user needs to map the tables of an underlying star schema to the respective dimension and cube structures. Once the cube dataset is registered and mapped to its underlying relational database, the system is ready to be queried. The user deals with cube queries, instead of using SQL, in a simple cube query language. The queries are passed from the front-end to the back-end of the engine for processing via an RMI connection. There, the following steps take place: (a) there is a model selection phase, where model extraction algorithms are selected to be applied on the results of the cube query, (b) there is a translation of the cube query into the respective SQL query that will actually be executed over the underlying database, (c) the resulting recordset is again translated back to cells and a cube is produced as an answer, (d) the selected model extraction algorithms are applied to the result of the query, (e) all the resulting cells, cubes, models, and model components are packaged and transferred back to the front-end for presentation (Fig.~\ref{fig:queryNmodels}).

To assess the practicality of the method we have worked with a cubified version of the PKDD99 schema. The setup of both the client and the server parts, as well as the execution of all experiments, have been performed on a rudimentary laptop with an Intel i5-7200U CPU at 2.5GHz, 8GB install RAM, and a 512GB disk.
We have worked with the original PKDD99 data set on bank loans, which is of small size, and artificially generated larger versions of it at 1M and 10M rows. The data set has a cube on loans; the amount loaned is the measure and the dimensions are accounts (generalized to geographical regions, with 3 levels), date, and status of the loan (practically single level). We execute a session that (1) starts by querying the entire cube, (2) focuses on the loan contracts that are running without problems, (3 and 4) compares drilled-down variants in terms of geography and date, and, finally, (5) re-focuses on a particular region where all contracts are queried. The execution of the queries is accompanied with the generation and execution of models: ranking, outlier detection, K-Means and KPI assessment.

\begin{figure*}[tbph]
	\centering
	\includegraphics[width=0.99\linewidth]{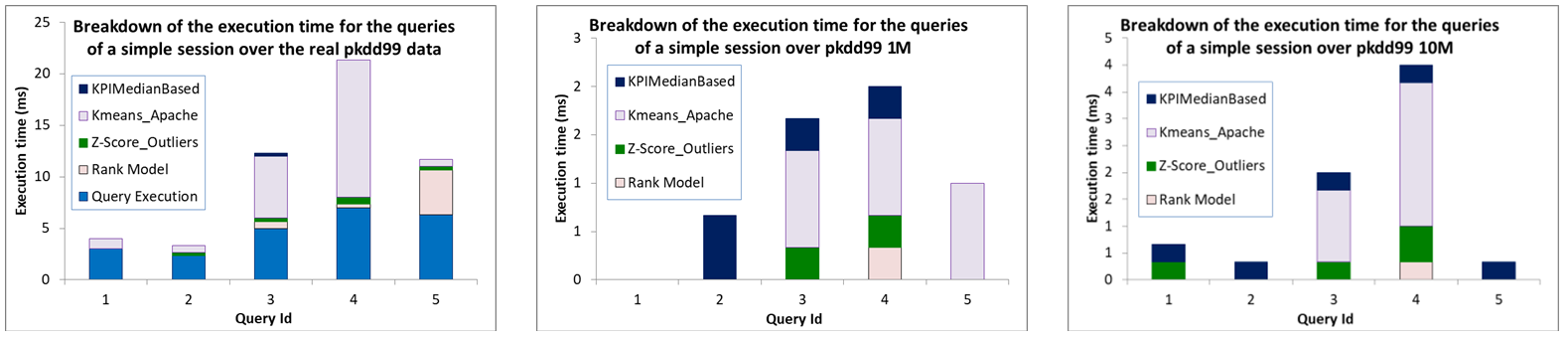}
	\caption{Breakdown of query and model execution time for the same query session, over different sizes for the same schema. All times are with hot cache.}\label{fig:modelGenerationTimes}
\end{figure*}

The querying time scales linearly with the cube size, and we depict them only in the first of the sub-Figures of Fig.~\ref{fig:modelGenerationTimes} to avoid visually suppressing the key message of the experiments which is found in the model generation and execution: in all cases, the model generation requires simply a few milliseconds (and frequently, fractions of them)! This is expected, as the models operate on the query result and not on the underlying cube. Thus, the psychological limit of 500 msec for producing an answer is not affected at all by the application of our models, in our experiment. We recognize that the visualizations employed are na\"{\i}ve, and thus fail to plug-in any time costs due to the visualization part ---however, we consider this to be a matter of an open, wide field of research for the future.

%% file: 60_RW.tex
\section{Related work}\label{sec:RW}

\subsection{Coupling data and models}

The idea of coupling data and analytical models is not completely new. Already in the mid-90's, inductive databases were proposed to couple data with patterns, i.e., generalizations extracted from the data. In this framework knowledge discovery is modelled as an interactive process in which users can query data as well as patterns using an ad hoc query language \cite{DBLP:journals/sigkdd/Raedt02}.

More specifically, in \cite{DBLP:series/anis/Pedersen09}, among the research challenges in BI the authors emphasized the need to achieve a unified view of data and models that describe data, so that this two components can be used and queried together. The intuition proposed to move from data to models and vice-versa is that of folding data into models and unfolding models into data. Our approach goes exactly in this direction, being folding/unfolding achieved through the definition of components.

Some degree of data-to-model unification is actually achieved in MauveDB \cite{DBLP:conf/sigmod/DeshpandeM06}, which provides language constructs for declaratively specifying \emph{model-based views} of data based on a variety of commonly used statistical models, meant as simplified descriptions of the underlying data. Though the authors recognize that the view definition has to be model-specific, they suggest to rely on the common aspects of different models to decrease the variation in the view-definition statements. They also indicate as the most promising approach to query processing over model-based views that of materializing an intermediate model-specific representation of the view. While their work is mostly focused on maintaining model-based views and transparently querying them in a SQL-like fashion, ours introduces intentional operators as a querying abstraction and uses highlights to emphasize relevant findings.

Most recently, the Northstar system has been proposed as a support to interactive data science \cite{DBLP:journals/pvldb/Kraska18}. Importantly, among the key requirements for interactive data science, the author mentions that of enabling users to seamlessly switch between data exploration and model building. To this end he developed the Alpine Meadow optimizer, which features a declarative language for machine learning tasks and a real-time strategy for hyper-parameter tuning.

\subsection{Exploratory querying and data exploration}\label{subsec:Explore}


Interactive Database Exploration (IDE) is the process of exploring a database by means of a sequence of queries aiming at answering an often imprecise user information need. 
Typically, an exploration includes several queries where the result of each query triggers the formulation of the next one. 
Many approaches have recently been developed to support IDE, as illustrated by a recent survey of the topic \cite{DBLP:conf/sigmod/IdreosPC15}.
In their survey, Idreos et al. adopt a top-down viewpoint and classify the existing approaches in three main categories: user interaction, middleware and database layer.
Techniques range from visual optimization (like query result reduction \cite{DBLP:conf/bigdataconf/BattleSC13}), automatic exploration (like query recommendation \cite{DBLP:journals/vldb/DrosouP13}), assisted query formulation (like data space segmentation \cite{DBLP:conf/cidr/SellamK13}), data prefetching (like result diversification \cite{DBLP:conf/ssdbm/KhanSA14}) and query approximation \cite{DBLP:conf/sigmod/HellersteinHW97}.

One of the conclusion of this survey is that  declarative “exploration” languages are still to be invented. We believe that the present work is a first step towards such a language.

OLAP exploration of data warehouses is a particular use case of database exploration that enables to work with simplifying assumptions, precisely, the multidimensional star schema or the regularities of multidimensional queries.
Many approaches have been specifically developed to support OLAP  exploration, as illustrated by the next subsection.

\subsection{OLAP models and operators}\label{subsec:advancedOLAPOperators}
The research on traditional models for OLAP and its operators (roll-up, drill-down, slice, drill-across) practically concluded around the turn of the millennium. We refer the interested reader to an excellent survey~\cite{DBLP:conf/dawak/RomeroA07}. 

Apart from the traditional operators, related research has explored the possibility of providing operators with more knowledge extraction results. In an emblematic paper in the area, Sarawagi introduces the \emph{DIFF} operator in \cite{DBLP:conf/vldb/Sarawagi99}, which returns a set of tuples that most successfully describe the difference of values between two cells of a cube that are given as input. The same author in \cite{DBLP:conf/vldb/Sarawagi00} describes a method that profiles the exploration of a user and uses the
Maximum Entropy principle to recommend which unvisited parts of the cube can be the most surprising in a subsequent query. In \cite{DBLP:conf/vldb/SatheS01}, Sathe and Sarawagi introduce the operator \emph{RELAX} which verifies whether a pattern observed at a certain level of detail is also present at a coarser level of detail, too.

The Cinecubes method, introduced in \cite{DBLP:conf/dolap/GkesoulisV13} and \cite{DBLP:journals/is/GkesoulisVM15} is aimed to provide automated reporting as a result to an original OLAP query. The proposed method enriches an original OLAP query with auxilliary queries to aid (a) the comparison and assessment of the result of the query to similar data and (b) the explanation of the result with values at the most detailed level. So, the result of the Cinecubes system can coarsely be grouped as the result of two operators: the first operator computes queries for values similar to ones defining the selection filters of the original query and the second one by drilling down into the dimensions of the result, one dimension at a time. 
The Cinecubes method also comes with the features of (a) (simple) highlight extraction, (b) packaging the result as a \emph{story}, presented as a Powerpoint story, with text commenting on the highlights, audio produced automatically from the text and visualization of the query results in slides. We refer the interested reader to \cite{DBLP:journals/is/GkesoulisVM15}  which also includes the discussion of \textbf{data narration and visualization}, that is not covered here.

The Shrink operator \cite{DBLP:journals/dke/GolfarelliGR14,DBLP:journals/jdwm/RizziGG15} goes in the direction of {\em approximate query answering}, whose main goal is to increase query efficiency by returning a reduced result while minimizing the approximation introduced. In particular, Shrink aims at balancing data precision with data size in cube visualization via pivot tables. To this end it takes as input a cube and compresses it to a given target size; this is done by fusing cube slices into a single representative slice, in such a way that the information loss is minimal. Other approaches to approximate query answering for OLAP are presented in \cite{AGP00,feng2002compressed,buccafurri2003quad}.

Finally, in \cite{DBLP:conf/vldb/ChenCLR05} the OLAP paradigm is reused to explore prediction cubes besides traditional data cubes. In a prediction cube, each cell summarizes a predictive model trained on the data corresponding to that cell. Specifically, each cell can measure either the accuracy of the model, or the similarity between two models, or the model predictiveness based on a test-set. This approach empowers decision making by supporting users in searching for interesting subsets of data in the light of a prediction model; however, differently from our approach, it does not identify highlights nor it provides a goal-oriented query language.

\subsection{Query Recommendations}\label{subsec:queryRec}
Query recommendation refers to the situation where (a) a user submits a query to the system and (b) the system follows up with suggesting subsequent queries to aid the user's exploration.  The suggestion can be based on the user's profile, history of queries, history of other users' queries, or other information.
Query recommendation has recently attracted many attentions for Interactive Database Exploration 
\cite{DBLP:conf/persdb/DrosouPS09,DBLP:journals/vldb/DrosouP13,DBLP:journals/tkde/EirinakiAPS14}
and particularly in the case of OLAP exploration of data cubes
\cite{DBLP:conf/dawak/JerbiRTZ09,DBLP:conf/dexa/BentayebF09,DBLP:journals/jdwm/GiacomettiMNS11,DBLP:conf/dawak/AufaureKMRV13,DBLP:journals/dss/AligonGGMR15,DBLP:conf/dexa/NegreRT18,DBLP:journals/is/DrushkuALMP19}, including approaches applied to spatial OLAP \cite{DBLP:conf/iimss/LayouniAA16,DBLP:journals/ijbis/BimonteN16}.
See \cite{DBLP:conf/persdb/DrosouPS09,DBLP:conf/eda/MarcelN11,DBLP:journals/dss/AligonGGMR15,DBLP:conf/iceis/Kozmina15} for a broader discussion.

As it is the case for general purpose recommender systems, query recommendation methods can be classified into content-based approaches, collaborative approaches, or combination thereof. 
Content-based approaches compute suggestions based on the data seen by the user and and unseen data. 
For instance, Cariou et al., in \cite{DBLP:conf/dawak/CariouCDGGK08} describe a method that mines the most interesting dimension for a user to explore, based on his history. As already mentioned, a similar problem has been addressed in \cite{DBLP:conf/vldb/Sarawagi00}. 
Jerbi et al. \cite{DBLP:conf/dawak/JerbiRTZ09}
propose an approach
where recommendations are computed based on
the navigation context and 
 preferences stated in the user profile. 
Collaborative approaches take advantage of the wisdom-of-the-crowd effect. The authors of 
\cite{DBLP:journals/jdwm/GiacomettiMNS11,DBLP:journals/dss/AligonGGMR15} use the query log of previous users to find similar queries which can give information to user who may not know it is available. 
Drushku et al. \cite{DBLP:journals/is/DrushkuALMP19}
use the query log of previous users to detect user intents
and recommend queries that fit the user's current intents.

As explained in Section \ref{sec:transitions}, our \textsf{suggest} intention encapsulates these  recommendation strategies in their variety.


\subsection{Intensional querying} 

Non-conventional query answering includes a variety of answering
mechanisms and happens when either the user has no clear formulation
of his needs (e.g., he does not know what he really wants) or has a
good understanding of his needs but is flexible enough to accept an
alternate, approximate or intensional answer. Among several kind of non-conventional answers, we distinguish intensional query answering as the more relevant to our work. According to
\cite{Motro2000}, an intensional query answer complements the
extensional one by including either a concise description of the
answer or some useful facts about it. 



Intensional query answering relies generally on knowledge like integrity constraints,
inference rules (in knowledge-based systems), ontology (and more
frequently a taxonomy), and user's preferences to either provide
more insight about the extensional answer or give an approximate
answer. 

Intensional query answering has been applied
in many area of computer science (e.g., object-oriented databases \cite{DBLP:conf/cikm/YoonSP94}, deductive
database \cite{DBLP:conf/dagstuhl/Flach98}, and question answering systems \cite{DBLP:conf/inlg/Benamara04,DBLP:journals/dke/CimianoRH10}) but, to the best of our
knowledge, the only work related to the OLAP area is \cite{DBLP:conf/dolap/MarcelMR12}, that proposes a framework
for computing an intensional answer to an OLAP query by leveraging the
previous queries in the current session. The idea is to use an intensional answer
to concisely characterize the cube regions whose data do not change along the
sequence, coupled with an extensional answer to describe in detail only the cube
regions whose data significantly differ from the expectation.

\subsection{Interestingness}

As explained in Section \ref{sec:interestingness}, our approach for highlight selection based on a significance score is inspired by the framework for subjective interestingness in exploratory data mining proposed by De Bie \cite{DBLP:conf/ida/Bie13}. The framework is based on the idea that the goal of the explorative pattern mining is to pick patterns that will result in the best updates of the user’s belief state, while presenting a minimal strain on the user’s resources. In this sense, an interestingness measure (IM) is subjective in that it depends on the belief of the explorer. A general definition for this IM is a real-valued function of a background distribution (that represents the belief) and a pattern (the artifact to be presented to the explorer). The belief is the probability P$(\omega)$ of the event  $x \in \omega$, i.e., the degree of belief the user attaches to a pattern (characterized by $\omega$) being present in the data $x$. In other words, if this probability is small, then the pattern is subjectively surprising for the explorer. The data mining process consists of extracting patterns and presenting first those that are subjectively surprising, and then refining the belief.
Our principle of highlight selection is inspired by this framework, in the sense that the highlights to be selected maximize the surprise caused by confronting a significance score computed for  the initial cube (that can be thought of as the user belief) with the significance score computed for the target cube to select a model component (that can be thought of as the pattern to be presented). \rem{In the future, we might try to add some stuff for the surveys of inter., as well as for the stuff that we once wrote for interest. }

\subsection{Principles behind the foundation of our operators}
The foundations of what are the basic operators that users perform during OLAP and data exploration have been a really difficult problem in our search. The related literature that we could find does not particularly help towards this direction.

We start by observing that our approach goes in the direction of reducing the number of interactions a user needs to achieve her analysis goals. Specifically, with reference to the Observation-Orientation-Decision-Action loop applied to BI \cite{DBLP:conf/dolap/Middelfart07}, intentional analytics empowers both the \emph{observation} phase (look at the data with an expectation of what it should be) and the \emph{orientation} phase (we look at the data in different ways depending on what it shows).

Interestingly, intent-driven query formulation in the OLAP context has been investigated to some extent in \cite{DBLP:conf/er/MiddelfartP11}, which proposes \emph{meta-morphing} as a way to have incomplete user queries completed by the system based on the previous interactions of that user. This approach is different from ours mainly in two ways: (i) an \emph{intention} for us is not an incomplete query but a high-level analytical goal, and (ii) query answers for us encompass both data and models.

In \cite{DBLP:journals/tvcg/KandelPHH12} the authors perform a semi-structured interview with 25 analysts to understand what data analysis and visualization entails as a process. The authors come up with 5 high-level tasks: discovery, wrangling, profiling, modeling and reporting of information. Expectedly (at least for the educated data warehouse experts)  discovery and wrangling of data are the most tedious of the tasks. However, when it comes to the environment of OLAP, which is performed over simply-but-neatly organized cubes, these two tasks, along with profiling (i.e., data quality assurance) have already been completed, either by the organization ETL workflow, or by a do-it-yourself data wrangling. The rest of the high-level tasks are too few and too high for our purpose here.

In a similar vein, in \cite{DBLP:journals/tvcg/BatchE18}, the authors propose a taxonomy of user tasks in exploratory data analysis that include (1)  discovery (hypothesis formulation and determination of the data source that can answer it) , (2) data acquisition (and preparation), (3) exploration of the data, (4) modeling, via the construction of a model that explains the data, and, (5) communication of the results to other people via reports and presentations. The discussed taxonomy is very close to the one presented in  \cite{DBLP:journals/tvcg/KandelPHH12} , but similarly suffers from the high-level of abstraction for the exploration part. Having said that, we also say that we are very close to the fundamental principle of working with data that this paper outlines: once the data are ready, users explore to find out what the status is and to understand the data, and construct models that explain the phenomena -- what we add is more detail on the exploration and, notably, automation of the process.

Another path that we followed was to search for foundations for on-line search in the web, which is a similar area. In  \cite{DBLP:journals/ipm/JansenBS09} the authors conduct a user study of 72 participants engaged in 426 user tasks and deduce it is worthwhile to deduce the intentions that drive the users to perform their searches. We quote from the abstract of the paper: "The implication of this research is that rather than solely addressing a searcher’s expressed information need, searching systems can also address the underlying learning need of the user." After a detailed survey of the literature on learning styles, the authors suggest that the best candidate to serve as the foundation of the learning tasks of users is the cognitive learning framework that was proposed by Anderson and Krathwohl as a refinement to Bloom’s taxonomy.

It is worthwhile to discuss Bloom's taxonomy and Anderson and Krathwohl's refinement to it \cite{Krathwohl2002}, \cite{Wilson2016}. The framework tries to categorize the different domains of human learning (mainly with a view to children education), and includes the following main cognitive tasks: (a)  remembering, (b) understanding (by extracting meaning out of messages or activities), (c) applying the material learned via a procedure, (d) analyzing by understanding how the different parts of artifacts or concepts relate to each other, (e) evaluating based on criteria and standards, (f) creating of new artifacts by composing individual pieces into a coherent or functional assembly. Our framework, tries to automate the remembering and application, but at the same time retains the understanding (via the description of the status in different ways and levels of abstraction), the analysis (via the explanations provided) and the evaluation parts. We cover the creation part at the future work section.


\subsection{Relationship to our previous work} 
A first version of this paper appears in \cite{DBLP:conf/dolap/VassiliadisM18}. The largest part of \cite{DBLP:conf/dolap/VassiliadisM18} is included in Section~\ref{sec:overview}. The rest of the contents of the paper are completely novel. The initial set of intentional operators of \cite{DBLP:conf/dolap/VassiliadisM18} has been abstracted further in this paper and replaced by the operators of Section~\ref{sec:transitions}. The parts on the model's definition and the interestingness evaluation are not covered in \cite{DBLP:conf/dolap/VassiliadisM18}.

%% file: 70_conclusions.tex


\section{Conclusions and paths for future research}\label{sec:conclusions}
This paper is a vision paper describing, in broad terms, a potential future for OLAP, to strengthen its place as the corner stone of BI. We are convinced that, after 50 years of query answering, it is now time to replace it with effortless, automatic insight gaining from the user. Instead of making the end user dig into sets of records, we can increase productivity and the understanding of the essence of the data by using two pillars, one devoted to querying (i.e., what an OLAP query is), and another devoted to answering (i.e., what the answer to an OLAP query is). Specifically, firstly, we want to allow the user to focus on \emph{high-level goals} of information acquisition, rather than details of what data to bring in, and secondly, to automatically suggest focus-points in the answers that will move the user effort from manual "jewel mining" to addressing the insights gained.

Beyond the proposed new viewpoint to OLAP, our call to arms to the research community involves several open roads for research.

\emph{New intentional operators}. In this paper we have proposed a set of fundamental operators for the OLAP tools of the close future. New operators, esp., as combinations of old ones can also be devised.  We would assume as a pre-requisite for each such operator to come with a graceful linkage to the overall model proposed here (in an attempt to be able to gracefully plug it in the respective BI tools under a uniform setting).

\emph{Alternatives for full automation}. Much like in traditional query processing, the intentional operators can come with alternative execution algorithms for the data collection and the model construction, in order to facilitate the optimization of the task. The optimization can be thought, not only in terms of performance, but also in terms of information content delivered. Of course, the fine tuning of any algorithms concerning their parameter fixing is also important. 

\emph{Optimization concerns}. In practice, intentional operators can be envisioned as sequences of  different logical OLAP and ML operators. Besides, the cube resulting from each logical operator can be enhanced with highlights obtained by applying different algorithms (e.g., clustering or classification). Thus, much like for the selection of an execution plan for a SQL query, executing an intentional operator requires an optimization phase to decide which logical operators and model mining algorithms to execute. How exactly this optimizer will be structured internally, and what optimization algorithms and tunings will be employed is clearly a topic of future research.\rem{RV1.PDF: Good visions. Be more specific} 

 

\emph{Packaging into data stories}. In this paper we have focused on the deeper layers of query answering, which involves data acquisition and mining. The visual representation of all these results, the automatic choice of graphical representation, the automatic generation of data stories and the overall packaging to the user are topics that, despite their importance, have traditionally been outside the walls of the database community, mainly due to the difficulty of experimentally verifying the effectiveness of any proposed method. Still, answers to the automation of the aforementioned tasks are highly valuable and pose open research topics.

\emph{Benchmarking and tools}. A free, open-source (FOSS) tool and a reference benchmark for the future BI (involving data, model, and highlight extraction requests and sessions) can be a really handy tool for the research community (otherwise, each new paper will need to improvise on its experimental assessment). A tool will also trigger other research directions like, e.g., the incorporation of research results on natural language processing to accept the user requests, new visualizations to show models and highlights, etc. \rem{all this for non-OLAP data?} \newline


\textbf{Acknowledgments}. We are particularly thankful to the reviewers of both this paper and its preliminary version in DOLAP 2018 for their comments that helped us enrich the breadth of the related work, and the clarity of concepts and terminology.

%% file: 91_data-queries-model.tex
\section{Formalizing data, dimension hierarchies cubes and cube queries}
An \emph{OLAP session} is a sequence of \emph{dashboards}  that the analyst sees, each with its own information, including data, charts and informative summaries of KPI performance. The sequence is produced by the actions of the analyst that changes the contents of the dashboard by requesting more information on the basis of a set of \emph{operations} made available to him by the tool. 

\subsection{Preliminaries on multidimensional modeling}
In this Section, we give the formal background of our modeling concerning multidimensional databases, hierarchies and queries. We closely follow the model of \cite{DBLP:journals/is/GkesoulisVM15} and slightly extend it.
As typically happens with multidimensional models, we assume that \emph{dimensions} provide a \emph{context} for facts
\cite{DBLP:series/synthesis/2010Jensen}. This is especially
important considering that dimension values come in
\emph{hierarchies}; every single fact can be
simultaneously placed in multiple hierarchically-structured
contexts, thus giving users the possibility of analyzing sets of facts from
different perspectives. The underlying data sets include \emph{measures} that are characterized with respect to these dimensions. \emph{Cube queries} involve measure aggregations at specific levels of granularity per dimension, along with filtering of data for specific values of interest.

\subsection{Domains, dimensions and underlying data}

\textbf{Domains}. We assume the following infinitely countable and
pairwise disjoint sets: a set of \emph{level names} (or simply
\result{levels}) $\mathcal{U_{L}}$, a set of \emph{measure names}
(or simply \result{measures}) $\mathcal{U_{M}}$, a set of
\emph{regular~data~columns} $\mathcal{U_{A}}$, a set of
\emph{dimension} names (or simply \emph{dimensions})
$\mathcal{U_{D}}$ and a set of \emph{cube} names (or simply
\emph{cubes}) $\mathcal{U_{C}}$. The set of \emph{data~columns}
 $\mathcal{U}$ is defined as $\mathcal{U}$ =
$\mathcal{U_{L}}$  $\cup$ $\mathcal{U_{M}}$ $\cup$
$\mathcal{U_{A}}$. For each $L$ $\in$ $\mathcal{U_{L}}$, we define
a countable totally ordered set $dom(L)$, the domain of $L$, which
is isomorphic to the integers. Similarly, for each $M$ $\in$
$\mathcal{U_{M}}$, we define an infinite set $dom(M)$, the domain
of $M$, which is isomorphic either to the real numbers or to the
integers. The domain for the regular data columns of
$\mathcal{U_{A}}$ is defined in a similar fashion to the one of
measures. We can impose the usual comparison operators to all the
values participating to totally ordered domains ${\{}<,
>, \le, \ge {\}}$.\\

\textbf{Dimensions and levels}.A \result{dimension} $D$ is a lattice ($\mathbf{L}$,$\prec$) such that:
\begin{itemize}
	\item $\mathbf{L}$ = $\{$$L_{1}$,$\ldots$ ,$L_{n}$$\}$, is a finite subset of $\mathcal{U_{L}}$.
	\item $dom(L_{i})$ $\cap$ $dom(L_{j})$= $\emptyset$ for every $i$ $\ne$ $j$.
	
	\item $\prec$ is a partial order defined among the levels of $\mathbf{L}$.
	\item With $D$ being a lattice, it follows that there is a highest and a lowest level in the hierarchy. The highest level of the hierarchy is the level $D$.$ALL$ with a domain of a single value, namely '$D$.$all$'. Moreover, there is also the \emph{lowest~level} in the dimension, $D.L_\perp$, for which there does not exist any other level $L$' in $\mathbf{L}$, such that $L$' $\prec$ $L_\perp$.
\end{itemize}
Each path in the dimension lattice, beginning from its upper bound and ending in its lower bound is called a \emph{dimension path}. The values that belong to the domains of the levels are called \result{dimension members}, or simply \emph{members} (e.g., the values $Paris$, $Rome$, $Athens$ are members of the domain of level $City$, and, subsequently, of dimension $Geography$).\rem{must sync if definition of level is updated with a different treatment of properties}

To ensure the consistency of the hierarchies, a family of functions $anc_{L_{1}}^{L_{2}}$ is defined, satisfying the following conditions:
\begin{enumerate}
	\item For each pair of levels $L_{1}$ and $L_{2}$ such that $L_{1}$ $\prec$ $L_{2}$,
	the function $anc_{L_{1}}^{L_{2}}$ maps each element of $dom(L_{1})$ to an element of
	$dom(L_{2})$.
	
	\item Given levels $L_{1}$, $L_{2}$ and $L_{3}$ such that $L_{1}$ $\prec$ $L_{2}$ $\prec$
	$L_{3}$, the function $anc_{L_{1}}^{L_{3}}$ equals to the composition $anc_{L_{1}}^{L_{2}}$ $\circ$ $anc_{L_{2}}^{L_{3}}$.
	This implies that:
	\begin{itemize}
		\item $anc_{L_{1}}^{L_{1}}(x)$ = $x$.
		
		\item if $y$ = $anc_{L_{1}}^{L_{2}}(x)$ and $z$ = $anc_{L_{2}}^{L_{3}}(y)$, then $z$ = $anc_{L_{1}}^{L_{3}}(x)$.
		
		\item for each pair of levels $L_1$ and $L_2$ such that $L_1$ $\prec$ $L_2$, the function $anc_{L_{1}}^{L_{2}}$ is monotone (preserves the ordering of values). In other words:
		
		$\forall$ $x$,$y$ $\in$ $dom(L_{1})$: $x$ $<$ $y$ $\Rightarrow$ $anc_{L_{1}}^{L_{2}}(x)$ $\le$
		$anc_{L_{1}}^{L_{2}}(y)$, $L_{1}$ $\prec$ $L_{2}$
	\end{itemize}
	
	\item
	For each pair of levels $L_1$ and $L_2$ such that $L_1\prec L_2$ the
	$anc_{L_{1}}^{L_{2}}$ function determines a set of finite equivalence classes $X_i$
	such that:
	\[(\forall x,y \in dom(L_1))\
	(anc_{L_{1}}^{L_{2}}(x)= anc_{L_{1}}^{L_{2}}(y)
	\Rightarrow x~and~y \mbox{ belong to the same } X_i).\]
	
	\item
	The relationship $desc^{L_2}_{L_1}$ is the inverse of the $anc_{L_{1}}^{L_{2}}$
	function, i.e.,
	\[desc^{L_2}_{L_1}(l)= \{x \in dom(L_1):anc_{L_{1}}^{L_{2}}(x)=l\}.\]
\end{enumerate}

%

\textbf{Level properties}. Levels are also annotated with properties. For each level $L$, we define a finite set of functions, which we call properties, that annotate the members of the level. So, for each level $L$, we define a finite set of functions $\mathcal{F}^L$ = $\{F^L_1, ~\ldots,~F^L_k\}$, with each such function $F^L_i$ mapping the domain of $L$ to a regular data column $A_i$, s.t., $A_i$ $\in$  $\mathcal{U_{A}}$, i.e., $F^L_i$: $dom(L)$ $\rightarrow$ $dom(A_i)$.
So, for example, for the level $City$, we can define the functions $population()$ and $area()$. Then, for the value $Paris$ of the the level $City$, one can obtain the value $2M$ for  $population(Paris)$ and $100Km^2$ for $area(Paris)$.\\

\textbf{Schemata}. First, we define what a \textbf{schema} is in a multidimensional space.\\

A \emph{schema} $\mathbf{S}$ is a finite subset of
$\mathcal{U}$.\\

A \emph{multidimensional schema} is divided in two parts:
$\mathbf{S}$ = [$D_{1}$.$L_{1}$, $\ldots$, $D_{n}$.$L_{n}$,
$M_{1}$, $\ldots$, $M_{m}$], where:
\begin{itemize}
	\item $\{$$L_{1}$,$\ldots$ ,$L_{n}$$\}$ are levels from a dimension set
	$\mathbf{D}$ = $\{$$D_{1}$,$\ldots$, $D_{n}$$\}$ and level $L_{i}$ comes from dimension
	$D_{i}$, for 1 $\le$ $i$ $\le$ $n$.
	
	\item  $\{$$M_{1}$,$\ldots$, $M_{m}$$\}$  are measures.
\end{itemize}

A \emph{detailed multidimensional schema} $\mathbf{S}^{0}$ is a
schema whose levels are the lowest in the respective dimensions.\\

\textbf{Facts and cubes}. Now we are ready to define what a fact is, expressed as a \textbf{cell}, or multidimensional tuple  in the multidimensional space.\\

A \emph{tuple} under a schema $\mathbf{S}$ = [$A_{1}$, $\ldots$,
$A_{n}$] is a point in the space formed by the Cartesian Product 
of the domains of the attributes $A_i$,
$dom(A_{1})$ $\times$ $\ldots$ $\times$ $dom(A_{n})$, such that
$t[A]$ $\in$ $dom(A)$ for each $A$ $\in$ $\mathbf{S}$.\\


A \emph{multidimensional tuple}, or equivalently, a \result{cell} or a \emph{fact}, $t$ is a tuple under a multidimensional schema $\mathbf{S}$ =
[$D_{1}$.$L_{1}$, $\ldots$, $D_{n}$.$L_{n}$, $M_{1}$, $\ldots$,
$M_{m}$].\\


Having expressed what individual pieces of data, or facts, are, we are now ready to define \textbf{data sets} and \textbf{cubes} .\\

A \emph{data set} $\mathbf{DS}$ under a schema $\mathbf{S}$ =
[$A_{1}$, $\ldots$, $A_{n}$] is a finite set of tuples under
$\mathbf{S}$.\\

A \emph{multidimensional data set} $\mathbf{DS}$, also referred to as a \result{cube}, under a schema
$\mathbf{S}$ = [$D_{1}$.$L_{1}$, $\ldots$, $D_{n}$.$L_{n}$,
$M_{1}$, $\ldots$, $M_{m}$] is a finite set of cells under
$\mathbf{S}$ such that:

\begin{itemize}
	\item $\forall$ $t_{1}$, $t_{2}$ $\in$ $\mathbf{DS}$, $t_{1}$[$L_{1}$,$\ldots$, $L_{n}$] = $t_{2}$[$L_{1}$, $\ldots$, $L_{n}$] $\Rightarrow$ $t_{1}$ = $t_{2}$.
	\item for no strict subset $X$ $\subset$ ${\{}L_{1},\ldots ,L_{n}{\}}$, the previous also holds.
\end{itemize}


In other words, $M_{1}$, $\ldots$, $M_{m}$ are functionally
dependent (in the relational sense) on levels
$\{$$L_{1}$,$\ldots$ ,$L_{n}$$\}$ of schema $\mathbf{S}$.\\

A \emph{detailed multidimensional data set} $\mathbf{DS}^{0}$ is a
data set under a detailed schema $\mathbf{S}^{0}$.\\

A \emph{star schema} ($\mathbf{D}$,$\mathbf{S}^{0})$ is a couple
comprising a finite set of dimensions $\mathbf{D}$ and a detailed
multidimensional schema $\mathbf{S}^{0}$ defined over (a subset
of) these dimensions.

\begin{example}\label{ex:cubeAdultDS}
	Consider the detailed data set $DS$ displayed in Figure~\ref{fig:adultDS}, coming from the well known Adult (a.k.a census income) dataset referring to data from 1994 USA census. There are 8 dimensions (Age, Native Country, Education, Occupation, Marital status, Work class,  Race and Gender) in the  data set and a single measure, Hours per Week. Each dimension comes with a lowest possible level, which we denote as $L_0$. Being a multidimensional data set, immediately makes $DS$ a detailed cube, so in the subsequent discussions, $DS$ will also be referred to as $C^0$. This detailed data set will be the basis  of our running example. 
	
	\begin{figure*}[tbh]
		\centering
		\includegraphics[width=\linewidth]{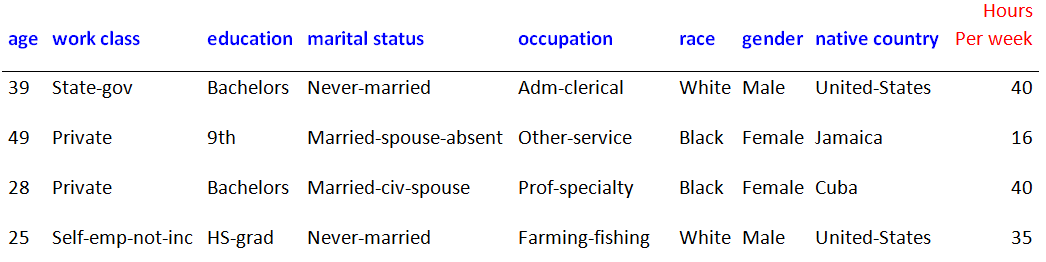}
		\caption{A subset of the detailed data set $DS$ (equiv., $C^0$), with its 8 dimensions at the lowest possible level of detail and its single measure (depicted in the last column).}\label{fig:adultDS}
	\end{figure*}
	
\end{example}

\subsection{Selections}

\textbf{Selection filters}. An \emph{atom} is $true$, $false$,
(with obvious semantics) or an expression of the form
$anc_{L_{0}}^{L_{1}}(L_1)$ $\theta$ $v$, or in shorthand, $L_1$ $\theta$ $v$,
with $v$ $\in$ $dom(L_1)$. $\theta$ is an operator from the set $\{ >, <, =, \ge, \le, \ne \}$.

A \emph{selection condition} $\phi$ is a formula involving atoms
and the logical connectives $\wedge$, $\vee$ and $\neg$. A
\emph{well-formed selection condition} is defined as a selection condition that is applied to a data set with all the level names that occur in it belonging to the schema of the data set. In the rest of our deliberations, we assume that all the selection conditions are well-formed, unless specifically mentioned otherwise.
The expression $\phi(\mathbf{DS})$ is a set of
tuples $\mathbf{X}$ belonging to $\mathbf{DS}$ such that when, for
all the occurrences of level names in $\phi$, we substitute the
respective level values of every $x$ $\in$ $\mathbf{X}$, the
formula $\phi$ becomes true.

A \emph{detailed selection condition} $\phi^{0}$ is a selection condition where all participating levels are the detailed levels of their dimensions.

\subsection{Cube queries}
\textbf{Cube queries}. The user can submit \emph{cube queries} to
the system. A cube query specifies (a) the detailed data set over which
it is imposed, (b) the selection condition that isolates the
records that qualify for further processing, (c) the aggregator
levels, that determine the level of coarseness for the result, and
(d) an aggregation over the measures of the underlying cube
that accompanies the aggregator levels in the final result. More
formally, a \result{cube query}, is an expression of the form:


\[c= \tuple{\ \mathbf{DS}^{0},\ \phi^{0},\ [L_1,\ldots,L_n,M_1,\ldots,M_m],\
[agg_1(M^0_1),\ldots,agg_m(M^0_m)]\ }\]
where
\begin{enumerate}
	\item
	$\mathbf{DS}^{0}$ is a detailed data set over the schema $\mathbf{S}$ =[$L_1^0$,
	$\ldots$, $L_n^0$, $M_1^0$, $\ldots$ ,$M_k^0$], $m$ $\le$ $k$.
	\item
	$\phi^{0}$ is a detailed selection condition,
	\item
	$L_1,\ldots,L_n$ are levels such that $L^0_i\prec L_i$, $1\le i\le n$,
	\item
	$M_1,\ldots,M_m$, $m\le k$, are aggregated measures (without loss of generality we assume that aggregation takes place over the first $m$ measures -- easily achievable by rearranging the order of the measures in the schema),
	\item
	$agg_1,\ldots,agg_m$ are aggregate functions from the set
	$\{sum,$ $min,max,count\}$.
\end{enumerate}

The semantics of a cube query in terms of SQL over a star schema are:

\vspace{6pt}\begin{tabular}{l}
	\small\texttt{SELECT $L_{1}$,\ldots ,$L_{n}$, $agg_{1}(M_1^0)$ AS $M_1$,\ldots,$agg_{1}(M_m^0)$  AS $M_m$}\\
	\small\texttt{FROM DS$^{0}$ NATURAL JOIN D$_{1}$ \ldots\ NATURAL JOIN D$_{n}$}\\
	\small\texttt{WHERE $\phi^{0}$}\\
	\small\texttt{GROUP BY $L_{1}$,\ldots ,$L_{n}$}\\
\end{tabular}

\vspace{6pt}
\noindent where $D_1$, $\ldots$, $D_n$ are the dimension tables of the underlying star schema and the natural joins are performed on the respective surrogate keys \footnote{This assumes identical names for the surrogate keys; in practice, we use INNER joins along with the appropriate columns of the underlying tables, which might have arbitrary names.}

A cube query specifies (a) the cube over which it is imposed, (b) a selection condition that isolates the
facts that qualify for further processing, (c) the grouping levels, which determine the coarseness of the result, and
(d) an aggregation over some or all measures of the cube that accompanies the grouping levels in the final result.  \\

\silence{
\hline
\patrick{somewhat confusing: we have cube names,
cube queries that are also cubes and therefore
sets of cells, and then c.cells for cube query
c that is a set of cell=cube=cube query.
Maybe notation c.cell should be dropped?}

\panos{We need somehow to speak about the results of a query as a set of cells, because then we have components to be linked to them. So we need them. But I don;t see exactly what the problem is. Would you prefer if we mv the definition of c.cells at the cube def., instead of here?} \sticky{Must resolve}
\hline

\patrick{Short:  this is quite minor stuff, I can perfectly live with it as it is.
Long: the problem is that c.cells is kind of redundant.
In classical DB theory, you use $q(I)$ to denote the tuples in
the answer of a query $q$ over an instance $I$.
This is because $q$ does not include $I$ in its expression.
Here a cube query $c$  includes in its expression
the instance $DS$ over which it is applied. 
This is why you can rightly say that a cube query 
is both a query and a cube. A cube is a set of cells.
Therefore a cube query $c$ is already $c.cells$.}
}
\result{Interestingly, a cube query carries the typical duality of views: it is, at the same time, both a query, as it involves a query expression imposed over the underlying data, but, also a cube, as it computes a set of cells as a result that obey the constraints we have imposed for cubes.}





\begin{example}\label{ex:cubeAdult2}
The following cube query produces the cube of Table \ref{tab:cubeAdult2}: 
$C^N=\tuple{DS,$

education.L3='Post-secondary' and work\_class.L2='With-Pay',

$\tuple{$ALL,ALL,L2,ALL,L0,ALL,ALL$},$

$Avg(Hours~per~Week)}$


\noindent where $DS$ is the detailed data set, the selection condition fixes Education to ’Post-Secondary’ (at level L3), and Work to ’With-Pay’ (at level L2), data is grouped by Education at level 2, and Work at level 1, and the Avg of Hours per Week is requested.

\begin{table}[ht]
\begin{center}
\begin{tabular}{|l||cccc|}
\hline
Hrs per Week&	Assoc&	Post-grad&	Some-college&	University\\
\hline
\hline
Federal-gov&	41.15&	43.86	&40.31&	43.38\\
Local-gov&	41.33&	43.96&	40.14&	42.34\\
State-gov&	39.09& 42.96&	34.73&	40.82\\
Private& 41.06&	45.19&	38.73&	43.06\\
Self-emp-inc& 48.68&	53.05&	49.31&	49.91\\
Self-emp-not-inc& 45.88&	43.39&	44.03&	44.44\\
\hline
\end{tabular}
\end{center}
\caption{A new cube $C^N$ as the output of the cube query of Example~\ref{ex:cubeAdult2} \label{tab:cubeAdult2}}
\end{table}

For the reader familiar with OLAP terminology, the new cube $C^N$ resulting from the query, is practically the result of a Drill-Down operation over the old cube $C^O$ of Example~\ref{ex:cubeAdult1}.
\end{example}

The expression characterizing a cube has the following formal semantics\footnote{With the kind help of Spiros Skiadopoulos}:
\[\begin{array}{l}
c=\mbox\{\ x\ |
\mbox(\exists y \in \phi^{0}(DS^0))\ (x=(anc^{L_1}_{L^0_1}(y[L^0_1]), \ldots, anc^{L_n}_{L^0_n}(y[L^0_n]),\ agg_1\{G_1\}, \ldots, agg_m\{G_m\})\mbox)\mbox\ \}
\end{array}\]
where for every $i$ ($1$ $\leq$ $i$ $\leq$ $m$) the set $G_i$ is defined as follows:
\[\begin{array}{l}
G_i=\mbox\{\ q\ | \mbox(\exists z \in \phi^{0}(DS^0))\ (x[L_1]=anc^{L_1}_{L^0_1}(z[L^0_1]), \ldots, x[L_n]=anc^{L_n}_{L^0_n}(z[L^0_n]),
       q=z[M^0_i]\mbox)\ \mbox\}
\end{array}\]

%% file: 92_algos-generatingData-model.tex
\section{Formalizing Algorithms}
To be able to compute the results of model and highlight extraction algorithms, \result{we resort to the modeling of algorithms as functions}. We will present a simple taxonomy of algorithms in the context of our cube model and then, we will proceed to define the formalization of algorithms as black-box functions. In our subsequent deliberations, we will discuss how algorithms facilitate the extension of database-generated query results with (a) simple computations over the query results, (b) model extraction data mining algorithms, and (c) highlight extraction algorithms isolating the important results of the previous computations for the dashboard state.

\subsection{Algorithms as functions and a taxonomy of functions} \label{sec:functions}

\textbf{Taxonomy.}  Specifically, we classify algorithms (remember: practically, functions) in three classes:
\begin{itemize}
	\item \emph{Cell-based algorithms}. These algorithms operate
	locally on a cell, independently of the rest of the contents of
	the input cube. Simple arithmetic computations
	belong to this category (e.g., $profit$ = $price$ -
	$cost$).
	
	\item \emph{Subcube-based algorithms}. These algorithms can operate
	by splitting the input cube to equivalence classes, which we call
	subcubes, according to some criterion. Then, the computation of
	the new attributes for the cells of the subcube, require the
	entire sub-cube to be fed to the algorithm as input. Take for example,
	(a) a time series decomposition algorithm producing three new attributes for each input
	cell (specifically, $trend$, $seasonality$ and $noise$), and, (b) a cube of $Sales$ that is
	grouped per $Product$ and $Month$. Each product, then, defines a
	time series. Thus, the algorithm can split the input to a
	set of subcubes, one per product and perform time series
	decomposition for each of its cells. Again, notice that although,
	ultimately, each cell gets its own values: (a) we cannot compute
	the new cell values by looking at each cell in isolation, and on the other hand, (b) we do not need the
	entire cube to compute them, but only the set of values that
	pertain to the same time series.
	
	\item \emph{Cube-based algorithms}. These algorithms require the
	entire cube to be present for the computation of a cell's extra
	attribute. Examples of this kind of algorithms range from very
	simple algorithms, like determining the bottom 3 values of a cube
	to highly sophisticated algorithms like outlier detection or Fourier analysis.
\end{itemize}
Given a data set $D$, and a tuple $x$ belonging to $D$, an algorithm $f$, depending on its category, induces an equivalence class relation $eqClass_{f}(x)$ producing a set of tuples that include (a) only $x$ for a cell-based algorithm $f$, (b) all $D$ for a cube-based $f$, and (c) a subset of $D$, depending on the semantics of $f$, in the case of subcube-based $f$.\\

\textbf{Algorithms as Functions}. We adopt a traditional
modeling of algorithms, which treats an \result{algorithm}, or equivalently, a \emph{function} as a triplet
involving the following \emph{signature} :
\begin{itemize}
	\item a algorithm name, say $f$
	\item a vector of input parameters, say $X$ = $\tuple{X_1, \ldots, X_I}$
	\item a vector of output parameters, say $Y$ = $\tuple{Y_1, \ldots, Y_O}$
\end{itemize}


To model algorithms and parameters, we assume a countable set of
data computation function names $\mathcal{U_{F}}$ and a countable set of parameter names $\mathcal{U_{P}}$, which is a subset of the names appearing in $\mathcal{U_{A}}$. Each parameter
$A$, input or output, is accompanied by a domain, $dom(A)$. We
will refer to $X$ as the \emph{input schema} of the algorithm and
to $Y$ as the \emph{output schema} of the algorithm. Then, the
algorithm is a relation mapping $dom(X_1)$ $\times$ $\ldots$
$\times$ $dom(X_I)$ to $dom(Y_1)$ $\times$ $\ldots$ $\times$
$dom(Y_O)$.

The call of an algorithm requires fixing a data set over which the algorithm will be applied and the assignment of the input parameters of the algorithm to columns of the data set or constants. The latter is done by assigning to them (a) constant values (without loss of
generality we assume constants belonging to $\mathbb{R}$), or, (b) data
columns from a data set. Given an underlying data set, a \emph{valid parameter binding} of an algorithm to the data set is a total mapping $B$: $X$
$\rightarrow$ $\mathcal{U}$ $\cup$ $\mathbb{R}$ that uses only constants and data columns of the
data set, respecting also the compatibility of the domains between
the algorithm signature and the members of the binding. Unless
explicitly mentioned, all our parameter bindings of algorithms to cube
queries are valid. We denote the \result{binding} $B$ of a algorithm $f$ to a data set $D$ and a valid binding of its input parameters $B(X)$ as $B(X|D)$.

Once such a binding has been done, the execution of a algorithm
produces a set of tuples abiding by the output schema $Y$.
To denote the \result{execution of the algorithm} we will use the notation $D^{+}$ = $f_D^{B(X)}$.

Assuming a data set $D$ under the schema $X$ and a algorithm $f$:
$X$ $\rightarrow$ $Y$, the \emph{extended data set} $D^{+}$
resulting from the execution of $f$ under a binding $B(X|D)$ is defined as follows:
\begin{itemize}
	\item The schema of $D^{+}$ is $S$ = $X$ $\cup$ $Y$
	\item For each tuple $x$ $\in$ $X$, there is exactly one tuple $s$ in
	$D^{+}$, such that $s[X_1,\ldots,X_I]$ = $x[X_1,\ldots,X_I]$ and
	$s[Y_1,\ldots,Y_O]$ = $f_D^{B(X)}$($eqClass_{f}(x))$ 
	\item No tuples other than the aforementioned belong to $D^{+}$
\end{itemize}

Observe, that we want to annotate \emph{each input tuple} with a set of output values, independently of whether the algorithm is cell-based or not. In the case of cell-based algorithms, $f_D^{B(X)}(x)$ ignores the other tuples of $D$ and uses only $x$, whereas in the case of subcube-based algorithms it uses only the subcube of $x$. This also means that each time, a tuple $x$ gets the same result with its equivalence class.

Whenever the details of the binding are not important, we will use a shorthand notation $D^{+}$ = $f(D)$.

An \result{extended cube query} $c^+$ produced by the application of
a algorithm $f$ to a cube query $c$ under the schema [$L_1$,
$\ldots$, $L_n$, $M_1$, $\ldots$, $M_m$], i.e., $c^+$ = $f(c)$, comes with the
same semantics as data sets.

\subsection{The generating data of a dashboard}
\label{sec:generatingData}

As we will demonstrate in the sequel, a dashboard includes a set of queries. \result{We can exploit the raw generating data of the dashboard (i.e., the data that come from the underlying database via database queries), to 
produce derived values for each of the cells of a cube query,
via the application of simple data computing functions. 
The produced cells are the generating data of the dashboard.}

We employ the term \emph{cube query set}, or for short, \emph{query set} for a finite set of queries $C$ = ($c_1$, $\ldots$, $c_k$). Assuming a query set $C$ = ($c_1$, $\ldots$, $c_k$) for a dashboard $S$, the results of the queries of $C$ are the \emph{raw generating data} of $S$.
Similarly to query sets, an \emph{extended query set} $C^+$ = ($c_1^+$, $\ldots$, $c_k^+$) is a finite set of extended queries, providing the \result{generating data of $S$}.\\



\subsection{Algorithm and function composition}

The \emph{composition} $f \circ g$ of two algorithms $f$ and $g$ carries
the same semantics and constraints as the composition of functions in mathematics. The
composition of a list of algorithms $f_1$, $\ldots$, $f_n$ is a
repetitive application of the composition operator (($\ldots$($f_1
\circ f_2$) $\ldots$ $\circ$ $f_{n-1}$) $\circ$ $f_n$).

For the sake of generality, we can also compute the composition of individual data computing algorithms over simple cube queries. Assume a cube query $c$ under the schema [$L_1$, $\ldots$, $L_n$,
$M_1$, $\ldots$, $M_m$]. Assume also the composition of a list of
algorithms $\mathcal{F}$ = $f_1$, $\ldots$, $f_n$. We say that the
composition of the algorithms' list has a valid binding to the cube
query $c$ when each algorithm $f_i$ has a valid binding to the
result of the composition $f_1$ $\circ$ $f_2$ $\circ$ $\ldots$
$\circ$ $f_{i-1}$ applied over $c$. The schema and the contents of
the extended cube query produced $c^+$ by the application of $f_1$
$\circ$ $f_2$ $\circ$ $\ldots$ $\circ$ $f_{n}$ to $c$ are produced
as defined above.

%

%% file: 93_models.tex
\section{Formalities for models}

\subsection{General Principles}


Having defined data and algorithms, we ca now proceed to discuss the computation of statistical models from the data. We are going to treat model construction algorithms as "black-box" functions without
probing into their internals, and, most importantly, without
assuming any specialized properties for their output. What does a
model construction algorithm do? Basically, the algorithm receives as input (a) a set of
input data, and, (b) a set of execution parameters that have to be
fixed for the algorithm's execution. Without loss of generality,
we can assume that a subset of these parameters will be bound to
string or numerical values and the rest will be mapped to
attributes of the input data. The output of a model construction algorithm is a
\emph{model} of the input data. Depending on the algorithm, the
result differs. For instance, a descriptive model built using unsupervised clustering  is basically just a labeling of each cube's cells, while a predictive one allows enriching the cube with predictions and comes with an accuracy score. 
In summary, the main properties of a model construction algorithm are outlined as follows:
\begin{enumerate}
	\item Input: a set of input data, which is the result of an extended cube query set of the dashboard, along with a binding of the algorithm's input parameters.
	\item Output: a (possibly complex) result composed of (a) a model of the input data, and, (b) several characterizations of it (precision, strength, p-value, etc.).
\end{enumerate}

Models are produced as instances of \emph{model types}.  A model is \emph{a concise representation of some knowledge about the data}. This knowledge can be some relationship between data attributes, some property or characterization of subsets of data,  or some computed value over the existing data. At the same time, despite its conciseness, typically a model also serves as \emph{an enrichment of the underlying data} -- in other words, conceptually, each record of the data can be extended,  annotated, or, in any case, enriched with extra information by the model. We do this by organizing models as sets of model components, with each model component having exactly one value for each of the cells of the model's generating data. This is what we refer to as data-to-model mappings.

\subsection{Model Types}

Every model construction algorithm has a result type: after the execution of the
algorithm, its output, i.e., the resulting model of the input data is bound to this
result type. To facilitate the management of models, we assume an infinitely countable domain of data type names $\mathcal{U_T}$, each member of which, say $T$, has a domain, $dom(T)$.\\


A \result{model type} $T$ is a tuple $T$ = $\tuple{S_I, S_O^\star, S_P}$, where  $S_I, S_O^\star, S_P$ are schemata with their members' names belonging to $\mathcal{U_{A}}$, with $S_I$ being the input schema,  $S_O^\star$ being the output schema and, $S_P$ being the model characterization schema. The output and characterization schemata are not to be in 1NF and can employ complex type constructors of the form \textsf{set} or \textsf{tuple}. $S_O^\star$ obligatorily includes a set-valued attribute $S_o: set\{A\}$, $A~\in~\mathcal{U_{A}}$, to be instantiated as a schema of components at the model level~\cite{DBLP:conf/oopsla/KhoshafianC86}. 




\subsection{Models, Model Components and Data-To-Model Mappings}
\result{A model $m$ is an instance of a model type $T$ and it is computed over a given cube $c$. To this end, we need a binding. The contents of the model, stored under the model's output schema, are structured along model components, which are practically annotations of the input cube cells with respect to the model being computed over them. This requires a mapping between the elements of the model contents and in the input cells.}  
\subsubsection{Models}
Given a model type $T$ = $\tuple{S_I, S_O^\star, s_P}$, a cube $c$, a valid binding $B(S_I~|~c)$ of $S_I$ to $c$ (i.e., assigning levels and measures to the type's input parameters, along with any needed constants for the tuning of the algorithms), then, a model $m$ is a named tuple $m$ = $\tuple{B(S_I~|~c),S_O,s_p}$, with $m$ acting as a (possibly automatically computed) name for the model, $S_O$ an output schema belonging to the domain of $S_O^\star$ and $s_p$ $\in$ $dom(S_P)$.\footnote{For the moment, we bind the parameters of a model type's input schema to a single cube, and leave the application of model construction algorithms to a combination of cubes as a generalization for future work.} By definition, a model's output schema $S_O$ includes $S_O.S_o$, which we call the \emph{output component schema} of the model and simplify its naming as simply $S_o$.\\


There are two necessary explanations here, on the output and the statistical characterization of a model. Let's start with the output schema. Assume a model type of decision trees. The output is a \emph{set~of~paths}, with each \emph{path} being characterized by an expression and a bitmap vector for the cells of the cube, on whether they belong to the path or not. So  at the model type level the output schema is a pair $S_O^\star$ = $\tuple{$ $Paths:~set\{Expr:String\}$, $S_O: set\{MC:Boolean\}}$. Then, at the model level, the model can contain an arbitrary number of paths, not a-priori known at the type level. Suppose then that a particular model $m$ has 6 paths, then we model the output schema of $m$ as a pair of (a) the set $Paths$ = $\{p_1, \ldots, p_6\}$, with each path $p$ defined as an expression, e.g., $p_3$ = $age > 10~and~weight > 50 \rightarrow class=overweight$ and (b) the components schema, $S_o$ being a set of components $S_o$ = $\{MC_1, \ldots, MC_6\}$, defined as Boolean attributes. Then, $m.S_O$ is the pair $S_O$ = $\tuple{Paths,~S_o}$. Naturally, $p_3$ refers to the model component $MC_3$ which is annotating the cube cells. In an exactly similar manner, a clustering algorithm has as the output of its model type a pair including a set of medoids corresponding to a set of clusters,  $S_O^\star$ = $\tuple{$ $Medoids:~set\{MD: vector\{coordinates\}\},~S_o:~set\{MC:Boolean\}}$. A particular cluster, say cluster $c_4$, can be reconstructed be the respective medoid $MD_4$ and the model component $MC_4$.  The statistical characterization of a model is an instance of the respective attribute of the model type, can follow arbitrary structures and can even avoid annotating the cells of the input cube.

\silence{
\hrule
\patrick{Some issues in the paragraph above. 
It is not clear if both $S_O$ and $S_o$ are attributes.
If I got it well, from the definition of type, $S_o$ should be
an attribute. But here you use it as a complex value of
attribute $S_O$. Maybe in the definition of type you meant $S_O$.
Then  $m$ is a model, it seems to be a complex value.
In your example, $m$ should be a triple (it needs the binding).
If $S_O$ is an attribute of the sort of $m$ then indeed
you can use $m.S_O$, but it is confusing to have again
the name $S_O$ in the sort of the value $m.S_O$.
Finally, if $p_3$ is associated with $MC_3$ then probably
Paths and $S_O$ should not be sets but tuples.
And then components should also be defined as tuples.
}\\
\panos{$S_O$ is a type generator. So it is still a type under OO conventions. Then, internally, $S_O$ includes a set-valued attribute $S_o$. \\
For the correspondence between $P$ and $S_o$ I avoided what you say. Instead, I made them distinct sets which allows me to treat $S_o$ as a \emph{set} of components (so that the rest of the paper is not changed). I agree that we could have used the tuple model that you suggest, but then (a) you must change the entire paper and (b) you cannot claim the \emph{exact same} attribute for \emph{all} model types, which is btw, very convenient. So, at the price of a semantic constraint, namely p3 corresponds to MC3, we get the nice common-to-all-models property of a set of components. I hope I understood the point well... }
\hrule
}
What is implied by the above definitions of model and model types is that model types can be data types of arbitrary complexity, in an object-oriented manner, and not restricted to be in 1NF. At the same time, the model schemata, can \emph{ultimately} be treated in a relational format, as a simple set of attributes. Even if the data type is a complex data type, it is always possible to un-nest it into a relational-like structure -- and, in any case, it is important that what matters here, i.e., the model components, are explicitly modeled as attributes. Attempts to relationaly code mining results already exist \cite{DBLP:conf/dasfaa/GiacomettiMS11}. 

\subsubsection{Model Components}
A model has an output, which includes a component schema of attributes. So practically speaking, the result of a model is a data set, with named attributes. These attributes we call model components.

Assume the aforementioned model $m$ = $\tuple{B(S_I~|~c),S_o,s_p}$. A \result{model component} $MC$ is an attribute belonging to $S_o$ (equiv., the output component schema $S_o$ is composed of model components). Assuming $S_o$ = $\{MC_1, MC_2, \dots, MC_m\}$, each component $MC$ of $S_o$ is instantiated with a list of values $\{v_1, v_2, \dots, v_n\}$, each $v_i$ $\in$ $dom(MC)$. We call the values that instantiate $MC$ as the \emph{model component elements} or simply \result{elements}. We denote the elements of a model component $MC$ as \emph{MC.elements}.


\silence{
\hrule
\patrick{check consistency with model}   \panos{Done, I think}

\patrick{Not clear, $S_o$ should not be schema,
but a relation instance? Is $dom(S_P)$ defined?}\panos{OK, if I get it right, you mean that So is the name of a relation instance, with the schema of $\tuple{MC_1, MC_2, \dots, MC_m}$. Then, we need to devise another name for the above $\tuple{MC_1, MC_2, \dots, MC_m}$ as a schema? }

\patrick{I forgot why we need to have
a schema $S_O$ for model type and a schema
$S_o$ for model. To be consistent with definition 2 (model), $S_o$ should be a set of tuples
that codes model components. 
Each component corresponds
to an attribute $MC$ of $S_O$ and its instance
$\{v_1, \ldots, v_n\}$  corresponds to
$\pi_{MC}(S_o)$
(under a bag semantics).} \panos{We need to have a common name for the components "relation". And I called $S_o$. The $S_O$ is a complex type that is specialized per model type. The $S_o$ is common to all models, no matter what, and the only thing that depends on the binding is how many columns it has -- 3 clusters? 5 clusters? depends on the binding...}
\hrule
}

\subsubsection{Data-To-Model Mappings}
Due to the inherent heterogeneity of models and model components, we need to devise a unifying model to cover them all. The unifying essence of all the plethora of diverse model types is  that, at the end of the day, all of them are annotations of the original data. 

We impose a \result{data-to-models} constraint that there is a bijective mapping between the cells of cube $c$ and the elements of each of its model components. Thus, we have two ways of viewing the computation of a model $m$ over a cube $c$: 
\begin{itemize}
    \item Extended data set computation: practically, the schema of $c$ is extended with $S_o$, and the instances are appropriately matched
    \item Data-to-model mapping: there is a bijection via the functions $f_c^{mc}$: $c.cells$ $\rightarrow$ $MC.elements$ and its inverse $f_{mc}^{c}$: $MC.elements$ $\rightarrow$ $c.cells$.
\end{itemize}

In other words, we can think of model components as a uniform mechanism for transforming statistical models to data, and at the same time, extending the input data with annotations concerning the respective models.

%% file: 94_highlights-dashboards.tex
\subsection{Highlight Production}
The set of
\emph{highlights} of the dashboard is a set of \emph{important
	findings} that accompany the dashboard. These can be findings of
any nature, e.g., important outliers in the contents of the dashboard's data,
all the tuples belonging to a certain class of a classification
scheme, the top or bottom values of a measure, etc. 

To define highlights, we need to introduce two more concepts.
\begin{itemize}
    \item We need a highlight selection criterion to allow us define which component is actually a highlight or not. We devote the entire Section~\ref{sec:interestingness} to this end. From the formal perspective, we assume the existence of a function (equiv., algorithm) $interestingness()$ that allows to annotate each component $MC$ with an interestingness score $MC.interestingness$. Then, a \result{highlight selection criterion} is a function that assigns true or false to the component of the output schema of a model, thus assigning to them the highlight property or not --i.e., $\phi_H$: $M.S_o$ $\rightarrow$ $Boolean$

    \item Since each highlight component annotates all the cells of a cube, we need to isolate only the elements of the component that are of particular importance. We call this subset, the \result{core data} of the highlight. Which elements qualify as core data is dependent (i) upon the model type (i.e., it is different if a clustering scheme devotes a bitmap per cluster, in which case we are interested in '1''s only, vs., the case of a classification scheme, where each element is assigned to a class, e.g., 'Low' or 'Unexpected'), and (ii) possibly, upon the criterion used (i.e., if the interestingness criterion is mostly focused towards outlierness, values like 'unexpected' are the highlight's core, whereas if the criterion is regularity, exactly the opposite holds). Since there is a 1:1 mapping between component elements and cube cells, we denote the core elements of a component as $MC.coreElements$ and their respective cube cells as $MC.coreCells$. 
\end{itemize}

Given an intentional query $q$ issued over a cube $C^{O}$ of a dashboard, and resulting to a new cube $c$, a new model $M$ over cube $c$, with components $M.MC_1, \ldots, M.MC_k$, a criterion for highlight selection $\phi_H$ (on the basis of a component scoring function $interestingness$), then, the triplet $h$ = $\tuple{MC_I$, $MC_I.coreElements$, $MC_I.coreCells}$ is a \result{highlight}, with: 

\begin{itemize}
    \item $MC_I$ being a specific component of $M$, s.t., $\phi_H$($MC_I.interestingness$) = $true$, i.e., it qualifies as highlight with respect to the selection criterion.
    \item $f_c^{MC_I}(MC_I.coreCells)$ = $MC_I.coreElements$, i.e., the core elements and the core cells fulfill the 1:1 mapping. 
\end{itemize}

The modeling that we adopt is open in many ways. First, the interestingness function can be defined in many ways. Second, the criterion $\phi_H$ is also open to alternative definitions: e.g., it can be whether the component has the top interestingness, or it is within the top-k, or it has some other property (e.g., it is in the skyline of interestingness aspects, if one defines a multi-aspect definition of interestingness in terms of a vector of scores). Finally, the definition of core cells is also open to different alternatives.\rem{Silenced the question on the deep nature of highlights...}\\

\silence{
\noindent\rule[0.5ex]{\linewidth}{1pt} 

\panos{PV: All this discussion, makes me think that maybe we need to \textbf{kill the triplet
and just define highlight as the component(s) that pass the criterion}. But ont he other hand, it's kinda nice to say, we have this component, these element and this data cells, so unless someone challenges it, I will opt for keeping the current triplet definition}\\

\patrick{I'm ok with one or the other.
But a thing that looks odd is to have a 1:1 mapping
and keep both coreElements and coreCells in the triplet.}

\noindent\rule[0.5ex]{\linewidth}{1pt} 
}

\subsection{Dashboards}

The triple of a cube $C$, its (set of) models $\mathbf{M}$, and its highlights $\mathbf{H}$ is called an \result{enhanced cube}. \\

A \emph{dashboard} $S$ is a finite set of enhanced cubes, $S$ = $\{c_1^\star, \ldots, c_S^\star\}$.\\




%% file: 95_operators.tex
\section{Operators and their Formal Definition}\label{sec:FormalOps}


In this section, we formally define the operators of the Intentional Analytics Model. In all our deliberations we assume:
\begin{itemize}
    \item A dashboard $S$ including a finite set of enhanced cubes, $S$ = $\{c_1^{\star}, \ldots, c_S^{\star}\}$
    \item An arbitrary enhanced cube $c^{\star}$ of $S$ is defined as a triplet with its generating cube data, its models and highlights, $c^{\star}$ = $\{c^{+}, \mathbf{M}, \mathbf{H}\}$, where the models of the enhanced cube are a finite set of models $\mathbf{M}$ = $\{M_1, \ldots, M_M\}$ and the highlights a finite set of highlights $\mathbf{H}$ = $\{h_1, \ldots, h_H\}$ and the extended $c^{\star}$ is produced from a cube query $c$ via a set of functions, where $c$ = ( $\mathbf{DS}^{0}$, $\phi^{0}$, $[L_1,\ldots,L_n,m_1,\ldots,m_m]$, $[agg_1(m^0_1),\ldots,agg_m(m^0_m)]$ )
    \item An arbitrary model $M$ comprises a finite set of model components in its output $M$ = $\{MC_1, \ldots, MC_c\}$
\end{itemize}

For each operator we assume a set of model types to be produced after the application of the operator. We will commonly refer to every such list as $\mathbf{T}$ and each time we will prescribe its components with an indicative set of models to be produced. 

After the execution of each operator, the dashboard $S$ is extended with a new enhanced cube $c^{\star~n}$ = $\{c^{+~n}, \mathbf{M}^n, \mathbf{H}^n\}$ 

The highlights $\mathbf{H}^n$ of 
$c^{\star~n}$ are automatically computed following the principle presented in Section \ref{sec:interestingness}, using a function called $selectHL()$. 
The cubes and models given as parameters of this function 
are the initial cube $c$, the new cube $c^n$ 
and the set of models $\mathbf{M}^n$ .
The other parameters, i.e., relation $proxies$ and functions
$significance$, ${\cal D}$, ${\cal A}^C$ and ${\cal A}^M$,
pertain to the subjective aspect of the interestingness
assessment and therefore are predetermined by both
the type of models used and by the user
history with the system. 
They are not detailed any further in what follows,
where we simply abbreviate the call to $selectHL()$ by $selectHL(c^n,\mathbf{M}^n)$.

The removal of cube from the dashboard is beyond the discussion of this paper --one can envision explicit removals by the user (user closes the respective window), or automatic caching-like replacements over a fixed screen, or any other scheme). 

\subsection{Describe}

The describe operator produces a new enhanced cube by focusing on a subcube at a possibly different aggregation level -- practically this is the operator to add new information to a dashboard. Remember that the general form of the operator was:\\

\sf{with} \emph{cube} \sf{describe} \emph{measure} \{, \emph{measure}\} [\sf{for} \emph{subcube}] [\sf{by} (\{\emph{level}\} $|$ \sf{size} \emph{integer})]\\

\textbf{Semantics}. The \sf{'by~level'} variant of the describe operator is formally defined as follows:\\

$c^{\star~n}$ = $Describe(c,~ m_1,\ldots,m_d,~\phi, ~D_i.L)$, with:

\begin{enumerate}
    \item First, a cube $c^n$ = ($\mathbf{DS}^{0}$, $\phi \wedge \phi^{0}$, $[L_1,\ldots,L_{i-1}, L, L_{i+1}, \ldots, L_n,m_1,\ldots,m_d]$, $[agg_1(m^0_1),\ldots,agg_m(m^0_d)]$ ) is computed, $d~\leq~m$
    \item Second, $\mathbf{M}^n$, a set of models that are computed over $c^n$, via the respective binding of the model types of $\mathbf{T}$=$\{ T_{topK}$, $T_{DomR}$, $T_{DomC}$, $T_{outl} \}$ is also obtained.
    \item Third, a set of highlights, $\mathbf{H}^n$, is automatically computed over $\mathbf{M}^n$ via an automatic highlight selection mechanism $\mathbf{H}^n$ = $selectHL(c^n,\mathbf{M}^n)$.
\end{enumerate}
The variant of the operator without the \sf{'by~level'} clause is a simplification of the aforementioned variant, where $L^n$ retains the level it had at $c$.\\

\silence{
\hrule 
\panos{An alternative construction for the semantics}
\begin{enumerate}[\arabic*,start=0]
\item If cube $c$ is not already available from a previous step in the session, we compute it
\item Once $c$ is available we project to the set of measures $\{m_1,~\ldots,~m_d\}$ prescribed in the query specification. 
\item If a \textsf{for} $subcube$ clause is also prescribed in the query specification (captured via the selection condition $\phi$ in the respective algebraic notation), we also apply the respective filter 
\item If a change of aggregation level is prescribed for dimension $D_i$ at level $L$, via the clause \textsf{by} $L$, we also change the aggregation level appropriately (this is uniformly modeling roll-Up and drill-down via a simple change of aggregation level in the list of levels of the query definition)
\item After all the above, a cube $c^n$ = ($\mathbf{DS}^{0}$, $\phi \wedge \phi^{0}$, $[L_1,\ldots,L_{i-1}, L, L_{i+1}, \ldots, L_n,m_1,\ldots,m_d]$, $[agg_1(m^0_1),\ldots,agg_m(m^0_d)]$ ) is computed, $d~\leq~m$
\item Then, .... the same ....
\end{enumerate}

\hrule 
}

The \sf{'size~integer'} variant of the describe operator is formally defined as follows:\\

$c^{\star~n}$ = $Describe(c,~ m_1,\ldots,m_d,~\phi, k)$, with:
\begin{enumerate}
    \item First, a cube $c^n$ = ($\mathbf{DS}^{0}$, $\phi \wedge \phi^{0}$, $[L_1,\ldots, L_n,m_1,\ldots,m_d]$, $[agg_1(m^0_1),\ldots,agg_m(m^0_d)]$) is computed, $d~\leq~m$
    \item Second, apply (a) a clustering's type $T_{clust}$ algorithm to the cube $c^n$ to produce $k$ clusters and (b) a shrink's type  $T_{shr}$ algorithm to produce $k$ cells, one per target summarizing value. The elements of $T_{clust}$'s output are bitmaps showing the participation or not of a cell to the respective cluster. The elements of $T_{shr}$'s output are bitmaps showing the participation or not of a cell to the respective shrunk cell. 
    \item Third, a set of highlights, $\mathbf{H}^n$, is automatically computed over $\mathbf{M}^n$ via an automatic highlight selection mechanism $\mathbf{H}^n$ = $selectHL(c^n,\mathbf{M}^n)$.
\end{enumerate}


\subsection{Assess}

The assess operator is all about comparing the results of a cube to "similar" or "reference" benchmark data that allow us to assess how good the situation presented by the cube is. Remember that the invocation of the \sf{ Assess } operator follows the syntax:\\ 

\sf{ with } \emph{cube} \sf{ assess } \emph{ measure }  \{,~\emph{measure}\}  [\sf{ for } \emph{subcube}]  \sf{ using } \emph{benchmark~model}  \{,~\emph{benchmark~model}\}\\

The formalization of the assess operator is as follows:\\

$c^{\star~n}$ = $Assess(c,~m_1,\ldots,m_k,~\phi,~\mathbf{B})$, with the set of benchmark types $\mathbf{B}$ = $\{T_1^b, \ldots, T_B^b\}$\\

\textbf{Assumptions}. We assume a set of benchmark model types, $\mathcal{U}_T^b$, subset of $\mathcal{U}_T$ that are used for the computation of the assess operator. Each such type $T$ must satisfy the following two constraints: (i) whenever bound to a cube $c$, it produces a single-component model, with $mc$ being the respective component, such that a bijection $f_c^{mc}$ can be defined (in other words, we can compute a total 1:1 mapping between the elements of the benchmark and the cells of the cube), and, (ii) it is accompanied by a computation algorithm $f_T$.\\

\textbf{Semantics}. The semantics of the operator are as follows. 

\begin{enumerate}

     \item First, we apply $\phi$ to $c$ producing a new base cube $c^a$ = ( $\mathbf{DS}^{0}$, $\phi$ $\wedge$ $\phi^{0}$, $[L_1,\ldots,L_n,m_1,\ldots,m_k]$, $[agg_1(m^0_1),\ldots,agg_m(m^0_k)]$ )
    \item Second, we apply the algorithms that pertain to the set of types of $\mathbf{B}$ over $c^a$ and obtain a set of single-component models $\mathbf{M^b}$ = $\{M_1^b, \ldots, M_M^b\}$.
    \item Third, for each component of the models of $\mathbf{M^b}$, we compute the difference of its elements with their respective cells of $c^a$ and populate an extra set of models $\mathbf{M^\delta}$ = $\{M_1^\delta, \ldots, M_M^\delta \}$. The union of $\mathbf{M^b}$ and $\mathbf{M^\delta}$ forms the set of models $\mathbf{M}^n$ for the operator's execution.
    \item Finally, a set of highlights, $\mathbf{H}^n$, automatically computed over $\mathbf{M}^n$ via an automatic highlight selection mechanism $\mathbf{H}^n$ = $selectHL(c^n,\mathbf{M}^n)$ is obtained. Unless otherwise tuned, the selection mechanism picks the members of $\mathbf{M^b}$ for which the respective member of $\mathbf{M^\delta}$ is maximized (i.e., the ones with maximum discrepancy from the benchmarks)
\end{enumerate}

The members of $\mathcal{U}_T^b$, i.e., the benchmarks along with their underlying computation algorithms is open and extensible (including, for example, having registered predefined goals for each cell, averaging of sibling cells, last $k$ values, etc). What is important is that for each cell of the cube, we can obtain (typically via its coordinates) the respective model element in a 1:1 fashion. This enables the assessment of each cell of the cube, by contrasting it to its respective component element!

\subsection{Explain}

The explain operator applies models to the results of cube queries that perform statistical (or other) analyses to them. For example, these models may test the correlation of the cube measures with other attributes, classify the data on the basis of a classifier, extract regression formulae for the measures, etc.

To apply the model construction algorithms that explain results over the cubes, we are in need to bind their execution to specific attributes. So, we need to define the binding in the invocation of the operator.

The simplest invocation of the \sf{Explain} operator follows the syntax 

\sf{ with } \emph{cube} \sf{ explain } \emph{measure}   [\sf{ for } \emph{subcube} ]  \sf{ using } \emph{explanation~model (attribute~list)} \{, \emph{explanation~model (attribute~list)}\}  \\

The formalization of the first variant of the explain operator is as follows:\\

$c^{\star~n}$ = $explain(c,~ m,~\phi,~\mathbf{T},~\mathbf{MB})$,\\

\noindent with a set of model types $\mathbf{T}$ = $\{T_1^e$, $\ldots$, $T_B^e \}$ and a set of bindings $\mathbf{MB}$ = $\{MB_1^e(\mathbf{A}_1)$, $\ldots$, $MB_B^e(\mathbf{A}_{B}) \}$ being bindings of the model types of $\mathbf{T}$ to the underlying data.\\

\textbf{Assumptions}. Much like the assess operator, we assume a set of explanatory model types, $\mathcal{U}_T^e$, subset of $\mathcal{U}_T$ that are used for the computation of the explanation operator. Each such type $T$ must satisfy the following two constraints: (i) whenever bound to a cube $c$, it produces valid models, such that for each of their components, say $mc$, a bijection $f_c^{mc}$ can be defined to the underlying cube $c$ (in other words, we can compute a total 1:1 mapping between the elements of the benchmark and the cells of the cube), and, (ii) it is accompanied by a computation algorithm $f$.

We assume that for the bindings to be valid, the members of each set of data columns $\mathbf{A}_i$ are either attributes of the underlying cube $c$, or properties of the levels of their dimensions.

\textbf{Semantics}. The semantics are as follows.
\begin{enumerate}
    \item First, we apply $\phi$ to $c$ producing a new base cube $c^a$ = ( $\mathbf{DS}^{0}$, $\phi$ $\wedge$ $\phi^{0}$, $[L_1,\ldots,L_n,m]$, $[agg(m^0)]$ )
    \item For each of the involved data types and bindings, we apply the model construction algorithms to $c^a$, i.e., we execute $f_{c^a}^{MB^e(\mathbf{A})}$ for each $MB$ $\in$ $\mathbf{MB}$. This computes the set of models $\mathbf{M}^n$ for the operator's execution.
    \item Third, a set of highlights, $\mathbf{H}^n$, is automatically computed over $\mathbf{M}^n$ via an automatic highlight selection mechanism $\mathbf{H}^n$ = $selectHL(c^n,\mathbf{M}^n)$
\end{enumerate}

The second variant of the explain operator does the aforementioned procedure over \emph{two} cubes (instead of one), which we compare:\\

\sf{ with } \emph{cube} \sf{ explain } \emph{measure}   [\sf{ for } \emph{subcube} ]  \sf{ using } \emph{explanation~model (attribute~list)} \{, \emph{explanation~model (attribute~list)}\} \sf{ against }  \emph{comparison~cube}\\

The essence of the operator is the demonstration to the user of the differences in the models of the antagonizing cubes. This is of course specific to the model type. For example, the difference in correlation is just a numerical value, whereas the difference in a decision tree is a set of paths, along with the change in the strength measures per path.

Practically, this entails the operator\\
$c^{\star~n}$ = $explain(c, ~c^c, ~ m,~\phi,~\mathbf{T},~\mathbf{MB})$\\

The semantics of the operator are:
\begin{enumerate}
    \item We perform steps (1) and (2) independently, for $c$ and $c^c$, obtaining $\mathbf{M}^n$ and $\mathbf{M}^{n^c}$ with the respective models for the input cubes
    \item For each model $M_i$ in $\mathbf{M}^n$ and its homologous model $M_i^c$ in $\mathbf{M}^{n^c}$, and for each pair of homologous model components $M_i.MC_j$ and $M_i^c.MC_j$, we compute the difference of their elements, resulting in a new model $M_i^\delta$. The union of this models is the set $\mathbf{M}^{n^\delta}$ which constitutes the set of models of the resulting $c^{\star~n}$.
    \item Third, a set of highlights are computed over $\mathbf{M}^{n^\delta}$ as usual.
\end{enumerate}

\subsection{Predict}
The operator Predict estimates a set of points for (a) a measure, evolving with respect to (b) a time dimension, via (c) a predictive model that computes the predicted value. The syntax of the operator is:\\

\sf{ with } \emph{cube} \sf{ predict next } \emph{k} \sf{ points of } \emph{measure}  [\sf{ for } \emph{subcube}] \sf{ over } \emph{time~dimension} \sf{ using } \emph{predictive~model}\\


The formalization of the operator is as follows:

$c^{\star~n}$ = $predict(c,~ m,~\phi,~k,~A,~T)$\\

We assume that the model type $T$ requires for its input a binding $B(m,k,A)$ for (a) a measure to be predicted, (b) the number of predicted points, $k$, and (c) an attribute $A$ (quite possibly a dimension level with time semantics) that plays the role of time dimension (a cube can have many of them). We also assume an algorithm $f$ for the computation of the prediction. As a side effect of the attribute $A$, the data of the input cube are sorted by $A$ internally in the execution of $f$.

\textbf{Semantics}. The semantics are as follows.
\begin{enumerate}
    \item First, we apply $\phi$ to $c$ producing a new base cube $c^a$ = ( $\mathbf{DS}^{0}$, $\phi$ $\wedge$ $\phi^{0}$, $[L_1,\ldots,L_n,m]$, $[agg(m^0)]$ )
    \item We bind the algorithm $f$ to $B(m,k,A)$ and execute $f^{B(m,k,A)}_{c^a}$. This computes the set of models $\mathbf{M}^n$ that depending on the algorithm may include (a) a single component model $M_P$ with a vector component for the projection of $k$ points later, for each point in the input cube, (b) a model with a component for the expected values on the basis of the regression -or other- model employed by $f$, (c) a model with components for trend, seasonality and noise, etc.
    \item Third, a set of highlights, $\mathbf{H}^n$, that is either assigned to include $M_P$ (default) or tuned to be automatically computed over $\mathbf{M}^n$ via an automatic highlight selection mechanism $\mathbf{H}^n$ = $selectHL(c^n,\mathbf{M}^n)$
\end{enumerate}

\newpage